\documentclass{SciPost}

\hypersetup{
    colorlinks,
    linkcolor={red!50!black},
    citecolor={blue!50!black},
    urlcolor={blue!80!black}
}
\usepackage{graphicx} 
\usepackage{amssymb}
\usepackage{amsmath}
\usepackage{amsbsy}
\usepackage{mathtools}
\usepackage{mathrsfs}
\usepackage{mathtools}
\usepackage{mathrsfs}
\usepackage[bitstream-charter]{mathdesign}
\usepackage{listings}
\usepackage{xcolor}
\usepackage[most]{tcolorbox}

\definecolor{commandcolor}{HTML}{2356c4}
\definecolor{filecolor}{HTML}{942831}
\definecolor{outcolor}{HTML}{8c898a}

\newcommand{\ket}[1]{\left\vert #1 \right\rangle}
\newcommand{\bra}[1]{\left\langle #1 \right\vert}
\newcommand{\Tr}{\mathrm{tr}}
\newcommand{\mathcomm}[1]{{\color{commandcolor} #1}}

\newcommand{\mathttt}[1]{\texttt{\color{commandcolor} #1}}
\newcommand{\filettt}[1]{\texttt{\color{filecolor} #1}}

\newenvironment{mathematicain}{\begin{quote}\color{commandcolor}}{\end{quote}}
\newenvironment{mathematicaout}{\begin{quote}\color{outcolor}}{\end{quote}}

\DeclareSymbolFont{usualmathcal}{OMS}{cmsy}{m}{n}
\DeclareSymbolFontAlphabet{\mathcal}{usualmathcal}

\fancypagestyle{SPstyle}{
\fancyhf{}
\lhead{\colorbox{scipostblue}{\bf \color{white} ~arXiv }}
\rhead{{\bf \color{scipostdeepblue}  }}

\fancyfoot[C]{\textbf{\thepage}}
}

\begin{document}

\pagestyle{SPstyle}

\begin{center}{\Large \textbf{\color{scipostdeepblue}{
MultiAtomLiouvilleEquationGenerator: A Mathematica package for
Liouville superoperators and master equations of multilevel atomic systems
\\
}}}\end{center}

\begin{center}\textbf{
Pablo Yanes-Thomas\textsuperscript{1$\star$},
Roc\'io J\'auregui\textsuperscript{1},
Santiago F. Caballero-Ben\'\i tez\textsuperscript{3},
Daniel Sahag\'un-S\'anchez\textsuperscript{1},
Alejandro Kunold\textsuperscript{2$\dagger$}
}\end{center}

\begin{center}
{\bf 1} Departamento de F\'\i sica Cu\'antica y Fot\'onica, Instituto de F\'isica, Universidad Nacional Aut\'onoma de M\'exico,
Ciudad de M\'exico 04510,  M\'exico
\\
{\bf 2} \'Area de F\'isica Te\'orica y
Materia Condensada, Universidad Aut\'onoma Metropolitana
Azcapotzalco, Av. San Pablo 180, Col. Reynosa-Tamaulipas,
02200 Cuidad de M\'exico, M\'exico
\\
{\bf 3} Departamento de F\'\i sica Cu\'antica y Fot\'onica, Instituto de F\'isica, LSCSC-LANMAC, Universidad Nacional Aut\'onoma de M\'exico,
Ciudad de M\'exico 04510, M\'exico
\\[\baselineskip]
$\star$ \href{mailto:email1}{\small akb@azc.uam.mx}\,,\quad
$\dagger$ \href{mailto:email2}{\small astralrequiem@gmail.com}
\end{center}

\section*{\color{scipostdeepblue}{Abstract}}
\textbf{\boldmath{%
MulAtoLEG (Multi-Atom Liouville Equation Generator)
is an open-source Mathematica package for generating
Liouville superoperators and Liouville equations, specialized
for multilevel atomic systems comprising an arbitrary number of atoms.
This scheme is based on an extension to multilevel atomic systems,
originally developed by Lehmberg
[R. H. Lehmberg, Phys. Rev. A 2, 883 (1970)]
as an adjoint master equation
for ensembles of two-level emitters
and later reformulated by Genes
[M. Reitz, C. Sommer and C. Genes, PRX Quantum 3, 010201 (2022)]
as a master equation.
The package facilitates the generation of equations for complex
transition configurations in alkali atoms.
Although primarily designed for atomic systems, it can also generate
the master and adjoint master equations for general Hamiltonians
and Lindbladians. In addition, it includes functionalities to construct
the differential equations in the dressed-state basis, where, in many cases,
the non-unitary evolution operator can be determined explicitly.
To maximize computational efficiency, the package leverages Mathematica's
vectorization and sparse linear algebra capabilities.
Since MulAtoLEG produces exact equations without approximations,
the feasible system size is naturally limited by the available
computational resources.
}}

\vspace{\baselineskip}



\vspace{10pt}
\noindent\rule{\textwidth}{1pt}
\tableofcontents
\noindent\rule{\textwidth}{1pt}
\vspace{10pt}


\section{Introduction}

Atoms and molecules naturally interact with electromagnetic fields whose sources have been available for about half a century. 
The technological improvements of these sources allowed experimental access to the  internal and external degrees of freedom of atoms by manipulating their electronic configuration. As a result, during the last couple of decades, atomic physics has yielded benchmarking research that provides the foundations of a new era in which quantum states can be prepared in laboratories worldwide. The scientific and technological relevance of these advances has been recognized by multiple Nobel prizes in physics for the field since the late 90s\footnote{ 1997: Chu, Cohen-Tanoudji,Phillips; 2001: Cornell, Ketterle, Wieman; 2005: Hall, Hänsch-, Glauber; 2012: Haroche, Wineland; 2018: Ashkin, Mourou, Strickland; 2022: Aspect, Clauser, Zeilenger; 2023: Agostini, Krausz, L'Huillier.}.

When an atom is excited the change in its electronic structure generates an electric dipole. Dipole-dipole interactions in an ensemble of atoms coupled to near-resonant electromagnetic fields give rise to cooperative, quantum interference phenomena such as superradiance, subradiance and frequency shifts \cite{PhysRevA.2.889,PhysRev.181.32,Gross_1982}. Such interactions become important when the characteristic distances of the system are comparable with the wavelength of the interacting fields. These distances have become more controllable throughout the years. In systems made of hot, cold or ultracold dilute gases the atomic position is not controllable. In most of the research carried out with such systems the characteristic distances have been large when compared to the interaction wavelength. Keeping the regime where cooperative effects are dominant practically unexplored.

Newer techniques, namely optical tweezers and lattices, are precise enough to individually manipulate atomic enselbes ranging from a single atom to a few thousand atoms at a time \cite{RevModPhys.90.031002,kaufman2021}. They have thus opened many experimental possibilities that have been increasing in complexity as technology advances. For example, in a seminal experiment  Darquie \textit{et al.} (2005) were able to control single-photon emission from a single two-level atom held in a tight optical tweezer \cite{Darquie:2005fr}; in 2022 emissions of Greenberger–Horne–Zeilinger (GHZ) single photons were induced through four-wave mixing (FWM) in single atoms confined in an optical lattice inside a high-finesse cavity \cite{Thomas2022}. The atom-light interfaces have become technically more sophisticated and the amount of optical frequencies simultaneously interacting with atoms have increased.     

With current technology it is possible to control the atomic position precisely enough to observe cooperative effects. Lamb shifts have been observed in nano-spectroscopy cells with atom-number densities much less than one inside the volume of interaction with light \cite{keaveney2012}. The fundamental dipole-dipole interaction has been observed and characterized for two Rydberg atoms held by optical tweezers \cite{chew2022,Mehaignerie2025}. With the current control of atomic positioning it is possible to create 2D and 3D atomic, or molecular, arrays with up to 2024 sites whose positions are dynamically controllable \cite{kim2016,lin2025}.

Atomic systems remain one of the pillars for the roadmap of quantum technologies \cite{Acin_2018}. In 2000 a seminal theoretical work identified atoms and coherent light as a plausible platform for entanglement distribution as required for quantum telecommunications \cite{Duan2000}. It was followed by the first proposal for a quantum repeater, also based in atomic ensembles \cite{Duan2001}. The later triggered research in a range of directions aimed at matching the requirements from photon pair sources and quantum memories \cite{sangouard2011}. Arrays of single atoms seem to be the dark horse of quantum computing and simulation with novel protocols based on optical tweezers and lattices, and manipulation of Rydberg states therein \cite{Henriet2020,cohen2021}. Atoms of the same element are identical, therefore all mentioned experimental techniques deliver data with an extremely high reproducibility. This makes these systems ideal for quantum control, which is dedicated to optimizing state preparation through electromagnetic fields \cite{Koch2022}.        

As complexity in the description needed for the development of quantum technology increases, the paradigm beyond unitary dynamics needs to be considered. There are useful formulations to solve time dependent unitary protocols, i.e. Wei-Norman methods\cite{WN1,WN2,WN3} usually applied to quantum optics problems for idealized quantum systems and few degrees of freedom. However, as one delves in the many-body problem more general tools are needed, specially when dealing with the problem of dissipation in realistic quantum systems for quantum technologies. Therefore, one needs to consider the open system dynamics and the many-body problem together. In the typical scenario of open quantum systems\cite{QO1,QO2}, for few body problems, requires  the verification of predictions that go beyond analytical methods and necessitates efficient computational tools. Exploring further as the quantum system grows in complexity motivates the use of efficient computational methods in order to develop new, effective theories, create optimized approximation schemes and possibly propose new means of simulation with reasonable computational resources both in terms of execution time and the classical representation of the Hilbert space of the problem, as this presents a memory bottleneck.

While the relevance of atomic systems continues to expand to a wider range of quantum technologies, their theoretical description has simultaneously become more intricate and computationally demanding. Modern quantum technologies frequently rely on multilevel atomic systems which feature complex interactions, including multiple decay channels and driving by several coherent fields. These characteristics make them ideal candidatesfor the open quantum systems approach.However, even for systems consisting of just a few atoms with a modest number of internal
levels, the construction of the necessary mathematical tools-specifically, Hamiltonians,
jump operators, and, most importantly,the master equation-becomes achallenging and time-consuming task.
This rising complexity has motivated us to create computational tools to systematically
generate the master equations that govern the dynamics of atomic and other relevant quantum systems.

In this paper, we introduce the \mathttt{MultiAtomLiouvilleEquationGenerator} (\mathttt{MulAtoLEG}).
\mathttt{MulAtoLEG} is a Mathematica\cite{Mathematica} package designed to obtain 
the exact master and adjoint master equations for multilevel
atomic systems with an arbitrary number of atoms, based on an extension to multilevel atoms of the adjoint master equation for two-level multi-atom systems originally derived by Lehmberg \cite{PhysRevA.2.883} and on the master equation formulated by Genes \cite{PRXQuantum.3.010201}. We now allow the possibility that the dipoles
related to different transitions interact among each other. Even though the extended Lehmberg model has appeared in previous works \cite{PhysRevResearch.7.013028}, for clarity 
we present here a thorough and rigorous exposition that elucidates
its foundations and how the present program broadens its range of applicability
with respect to previous analysis.
Efforts have been made to simplify the usage of the package
so that users only need to provide the basic multi level structure,
the allowed transitions and the external fields.

The package includes functionalities that simplify the generation of equations for complex transition configurations in alkali atoms. Although \mathttt{MulAtoLEG} is primarily aimed at atomic systems,
it is also capable of generating master and adjoint master equations
for arbitrary Hamiltonians and Lindbladians.
Additionally, it provides tools to construct differential equations
in the dressed-state basis, which, in many cases, allow one to compute
the non-unitary evolution operator explicitly.
The design philosophy of \mathttt{MulAtoLEG} is to exploit
Mathematica’s vectorization and sparse linear algebra capabilities
to optimize the use of computational resources. Since the package produces full equations without approximations,
the achievable system size is, in practice, limited by the available computational resources.

The paper is organized as follows. In Section \ref{sec:model}, we develop
an extension of the adjoint master equation originally derived by
Lehmberg \cite{PhysRevA.2.883, PhysRevA.2.889}
to the case of multilevel atoms.
The corresponding master equation is also derived.
The special case in which lasers or other coherent
sources are introduced is discussed in Section~\ref{sec:lasers}.
Readers interested exclusively in constructing the master equation
for an atomic system driven by lasers or coherent sources may
skip the preceding sections and proceed directly to this one.
As mentioned earlier, \mathttt{MulAtoLEG} is not limited to
constructing master equations for atomic systems,
but can also be used to generate the most general
form of Lindblad master equations.
In Section~\ref{sec:mastered}, we present the general
master and adjoint master equations that can be solved using \mathttt{MulAtoLEG}.
The method used to vectorize these equations into the Liouville space
through a matrix-basis decomposition is described in Section \ref{sec:solmastereq}.
Section \ref{sec:rotframe} is devoted to the rotating-frame
quantum state basis and the corresponding unitary transformations
in both the standard and Liouville spaces.
These transformations remove the time dependence from the Liouville
equations, yielding the explicit form of the evolution operator.
The expressions for the far-field emission are obtained in Section \ref{sec:farfield}.
The design principles underlying \mathttt{MulAtoLEG} are described
in Section~\ref{sec:desprin}. Installation instructions are provided in
Section~\ref{sec:installation}. The general workflow for specifying configuration
variables and generating master equations is explained in Section~\ref{sec:usage}. 
Section~\ref{sec:comparison} is devoted to a comparison
of \mathttt{MulAtoLEG} with related Python and Mathematica packages.
Finally, seven examples illustrating most of the
\mathttt{MulAtoLEG} functions are presented in
Section \ref{sec:examples},
and conclusions are given in Section \ref{sec:conclusions}.

\section{General model for atomic systems}\label{sec:model}
The purpose of this section is to establish the theoretical framework
from which the master equations for systems of multilevel emitters
considered in this work are derived.
Readers whose primary interest lies in the automated generation and
solution of these equations may safely skip this section without loss
of continuity and proceed directly to the implementation-oriented
discussion.

Building on the adjoint master equation originally derived by
Lehmberg \cite{PhysRevA.2.883, PhysRevA.2.889} and subsequently
reformulated in the density-matrix picture by Genes \cite{Genes2022},
we extend the formalism to systems of multilevel emitters.
The present treatment is closely related to the approach
developed in our previous work \cite{PhysRevResearch.7.013028}
on driven multilevel atoms
and collective light–matter interactions, on which several aspects
of the modeling strategy are based. For the sake of completeness
and to make the present paper self-contained,
we briefly re-derive the underlying model and key equations here,
while adapting them to the broader class of systems considered in this work.
At the same time, the present study goes beyond
Ref. \cite{PhysRevResearch.7.013028} by introducing new
calculations and extensions that are specific to the problems
addressed below, which have not been presented elsewhere.

We analyze an ensemble of $n_a$ identical, non-overlapping emitters
interacting with an external light source
and with each other through dipole-dipole electromagnetic coupling.
The emitters either have
fixed positions $\boldsymbol{r}_1$, $\boldsymbol{r}_2$, $\dots ,$
$\boldsymbol{r}_{n_a}$ or move slowly enough
that no appreciable changes are observed in
their quantum levels.
Each emitter is assumed to have $n_l$ levels with energies
$\varepsilon_1$, $\varepsilon_2$, $\dots ,$ $\varepsilon_{n_l}$.
The standard lowering operators $\sigma_{i,\alpha}$
mediate the transitions between the energy levels of the $\alpha$-th
emitter.
The index $i$ runs over all available transitions between
ordered pairs of energy levels, taking the form
$i=\{i_1,i_2\}$.

The lowering atomic operator,
associated to the transition energy
$\Delta_i=\varepsilon_{i_2}-\varepsilon_{i_1}$,
may be defined as
\begin{equation}
\sigma_{i, \alpha} 
= \ket{i_1}_\alpha \prescript{}{\alpha}{\bra{i_2}}.
\end{equation}
The assumption of no atomic overlap
implies that
\begin{equation}
\left[\sigma_{i,\alpha}, \sigma_{j,\beta}\right]=
\left[\sigma_{i,\alpha}^\dagger, \sigma_{j,\beta}\right]=
\left[\sigma_{i,\alpha}^\dagger, \sigma_{j,\beta}^\dagger\right]=0
\end{equation}
if $\alpha \ne \beta$.

The electromagnetic field operator is written as
\begin{equation}
\boldsymbol{E}(\boldsymbol{r})=
\boldsymbol{E}^{+}(\boldsymbol{r})+\boldsymbol{E}^{-}(\boldsymbol{r}),
\end{equation}
where $\boldsymbol{E}^{\pm}(\boldsymbol{r})$ denote the positive- and
negative-frequency components, respectively.
In the present treatment, $\boldsymbol{E}(\boldsymbol{r})$ represents the total field,
 which contains both the incident (free) electromagnetic field and the field scattered by the emitters.
These two contributions are not
distinguished explicitly until a subsequent analysis.
The positive-frequency part of the total field can be modeled, as usual, by a plane-wave
expansion of the quantized field operator,
\begin{equation}
\boldsymbol{E}^+(\boldsymbol{r})
=\sum_{q,\epsilon}\mathcal{E}_{q,\epsilon}
\boldsymbol{e}_{q,\epsilon}
\mathrm{e}^{i\boldsymbol{k}_q\cdot \boldsymbol{r}}
a_{q,\epsilon},
\label{eq:posfreqfield0}
\end{equation}
Here
$\mathcal{E}_{q,\epsilon}
\sqrt{\hbar\omega_q/2\epsilon_0 V}$
is the single-photon field amplitude and $V$ is the quantization volume.
with the negative-frequency part given by
$\boldsymbol{E}^-(\boldsymbol{r})
=[\boldsymbol{E}^+(\boldsymbol{r})]^\dagger$.
Within the input–output or source-field formalism, this operator implicitly includes
both the free (incident) field and the field radiated by the atomic dipoles.

The plane waves are characterized by an
angular frequency $\omega_q$, a wave vector $\boldsymbol{k}$
and the orthogonal polarization unit vectors $\boldsymbol{e}_{q,\epsilon}^*$
where $\epsilon$ labels the two orthogonal polarization directions.
The annihilation and creation operators,
$a_{q,\epsilon}$ and $a_{q,\epsilon}^\dagger$ respectively,
follow the standard commutation relation
$[a_{q,\epsilon},a_{q^\prime, \epsilon^\prime}^\dagger]=
\delta_{q,q^\prime}\delta_{\epsilon,\epsilon^\prime}$.

In the dipole approximation the interaction of the emitters
with a quantized photon field is described by the Hamiltonian
\begin{equation}
H=H_S
+H_P
+H_{SP},
\end{equation}
where the unperturbed emitter and photon Hamiltonians are
\begin{eqnarray}
H_S &=& \sum_{l=1}^{n_l}\sum_{\alpha=1}^{n_a}
\hbar \varepsilon_l\ket{l}_\alpha \prescript{}{\alpha}{\bra{l}},
\label{eq:freehamatom}\\
H_P &=& \sum_q\hbar \omega_q a_{q,\epsilon}^\dagger a_{q,\epsilon},
\end{eqnarray}
respectively,
and the light-matter interaction Hamiltonian is
\begin{equation}
H_{SP}
= -\sum_{i=1}^{n_l}\sum_{\alpha=1}^{n_a}\sum_{q,\epsilon}
 \mathcal{E}_{q,\epsilon}
\left(
\boldsymbol{e}_{q,\epsilon}
\mathrm{e}^{i\boldsymbol{k}_{q}\cdot\boldsymbol{r}_\alpha}
a_{q,\epsilon}
+\boldsymbol{e}_{q,\epsilon}^*
\mathrm{e}^{-i\boldsymbol{k}_{q}\cdot\boldsymbol{r}_\alpha}
a_{q,\epsilon}^\dagger\right)
\cdot
\left(\boldsymbol{p}_i\sigma_{i,\alpha}
+\boldsymbol{p}_i^*\sigma_{i,\alpha}^\dagger\right),
\end{equation}
where
$\boldsymbol{p_{i}}=\left\langle i_1 \left\vert
e\boldsymbol{x}\right\vert i_2 \right\rangle$
is the $i$-th transition
matrix element of the atomic dipole operator.
Defining
\begin{equation}
\boldsymbol{P}_{i,\alpha}
=\boldsymbol{p}_i\sigma_{i,\alpha}
+\boldsymbol{p}_i^*\sigma_{i,\alpha}^\dagger 
\end{equation}
the  Hamiltonian takes the form
\begin{equation}
H = H_S+H_P
 -\sum_{i}\sum_{\alpha}\sum_{q,\epsilon}
 \mathcal{E}_{q,\epsilon}
\left(
\boldsymbol{e}_{q,\epsilon}
\mathrm{e}^{i\boldsymbol{k}_{q}\cdot\boldsymbol{r}_\alpha}
a_{q,\epsilon}
+\boldsymbol{e}_{q,\epsilon}^*
\mathrm{e}^{-i\boldsymbol{k}_{q}\cdot\boldsymbol{r}_\alpha}
a_{q,\epsilon}^\dagger\right)
\cdot \boldsymbol{P}_{i,\alpha}.
\end{equation}
The dynamical
equation for the photon annihilation operator
takes the form
\begin{equation}
\dot{a}_{q,\epsilon}(t)=\frac{i}{\hbar}\left[H(t), a_q(t)\right]=
-i\omega_q a_{q,\epsilon}(t)
-\frac{i}{\hbar}\sum_{i}\sum_{\alpha}
\mathcal{E}_{q,\epsilon}
\mathrm{e}^{-i\boldsymbol{k}_{q}\cdot\boldsymbol{r}_\alpha}
\boldsymbol{e}_q^*\cdot\boldsymbol{P}_{i,\alpha}(t),\label{eq:aqdifeq}
\end{equation}
where the time-dependence
indicate that the operators are expressed in the
Heisenberg picture. In this picture,
$a_{q,\epsilon}(t)=U_H^\dagger(t) a_{q,\epsilon}U_H(t)$
where $i\hbar \partial_tU_H(t)=H(t)U_H(t)$.
The formal solution of \eqref{eq:aqdifeq} is
\begin{equation}
a_{q,\epsilon}(t)
=a_{q,\epsilon}(0)\mathrm{e}^{-i\omega_q t}
-\frac{i}{\hbar}\sum_{i}\sum_{\alpha}
\mathcal{E}_{q,\epsilon}
\mathrm{e}^{-i\boldsymbol{k}_{q}\cdot\boldsymbol{r}_\alpha}
\int_0^t ds \mathrm{e}^{i\omega_q (s-t)}
\boldsymbol{e}_{q,\epsilon}^*\cdot\boldsymbol{P}_{i,\alpha}(s).
\label{eq:aq1}
\end{equation}
Similarly, the dynamical equation
for any  purely atomic Hermitian operator $Q$ in the
Heisenberg picture is
\begin{equation}
\dot{Q}_H(t)=\frac{i}{\hbar}\left[H(t),Q_H(t)\right].
\label{eq:Qdot}
\end{equation}
Substituting the explicit forms of $a_{q,\epsilon}(t)$
and $a_{q,\epsilon}^\dagger(t)$ from  \eqref{eq:aq1}
into \eqref{eq:Qdot} we obtain
\begin{multline}
\dot{Q}_H(t)=
\frac{i}{\hbar}
    \left[H_S(t)+H_E(t),Q_H(t)\right]\\
+\frac{1}{\hbar^2}\sum_{i,j}\sum_{\alpha,\beta}\sum_{q,\epsilon}
\mathcal{E}_{q,\epsilon}^2
 \bigg\{\mathrm{e}^{i\boldsymbol{k}_{q}\cdot\boldsymbol{r}_{\alpha,\beta}}
 \int_0^t ds \mathrm{e}^{i\omega_q (s-t)}
\boldsymbol{e}_{q,\epsilon}\cdot
\left[\boldsymbol{P}_{i,\alpha}(t),Q_H(t)\right]
\boldsymbol{e}_{q,\epsilon}^*\cdot \boldsymbol{P}_{j,\beta}(s)\\
-\mathrm{e}^{-i\boldsymbol{k}_{q}\cdot\boldsymbol{r}_{\alpha,\beta}}
 \int_0^t ds \mathrm{e}^{-i\omega_q (s-t)}
\boldsymbol{e}_{q,\epsilon}\cdot \boldsymbol{P}_{j,\beta}(s)
\left[Q_H(t),\boldsymbol{P}_{i,\alpha}(t)\right]
\cdot\boldsymbol{e}_{q,\epsilon}^*
\bigg\},
\label{eq:Qdot2}
\end{multline}
where $\boldsymbol{r}_{\alpha,\beta}
=\boldsymbol{r}_{\alpha}-\boldsymbol{r}_{\beta}$ and
\begin{equation}
H_{E}(t)=-\sum_\alpha\sum_{i}
    \left[\boldsymbol{E}_0^+(\boldsymbol{r}_\alpha,t)
    + \boldsymbol{E}^-_0(\boldsymbol{r}_\alpha,t)
    \right]\cdot \boldsymbol{P}_{i,\alpha}(t)
\end{equation}
describes the coupling of the emitters to the incident (free)
electromagnetic field, represented by the
positive- and negative-frequency components
\begin{equation}
\boldsymbol{E}_0^+(\boldsymbol{r}_\alpha,t) = \sum_{q,\epsilon}
\mathcal{E}_{q,\epsilon}
\boldsymbol{e}_{q, \epsilon}
\mathrm{e}^{i\left(
  \boldsymbol{k}_{q}\cdot\boldsymbol{r}_\alpha
  -i\omega_q t
\right)}a_{q,\epsilon}(0),
\end{equation}
and $\boldsymbol{E}_0^-(\boldsymbol{r}_\alpha,t)
=[\boldsymbol{E}_0^+(\boldsymbol{r}_\alpha,t)]^\dagger$.
In contrast, the time-nonlocal terms in Eq.~\eqref{eq:Qdot2},
involving integrals over the emitter operators at earlier times $s<t$,
arise from the field radiated by the emitters themselves and account
for the scattered (or self-generated) electromagnetic field.
These contributions mediate both self-interaction effects and
photon-exchange processes between distinct emitters.

To put this equation into a more useful form,
we move into the continuous limit of the
photon spectrum by taking the limits $V\rightarrow \infty$,
$\boldsymbol{k}_q\rightarrow \boldsymbol{k}$ and
$\boldsymbol{e}_{q,\epsilon}\rightarrow \boldsymbol{e}_{\boldsymbol{\kappa},\epsilon}$,
so that
\begin{equation}
\frac{1}{V}\sum_{q,\epsilon} 
\rightarrow \int \frac{dk^3}{(2\pi)^3}\sum_{\epsilon}
=\frac{1}{(2\pi c)^3}\int d\omega \, \omega^2
\int d\Omega_{\boldsymbol{\kappa}}\sum_{\epsilon}\,\,\,,
\end{equation}
where $\boldsymbol{\kappa}=\boldsymbol{k}/|\boldsymbol{k}|$,
$d\Omega_{\boldsymbol{\kappa}}$ is the differential solid
angle around the origin of the
wave vector space, $V/(2\pi)^3$ is the density of optical modes
inside $V$ and the sum over $\epsilon$
accounts for the two orthogonal polarizations.
In this limit, Eq. \eqref{eq:Qdot2} takes the form
\begin{multline}
\dot{Q}_H(t)=
\frac{i}{\hbar}
    \left[H_S(t)+H_E(t),Q_H(t)\right]
+\frac{1}{16\pi^3c^3\epsilon_0\hbar}\sum_{i,j}\sum_{\alpha,\beta}
\int d\omega \omega^3\\
\times\int d\Omega_{\boldsymbol{\kappa}}\sum_{\epsilon}
\bigg\{\mathrm{e}^{i\boldsymbol{k}\cdot\boldsymbol{r}_{\alpha,\beta}}
\int_0^t ds \mathrm{e}^{i\omega_q (s-t)}
\boldsymbol{e}_{\boldsymbol{\kappa},\epsilon}\cdot
\left[\boldsymbol{P}_{i,\alpha}(t),Q_H(t)\right]\,\,
\boldsymbol{e}_{\boldsymbol{\kappa},\epsilon}^*\cdot \boldsymbol{P}_{j,\beta}(s)\\
+\mathrm{e}^{-i\boldsymbol{k}\cdot\boldsymbol{r}_{\alpha,\beta}}
 \int_0^t ds \mathrm{e}^{-i\omega_q (s-t)}
\boldsymbol{e}_{\boldsymbol{\kappa},\epsilon}\cdot \boldsymbol{P}_{j,\beta}(s)
\,\,\boldsymbol{e}_{\boldsymbol{\kappa},\epsilon}^*\cdot
\left[\boldsymbol{P}_{i,\alpha}(t),Q_H(t)\right]
\bigg\},
\label{eq:Qdot3}
\end{multline}

The dipole approximation ceases to be valid
as $\omega$ approaches $c/a_B$ where
$a_B$  is the Bohr radius.
Thus, in the evaluation of the time integral 
in \eqref{eq:Qdot3}, the most significant contributions
are those within the region on the order $a_B/c$
around $t-\boldsymbol{\kappa}\cdot \boldsymbol{r}_{\alpha,\beta}/c$.
We assume that we are in the Markovian regime since the
time scales defined by the
 energy separation
between any two states
and the typical size of an atom satisfy
$\hbar\Delta_j=\varepsilon_{j_2}-\varepsilon_{j_1}\ll \hbar a_B/c$.
We can then make the approximation
\begin{equation}
\sigma_{j,\beta}(s)\approx 
\sigma_{j,\beta}
\left(t-\boldsymbol{\kappa}\cdot \boldsymbol{r}_{\alpha,\beta}/c\right)\\
\exp\left[-i\Delta_j
\left(s -t+\boldsymbol{\kappa}\cdot \boldsymbol{r}_{\alpha,\beta}/c\right)
\right].\label{eq:markov1}
\end{equation}
Moreover, retardation effects can be neglected if
the ensemble is situated in a space small enough
as to be traversed by photons much faster
than changes can occur in the quantum levels
of emitters then $\max\left(r_{\alpha,\beta}\right)\ll c\Delta t$,
so that
\begin{equation}
\sigma_{j,\beta}(s)\approx
\sigma_{j,\beta}(t)
\exp\left[-i\Delta_{j}\left(s-t\right)\right].\label{eq:markov2}
\end{equation}

Substituting Eq. \eqref{eq:markov2} into Eq. \eqref{eq:Qdot3},
eliminating rapidly oscillating terms and summing over the
polarization index $\epsilon$,
the second line of Eq. \eqref{eq:Qdot3} takes the form
\begin{multline}
\int \omega^3d\omega
\int d\Omega_{\boldsymbol{\kappa}}\sum_{\epsilon}
\mathrm{e}^{i\boldsymbol{k}\cdot\boldsymbol{r}_{\alpha,\beta}}
\int_0^t ds \mathrm{e}^{i\omega_q (s-t)}
\boldsymbol{e}_{\boldsymbol{\kappa},\epsilon}\cdot
\left[\boldsymbol{P}_{i,\alpha}(t),Q_H(t)\right]
\boldsymbol{e}_{\boldsymbol{\kappa},\epsilon}^*
\cdot \boldsymbol{P}_{j,\beta}(s)\\
=\int \omega^3d\omega
\int d\Omega_{\boldsymbol{\kappa}}\\
\times\bigg\{
\left[\boldsymbol{p}_i\cdot\boldsymbol{p}_j^*
-\left(\kappa\cdot \boldsymbol{p}_i\right)
\left(\kappa\cdot \boldsymbol{p}_j^*\right)\right]
\mathrm{e}^{i\boldsymbol{k}\cdot\boldsymbol{r}_{\alpha,\beta}}
\int_0^t ds \mathrm{e}^{i(\omega+\Delta_j) (s-t)}
\left[\sigma_{i,\alpha}(t),Q_H(t)\right]\sigma_{j,\beta}^\dagger(t)
\\
+\left[\boldsymbol{p}_i^*\cdot\boldsymbol{p}_j
-\left(\kappa\cdot \boldsymbol{p}_i^*\right)
\left(\kappa\cdot \boldsymbol{p}_j\right)\right]
\mathrm{e}^{i\boldsymbol{k}\cdot\boldsymbol{r}_{\alpha,\beta}}
\int_0^t ds \mathrm{e}^{i(\omega-\Delta_j) (s-t)}
\left[\sigma_{i,\alpha}^\dagger(t),Q_H(t)\right]\sigma_{j,\beta}(t)
\bigg\},
\end{multline}
where the dipole matrix elements have been expanded as
\begin{equation}
\boldsymbol{p}_{i}
=\sum_\epsilon
\left(\boldsymbol{e}_\epsilon^*\cdot \boldsymbol{p}_{i}\right)
\boldsymbol{e}_\epsilon
+\left(\boldsymbol{\kappa}\cdot \boldsymbol{p}_{i}\right)
\boldsymbol{\kappa}.
\end{equation}

Performing the integrals over
the solid angle $d\Omega_{\boldsymbol{\kappa}}$
and over time $s$ we are left with
\begin{multline}
\int_0^\infty \omega^3d\omega
\int d\Omega_{\boldsymbol{\kappa}}\sum_{\epsilon}
\mathrm{e}^{i\boldsymbol{k}\cdot\boldsymbol{r}_{\alpha,\beta}}
\int_0^t ds \mathrm{e}^{i\omega_q (s-t)}
\boldsymbol{e}_{\boldsymbol{\kappa},\epsilon}\cdot
\left[\boldsymbol{P}_{i,\alpha}(t),Q_H(t)\right]
\boldsymbol{e}_{\boldsymbol{\kappa},\epsilon}^*
\cdot \boldsymbol{P}_{j,\beta}(s)\\
=i\frac{8\pi}{3} \int_0^\infty \omega^3d\omega
W_{i,j}\left(k r_{\alpha,\beta}\right)
\frac{\mathrm{e}^{-i(\omega+\Delta_j)t}-1}{\omega+\Delta_j}
\left[\sigma_{i,\alpha}(t),Q_H(t)\right]\sigma_{j,\beta}^\dagger(t)\\
i\frac{8\pi}{3} \int_0^\infty \omega^3d\omega
W_{i,j}^*\left(k r_{\alpha,\beta}\right)
\frac{\mathrm{e}^{-i(\omega-\Delta_j)t}-1}{\omega-\Delta_j} 
\left[\sigma_{i,\alpha}^\dagger(t),Q_H(t)\right]\sigma_{j,\beta}(t)
\end{multline}
and
\begin{multline}
W_{i,j}\left(\zeta\right)
=\frac{3}{2}\left[\boldsymbol{p}_i\cdot \boldsymbol{p}_j^*
-\left(\boldsymbol{p}_i\cdot \hat{\boldsymbol{r}}_{\alpha,\beta}\right)
\left(\boldsymbol{p}_j^*\cdot \hat{\boldsymbol{r}}_{\alpha,\beta}\right)
\right]\frac{\sin\zeta}{\zeta}\\
+\frac{3}{2}\left[\boldsymbol{p}_i\cdot \boldsymbol{p}_j^*
-3\left(\boldsymbol{p}_i\cdot \hat{\boldsymbol{r}}_{\alpha,\beta}\right)
\left(\boldsymbol{p}_j^*\cdot \hat{\boldsymbol{r}}_{\alpha,\beta}\right)
\right]\left(\frac{\cos\zeta}{\zeta^2}-\frac{\sin\zeta}{\zeta^3}\right).
\end{multline}

Integrating over $\omega$ in the previous expression, we obtain
\begin{multline}
\int_0^\infty \omega^3d\omega
\int d\Omega_{\boldsymbol{\kappa}}\sum_{\epsilon}
\mathrm{e}^{i\boldsymbol{k}\cdot\boldsymbol{r}_{\alpha,\beta}}
\int_0^t ds \mathrm{e}^{i\omega_q (s-t)}
\boldsymbol{e}_{\boldsymbol{\kappa},\epsilon}\cdot
\left[\boldsymbol{P}_{i,\alpha}(t),Q_H(t)\right]
\boldsymbol{e}_{\boldsymbol{\kappa},\epsilon}^*
\cdot \boldsymbol{P}_{j,\beta}(s)\\
=-i\frac{8\pi c^3}{3}
\mathcal{P}\int_0^\infty \frac{k^3}{k+\kappa_j}
W_{i,j}\left(k r_{\alpha,\beta}\right)
dk\,\,\,
\left[\sigma_{i,\alpha}(t),Q_H(t)\right]\sigma_{j,\beta}^\dagger(t)\\
-i\frac{8\pi c^3}{3}
\mathcal{P}
\int_0^\infty \frac{k^3}{k-\kappa_j}
W_{i,j}^*\left(k r_{\alpha,\beta}\right)
dk\,\,\,
\left[\sigma_{i,\alpha}^\dagger(t),Q_H(t)\right]\sigma_{j,\beta}(t)
\\
+\frac{8\pi^2}{3} \Delta_j^3
W_{i,j}^*\left(\kappa_j r_{\alpha,\beta}\right) 
\left[\sigma_{i,\alpha}^\dagger(t),Q_H(t)\right]\sigma_{j,\beta}(t).
\label{eq:Qdot4}
\end{multline}
where $\kappa_j=\Delta_j/c$.
When integrating over $\omega$
we have used the Sokhotski–Plemelj theorem
\begin{equation}
\lim_{\omega_0\rightarrow 0^+}
 \frac{1}{\omega \pm \omega_0}
=\mp i\pi \delta\left(\omega\right)
+\mathcal{P} \left(\frac{1}{\omega}\right),
\end{equation}
where $\mathcal{P}$ denotes the Cauchy principal value.

Substituting \eqref{eq:Qdot4} and its Hermitian
conjugate into \eqref{eq:Qdot3}, we obtain
\begin{multline}
\dot{Q}_H(t)=
\frac{i}{\hbar}
    \left[H_S(t)+H_E(t),Q_H(t)\right]
\\
+\sum_{i,j}\sum_{\alpha\ne \beta}
\bigg\{-i\Omega_{i,j,\alpha,\beta}^+
\left[\sigma_{i,\alpha}(t),Q_H(t)\right]\sigma_{j,\beta}^\dagger(t)
-i\Omega_{i,j,\alpha,\beta}^{-*}
\left[\sigma_{i,\alpha}^\dagger(t),Q_H(t)\right]\sigma_{j,\beta}(t)
\\
-i\Omega_{i,j,\alpha,\beta}^{+*}
\sigma_{j,\beta}(t)\left[\sigma_{i,\alpha}^\dagger(t),Q_H(t)\right]
-i\Omega_{i,j,\alpha,\beta}^{-}
\sigma_{j,\beta}^\dagger(t)\left[\sigma_{i,\alpha}(t),Q_H(t)\right]
\bigg\}
\\
+\sum_{i,j}\sum_{\alpha, \beta}
\bigg\{
\frac{1}{2}\gamma_{i,j,\alpha,\beta}^*
\left[\sigma_{i,\alpha}^\dagger(t),Q_H(t)\right]\sigma_{j,\beta}(t)
-\frac{1}{2}\gamma_{i,j,\alpha,\beta}
\sigma_{j,\beta}^\dagger(t)\left[\sigma_{i,\alpha}(t),Q_H(t)\right]
\bigg\}.
\label{eq:Qdot5}
\end{multline}
where
\begin{eqnarray}
\Omega_{i,j,\alpha,\beta}^\pm &=& \frac{1}{6\pi^2\epsilon_0\hbar}
\mathcal{P}\int_0^\infty \frac{k^3}{k\pm\kappa_j}
W_{i,j}\left(k r_{\alpha,\beta}\right)dk,
\label{eq:defOmega1}\\
\gamma_{i,j,\alpha,\beta} &=& \frac{1}{6\pi\epsilon_0\hbar}
\kappa_j^3 W_{i,j}\left(\kappa_j r_{\alpha,\beta}\right).
\label{eq:defgamma1}
\end{eqnarray}
These are usually referred to as the real and imaginary parts of the
Green’s function that govern the propagation of the electromagnetic field.
However, in the most general case - where the dipole matrix elements
 are complex - they are themselves complex numbers. More specifically, when
$W_{i,j}(\zeta)$ couples two distinct transitions with dipole matrix elements
$\boldsymbol{p}_i$ and $\boldsymbol{p}_j$, it can also be complex.
This is accounted for by the code of the package presented here.
Note that we have restricted the summation of the terms proportional
to $\Omega_{i,j,\alpha,\beta}^\pm$ to $\alpha\ne\beta$. The terms
where $\alpha=\beta$ originally yield Lamb shifts of the enegy levels; however,
the dipole approximation breaks down for $\kappa_jr_{\alpha,\beta}=0$, making it 
impossible to calculate the Lamb shifts within this approximation.
We therefore assume that the Lamb shifts are already incorporated
in the level energies of Eq. \eqref{eq:freehamatom}.

For the terms in Eq. \eqref{eq:Qdot5} to have
a significant contribution, it is required that
$\Delta_i\approx \Delta_j$ or $\kappa_i\approx \kappa_j$.
In this case, we can replace $\kappa_j$ by
$\kappa_{i,j}=(\kappa_i+\kappa_j)/2$, resulting
$\Omega_{i,j,\alpha,\beta}^\pm\approx \Omega_{j,i,\alpha,\beta}^{\pm*}$ and
$\gamma_{i,j,\alpha,\beta}\approx \gamma_{j,i,\alpha,\beta}^*$.
Due to this, Eqs. \eqref{eq:defOmega1} and \eqref{eq:defgamma1}
can be approximated by
\begin{eqnarray}
\Omega_{i,j,\alpha,\beta}^\pm &=& \frac{1}{6\pi^2\epsilon_0\hbar}
\mathcal{P}\int_0^\infty \frac{k^3}{k\pm \kappa_{i,j}}
W_{i,j}\left(k r_{\alpha,\beta}\right)dk,
\label{eq:defOmega2}\\
\gamma_{i,j,\alpha,\beta} &=& \frac{1}{6\pi\epsilon_0\hbar}
\kappa_{i,j}^3 W_{i,j}\left(
\kappa_{i,j} r_{\alpha,\beta}\right),
\label{eq:defgamma2}
\end{eqnarray} 
rendering them completely symmetric.
This simplifies Eq. \eqref{eq:Qdot5} into
\begin{multline}
\dot{Q}_H(t)=
\frac{i}{\hbar}
    \left[H_S(t)+H_E(t),Q_H(t)\right]
\\
+\sum_{i,j,\kappa_i=\kappa_j}\sum_{\alpha\ne \beta}
\bigg\{-i\Omega_{i,j,\alpha,\beta}^+
\left[\sigma_{j, \beta}(t),Q_H(t)\right]\sigma_{i,\alpha}^\dagger(t)
-i\Omega_{i,j,\alpha,\beta}^{-}
\left[\sigma_{i,\alpha}^\dagger(t),Q_H(t)\right]\sigma_{j,\beta}(t)
\\
-i\Omega_{i,j,\alpha,\beta}^{+}
\sigma_{j,\beta}(t)\left[\sigma_{i,\alpha}^\dagger(t),Q_H(t)\right]
-i\Omega_{i,j,\alpha,\beta}^{-}
\sigma_{i,\alpha}^\dagger(t)\left[\sigma_{j,\beta}(t),Q_H(t)\right]
\bigg\}\\
+\sum_{i,j,\kappa_i=\kappa_j}\sum_{\alpha, \beta}\bigg\{
\frac{1}{2}\gamma_{i,j,\alpha,\beta}
\left[\sigma_{i,\alpha}^\dagger(t),Q_H(t)\right]\sigma_{j,\beta}(t)
-\frac{1}{2}\gamma_{i,j,\alpha,\beta}
\sigma_{i,\alpha}^\dagger(t)\left[\sigma_{j,\beta}(t),Q_H(t)\right]
\bigg\},
\label{eq:Qdot6}
\end{multline}
where we have discarded the rapidly
oscillating terms, retaining only those whose energy differences
are identical, i.e.,$\Delta_i=\Delta_j$
or $\kappa_i=\kappa_j$.

To put Eq. \eqref{eq:Qdot6} in the more familiar Lindblad form,
we use the commutator relations
\begin{eqnarray}
A\left[B, C\right] + \left[A, C\right]B
&=& \left[AB, C\right], \\
A\left[B, C\right] - \left[A, C\right]B
&=& \left\{AB, C\right\}-2ACB.
\end{eqnarray}
Eq. \eqref{eq:Qdot6} becomes
\begin{multline}
\dot{Q}_H(t)=
\frac{i}{\hbar}
    \left[H_S(t)+H_E(t),Q_H(t)\right]
-i\sum_{i,j,\kappa_i=\kappa_j}\sum_{\alpha\ne \beta}
\Omega_{i,j,\alpha,\beta}
\left[\sigma_{i,\alpha}^\dagger(t) \sigma_{j,\beta}(t), Q_H(t)\right]\\
-\frac{1}{2}\sum_{i,j,\kappa_i=\kappa_j}\sum_{\alpha,\beta}
\gamma_{i,j,\alpha,\beta}\left[
\left\{\sigma_{i,\alpha}^\dagger(t)\sigma_{j,\beta}(t), Q_H(t)\right\}
-2\sigma_{i,\alpha}^\dagger(t) Q_H(t)\sigma_{j,\beta}(t)
\right]
\label{eq:Qdot7}
\end{multline}
where
\begin{multline}
\Omega_{i,j,\alpha,\beta}=\Omega_{i,j,\alpha,\beta}^++\Omega_{i,j,\alpha,\beta}^-
=\frac{1}{6\pi^2\epsilon_0\hbar}
\mathcal{P}\int_{-\infty}^\infty \frac{k^3}{k+ \kappa_{i,j}}
W_{i,j}\left(k r_{\alpha,\beta}\right)dk\\
=\frac{1}{6\pi^2\epsilon_0\hbar}\kappa_{i,j}^3M_{i,j}(\kappa_{i,j}r_{\alpha,\beta})
\end{multline}
and
\begin{multline}
M_{i,j}(\zeta)=-\frac{3}{4}\left[\boldsymbol{p}_i\cdot \boldsymbol{p}_j^*
-\left(\boldsymbol{p}_i\cdot \hat{\boldsymbol{r}}_{\alpha,\beta}\right)
\left(\boldsymbol{p}_j^*\cdot \hat{\boldsymbol{r}}_{\alpha,\beta}\right)
\right]\frac{\cos\zeta}{\zeta}\\
+\frac{3}{4}\left[\boldsymbol{p}_i\cdot \boldsymbol{p}_j^*
-3\left(\boldsymbol{p}_i\cdot \hat{\boldsymbol{r}}_{\alpha,\beta}\right)
\left(\boldsymbol{p}_j^*\cdot \hat{\boldsymbol{r}}_{\alpha,\beta}\right)
\right]\left(\frac{\sin\zeta}{\zeta^2}-\frac{\cos\zeta}{\zeta^3}\right).
\end{multline}
We can define the dipole-dipole interaction Hamiltonian
\begin{equation}
H_{dd}(t)=-\sum_{i,j,\kappa_i=\kappa_j}\sum_{\alpha,\beta}
\Omega_{i,j,\alpha,\beta}
\sigma_{i,\alpha}^\dagger(t)\sigma_{j,\beta}(t)
\end{equation}
and the adjoint Lindbladian which is given by the linear map
\begin{equation}
\mathcal{L}_H^\dagger(t) Q_H(t)=
-\frac{1}{2}\sum_{i,j,\kappa_i=\kappa_j}\sum_{\alpha,\beta}
\gamma_{i,j,\alpha,\beta}
\left[
\left\{\sigma_{i,\alpha}^\dagger(t)\sigma_{j,\beta}(t), Q_H(t)\right\}
-2\sigma_{i,\alpha}^\dagger(t) Q_H(t)\sigma_{j,\beta}(t)
\right],
\end{equation}
Using these definitions, we can simplify Eq. \eqref{eq:Qdot7}
\begin{tcolorbox}[colframe=commandcolor, colback=white, boxrule=1pt]
\begin{equation}
\dot{Q}_H(t)=\frac{i}{\hbar}
\left[H_S(t)+H_E(t)+H_{dd}(t), Q_H(t)\right]
+\mathcal{L}_H^\dagger(t) Q_H(t).
\label{eq:Qdot8}
\end{equation}
\end{tcolorbox}
\begin{flushleft}
This equation is referred to as the adjoint quantum master equation;
it governs the evolution of any atomic system. Once it is solved
it can be used to infer the time evolution of the electromagnetic field
using Eq. \eqref{eq:aq1}.
\end{flushleft}

Unlike the standard quantum master equation, which governs the evolution
of the density matrix, this equation describes the dynamics of the expectation
values of operators $Q_H(t)$,
analogous to the Heisenberg picture of closed
quantum systems\cite{breuer323}. The subscript $S$ indicates
that the density matrix as well as other operators
are in the Schr\"odinger picture.
For completeness, we now recall that the corresponding
Schrödinger-picture equation - the quantum master
equation - governs the time evolution of the density matrix
\cite{lindblad1976generators,breuer322,manzano2020}
and reads
\begin{tcolorbox}[colframe=commandcolor, colback=white, boxrule=1pt]
\begin{equation}
\dot{\rho}_S(t) 
= -\frac{i}{\hbar}\left[H_S+H_{E,S}+H_{dd,S},\rho_S(t)\right]
+\mathcal{L}_S\rho_S(t)
\label{eq:rhodot1}
\end{equation}
\end{tcolorbox}
\begin{flushleft}
where the Lindbladian in the Schr\"odinger picture is given by
\end{flushleft}
\begin{equation}
\mathcal{L}_S\rho_S(t)=
-\frac{1}{2}\sum_{i,j,\kappa_i=\kappa_j}\sum_{\alpha,\beta}
\gamma_{i,j,\alpha,\beta}\left[
\left\{\sigma_{i,\alpha}^\dagger\sigma_{j,\beta},\rho_S(t)\right\}
-2\sigma_{j,\beta} \rho_S(t)\sigma_{i,\alpha}^\dagger
\right].
\end{equation}
Note that in this case, the operators are in the Schr\"odinger picture
and therefore do not exhibit the time dependence typical of the Heisenberg picture.

Together, the adjoint quantum master equation \eqref{eq:Qdot8} and the Schrödinger-picture
master equation \eqref{eq:rhodot1} form a dual description of the open-system dynamics.
These two equations are the central objects that \mathttt{MulAtoLEG} aims to construct
and solve, allowing one to work interchangeably in either picture depending on the
physical quantities of interest.

\section{Incident laser fields}\label{sec:lasers}

In most experimentally relevant situations, the atomic system of interest
is driven by one or several coherent electromagnetic sources, such as
lasers, characterized by well-defined amplitudes and phases and
conveniently described through Rabi parameters.
We therefore specialize the general formalism introduced above to this
laser-driven regime, which is the primary focus of \mathttt{MulAtoLEG}.

In order to solve either Eq.~\eqref{eq:Qdot8} or Eq.~\eqref{eq:rhodot1},
we introduce two additional simplifications, which are more conveniently
implemented at the level of Eq.~\eqref{eq:rhodot1}, written in
the Schr\"odinger picture. These simplifications consist of
(i) eliminating the explicit photonic degrees of freedom associated
with the incident (driving) field, and
(ii) moving to the interaction picture with respect to the
free system Hamiltonian.
At this stage, the Hamiltonian $H_{E,S}$ still contains operators acting on
the photonic space, namely the free-field operators
$a_{q,\epsilon}(0)$ and $a_{q,\epsilon}^\dagger(0)$.
These operators correspond exclusively to the free electromagnetic
field, i.e., the incident field that drives the emitters.
The effects of the scattered field, generated by the emitters
themselves, have already been incorporated into the nonunitary
and coherent interaction terms appearing in Eqs.~\eqref{eq:Qdot8}
and \eqref{eq:rhodot1}.
To eliminate the remaining photonic operators, we now assume
that the system is illuminated by $n_c$ coherent classical light sources.
Specifically, $n_c$ selected modes with wave vectors 
$\boldsymbol{k}_1,\boldsymbol{k}_2,\dots,\boldsymbol{k}_{n_c}$
and frequencies $\omega_i=c k_i$ are externally and continuously
driven into coherent states
$\left\vert A_1\right\rangle\otimes\left\vert A_2\right\rangle
\otimes \dots \otimes \left\vert A_{n_c}\right\rangle$
with amplitudes
$A_1$, $A_2$, $\dots$, $A_{n_c}$ satisfying
$a_i(0)\left\vert A_i\right\rangle = A_i \left\vert A_i\right\rangle$.
With all other modes being in the vacuum state 
$\left\vert 0\right\rangle\otimes \left\vert 0\right\rangle
\otimes\dots \otimes \left\vert 0\right\rangle$, 
we can write the initial photonic density matrix as
$\rho_p(0)=\left\vert \Psi_0\right\rangle
\left\langle \Psi_0\right\vert$
where
$\left\vert \Psi_0\right\rangle$
$=\left\vert A_1\right\rangle\otimes\left\vert A_2\right\rangle
\otimes \dots \otimes \left\vert A_{n_c}\right\rangle$
$\otimes \left\vert 0\right\rangle\otimes \left\vert 0\right\rangle
\otimes\dots \otimes \left\vert 0\right\rangle$.
Under these assumptions, the action of $H_{E,S}$ on the initial
photonic state reduces to an effective Hamiltonian acting solely
on the emitter degrees of freedom,
$H_{E,S}\rho_P(0)=H_{C,S}\rho_P(0)$,
where, after dropping rapidly oscillating terms, the classical driving Hamiltonian
\begin{equation}
H_{C,S} = -\sum_\alpha\sum_{i=1}^{n_c}
    \left(\Lambda_i\mathrm{e}^{i\boldsymbol{k}_i
    \cdot \boldsymbol{r}_\alpha}\mathrm{e}^{-i\omega_it}
    \sigma^\dagger_{i,\alpha}
    + \Lambda_i^*\mathrm{e}^{-i\boldsymbol{k}_i
    \cdot \boldsymbol{r}_\alpha}\mathrm{e}^{i\omega_it}
    \sigma_{i,\alpha}
    \right),
\end{equation}
is devoid of photonic operators.
The corresponding Rabi frequencies are
\begin{equation}
\Lambda_i =\sqrt{\frac{\hbar \omega_i}{2\epsilon_0 V}}
\boldsymbol{e}_i\cdot \boldsymbol{p}_i^*A_i \,\, ,
\end{equation}
where $\boldsymbol{e}_i$ denotes the polarization of the $i$-th coherent driving field.
From this point on we may use the part of the density matrix
that contains the atomic degrees of freedom.
From this point onward, the photonic degrees of freedom associated
with the incident field are fully absorbed into classical driving terms,
and the system dynamics is completely described in terms of the
reduced density matrix of the emitters alone. The influence of the scattered
electromagnetic field survives implicitly through the effective coherent
and dissipative interactions previously derived.

The next simplification consists in transforming
the density matrix $\rho_S(t)$ in
to the interaction picture, $\rho(t)=U_S\rho_S(t)U_S^\dagger(t)$,
via the unitary transformation
$U_S=\exp(i H_S t/\hbar)$. This procedure transforms
the master equation \eqref{eq:rhodot1} into the following
interaction picture master equation 
for the atomic degrees of freedom
\begin{equation}
\dot{\rho}(t) 
= -\frac{i}{\hbar}\left[H_{C}+H_{dd},\rho(t)\right]
+\mathcal{L}\rho(t).
\label{eq:atomicrhodot1}
\end{equation}
The interaction picture Hamiltonians are
\begin{eqnarray}
H_{C} &=& -\sum_\alpha\sum_{i=1}^{n_c}
    \left(\Lambda_i\mathrm{e}^{i\boldsymbol{k}_i
    \cdot \boldsymbol{r}_\alpha}\mathrm{e}^{i\delta_it}
    \sigma^\dagger_{i,\alpha}
    + \Lambda_i^*\mathrm{e}^{-i\boldsymbol{k}_i
    \cdot \boldsymbol{r}_\alpha}\mathrm{e}^{-i\delta_it}
    \sigma_{i,\alpha}
    \right),\\
H_{dd} &=& -\sum_{i,j,\kappa_i=\kappa_j}\sum_{\alpha,\beta}
\Omega_{i,j,\alpha,\beta}\mathrm{e}^{i\delta_{i,j}t}
\sigma_{i,\alpha}^\dagger\sigma_{j,\beta},
\label{eq:hdd1}
\end{eqnarray}
where $\delta_i=\Delta_i-\omega_i$ is the detuning of the
$i$-th transition and $\delta_{i,j}=\Delta_i-\Delta_j$
is the energy mismatch between the $i$-th and $j$-th
transition energies.
Similarly, the Lindbladian in the interaction picture is given by
\begin{equation}
\mathcal{L}\rho(t)=
-\frac{1}{2}\sum_{i,j,\kappa_i=\kappa_j}\sum_{\alpha,\beta}
\gamma_{i,j,\alpha,\beta}\mathrm{e}^{i\delta_{i,j}t}
\left[
\left\{\sigma_{i,\alpha}^\dagger\sigma_{j,\beta},\rho(t)\right\}
-2\sigma_{j,\beta} \rho(t)\sigma_{i,\alpha}^\dagger
\right].
\label{eq:lindblad1}
\end{equation}
From Eqs. \eqref{eq:atomicrhodot1}-\eqref{eq:lindblad1}
it is straightforward to derive the adjoint master equation
in the interaction picture for
for $Q(t)=U_S(t)Q_H(t)U_S^\dagger(t)$.
It takes the form
\begin{equation}
\dot{Q}(t)=\frac{i}{\hbar}
\left[H_C(t)+H_{dd}(t), Q(t)\right]+\mathcal{L}^\dagger(t) Q(t).
\label{eq:Qdot9}
\end{equation}
where the transition operators
in the Hamiltonians
\begin{eqnarray}
H_{C}(t) &=& -\sum_\alpha\sum_{i=1}^{n_c}
    \left[\Lambda_i\mathrm{e}^{i\boldsymbol{k}_i
    \cdot \boldsymbol{r}_\alpha}\mathrm{e}^{i\delta_it}
    \sigma^\dagger_{i,\alpha}(t)
    + \Lambda_i^*\mathrm{e}^{-i\boldsymbol{k}_i
    \cdot \boldsymbol{r}_\alpha}\mathrm{e}^{-i\delta_it}
    \sigma_{i,\alpha}(t)
    \right],\label{eq:cohesource01}\\
H_{dd}(t) &=& -\sum_{i,j,\kappa_i=\kappa_j}\sum_{\alpha,\beta}
    \Omega_{i,j,\alpha,\beta}\mathrm{e}^{i\delta_{i,j}t}
    \sigma_{i,\alpha}^\dagger(t)\sigma_{j,\beta}(t),
    \label{eq:hdd2}
\end{eqnarray}
and the Lindbladian
\begin{multline}
\mathcal{L}^\dagger(t)Q(t) =
-\frac{1}{2}\sum_{i,j,\kappa_i=\kappa_j}\sum_{\alpha,\beta}
\gamma_{i,j,\alpha,\beta}\mathrm{e}^{i\delta_{i,j}t}
\big[
\left\{\sigma_{i,\alpha}^\dagger(t)\sigma_{j,\beta}(t),
Q(t)\right\}\\
-2\sigma_{j,\beta}(t) Q(t)\sigma_{i,\alpha}^\dagger(t)
\big],
\label{eq:lindblad2}
\end{multline}
are now in the Heisenberg picture.

For any of the terms in
Eqs. \eqref{eq:atomicrhodot1}-\eqref{eq:lindblad1}
(master equation)
or Eqs. \eqref{eq:Qdot9}-\eqref{eq:lindblad2}
(adjoint master equation)
to be relevant, the energy mismatches should be small
compared to the energy differences, i.e.,
$\delta_{i,j}\ll \Delta_i, \Delta_j$.
In this respect three possible situations
can be identified.
First, when the energy mismatches ($\delta_{i,j}=\Delta_i-\Delta_j$)
between different transitions ($i\ne j$) are large compared to
$\Delta_i$ and $\Delta_j$, only the terms with $i=j$
contribute significantly. In this regime, cross terms between 
different transitions are effectively suppressed.
Second, when the full internal structure of the atom is
taken into account, for example, by including all possible
values of the total angular momentum $F$-many
levels can become degenerate. In such cases, transitions
between different pairs of states ($i\ne j$) can have the
same energy, resulting in  $\delta_{i,j}=0$ even though the
transitions occur between states with distinct quantum numbers.
This leads to multiple, physically different transitions that are 
nonetheless energetically degenerate and therefore have
a significant role in the master equation.
Third, similar to the second case, the full atomic structure
is taken into account; however, an external perturbation—such
as a magnetic field—lifts the degeneracy of the energy levels through
mechanisms like Zeeman splitting. As a result, transitions that were 
originally degenerate now acquire slightly different energies,
leading to small but nonzero mismatches $\delta_{i,j}\ne 0$
for $i\ne j$. These small energy differences may still be
much smaller than $\Delta_i$ and $\Delta_j$,
and can thus play a non-negligible role in the dynamics
described by the master equation.

A general form of expressing the density matrix dynamics
is the linear map given by the Kraus operator sum
\cite{KRAUS1971311,lindblad1976generators,havel2003,PhysRevA.74.062113,breuer321}
\begin{equation}
\rho(t)=V(t)\rho=\sum_{m}S_m(t)\rho S_{m}^\dagger(t)
\label{eq:kraussum1}
\end{equation}
where $\rho=\rho(0)$ and $V^\dagger(0)=1$. The, $S_m(t)$ are a set of matrices such that \eqref{eq:kraussum1} is fulfilled. $V(t)$ can be understood as the
non-unitary evolution operator of the open quantum system
under consideration.
Correspondingly, the dynamics of the operators in the
Heisenmberg picture can be expressed in terms
of the adjoint map
\begin{equation}
Q(t)=V^\dagger(t)Q=\sum_{m}S_m^\dagger(t)QS_{m}(t),
\label{eq:kraussum2}
\end{equation}
where $Q=Q(0)$ and $V(0)=1$.
The code presented in this paper is devoted to finding
these two maps as the solutions of the quantum master
equation \eqref{eq:atomicrhodot1} and the adjoint master
equation \eqref{eq:Qdot9}.

\section{The general master and adjoint master equations}
\label{sec:mastered}

The package \mathttt{MulAtoLEG} was originally conceived
to generate the Liouville equations for atomic systems.
However, it is also capable of computing the differential
equations for the density matrix coefficients corresponding
to a general master equation \cite{lindblad1976generators,manzano2020} of the form
\begin{equation}
\dot{\rho}(t) = -\frac{i}{\hbar}\left[H(t), \rho(t)\right] + \mathcal{L}\rho(t),
\label{eq:genmastereq}
\end{equation}
where $H(t)$ is an arbitrary time-dependent Hamiltonian and the action of the Lindbladian on the density matrix is given by
\begin{equation}
\mathcal{L}\rho(t) = -\frac{1}{2}\sum_{i,j}\gamma_{i,j}
\left(\left\{F_i^\dagger F_j, \rho(t)\right\} - 2F_j\rho(t)F_i^\dagger\right),
\label{eq:genlindbladian}
\end{equation}
with $\gamma_{i,j}$ denoting the decay rates and $F_i$ the jump operators.

Similarly, it can generate the differential equations for the adjoint master equation
\begin{equation}
\dot{Q}(t) = \frac{i}{\hbar}\left[H(t), Q(t)\right] + \mathcal{L}^\dagger Q(t),
\label{eq:genadjointmastereq}
\end{equation}
where the action of the adjoint Lindbladian on $Q(t)$ is defined as
\begin{equation}
\mathcal{L}^\dagger Q(t) = -\frac{1}{2}\sum_{i,j}\gamma_{i,j}
\left(\left\{F_i^\dagger F_j, Q(t)\right\} - 2F_i^\dagger Q(t) F_j\right).
\label{eq:genadjointlindbladian}
\end{equation}

\section{Solution of the master and
the adjoint master equations}\label{sec:solmastereq}
In this section we follow a general approach to
to solving the quantum master equations \eqref{eq:atomicrhodot1}
or \eqref{eq:genmastereq}
and the adjoint quantum mater equations \eqref{eq:Qdot9}
or \eqref{eq:genadjointmastereq}. If the reader is only interested in using the package to solve physical problems, it is not necessary to understand this section. However, readers interested in the mathematical particulars of the solution method may find it illuminating.

For the sake of brevity, we write the adjoint equation as
\begin{equation}
\dot{Q}(t)=L^\dagger\left(t\right)Q(t)=V^\dagger(t)LQ,
\label{eq:Qdot10}
\end{equation}
where $L^\dagger\left(t\right)Q(t)$ represents the
adjoint Liouville linear map acting on any atomic operator $Q(t)$
on the right-hand side of Eq. \eqref{eq:Qdot8}.
Similarly the the master equation for the density matrix
follows
\begin{equation}
\dot{\rho}(t)=L\rho(t)= L V(t) \rho,
\label{eq:rhodot2}
\end{equation}
where $L\rho(t)$ corresponds to the right-hand side
of Eq. \eqref{eq:atomicrhodot1}.
To solve the differential equations \eqref{eq:Qdot9} and \eqref{eq:rhodot2}
we resort to the so-called direct method \cite{alicki2007quantum,schlosshauer2007decoherence}
which has been extensively used to solve quantum master equations associated
with magnetic resonance problems.
This method consists of projecting both sides of 
Eqs. \eqref{eq:Qdot9} and \eqref{eq:rhodot2} onto
a complete set of matrices that span any atomic operator $Q(t)$
or density matrix $\rho(t)$, explicitly expressing the Liouville 
operator as a matrix.
Such a set of matrices, $\mathcal{H}_n=\{h_1, h_2,\dots, h_n\}$,
is considered complete if any atomic operator or density matrix of a
system with $n_a$ emitters with $n_l$ levels can be expressed
as a linear combination of its elements.
The total number of states for such a system of emitters
is $n_t=n_l^{n_a}$. Hence, the set $\mathcal{H}_n$ must
be composed of $n=n_t^2=n_l^{2n_a}$ matrices.
It can be shown that any such group of matrices
forms an internal vector space and can be
arranged so they are orthogonal under the trace
\begin{equation}
\mathrm{Tr}\left[h_i^\dagger h_j\right]=\delta_{i,j},
\label{eq:orthogonal0}
\end{equation}
assuming that they are normalized.
Thus, any atomic operator $O$
can be expanded in terms of  the
elements of $\mathcal{H}_n$ as
\begin{equation}
O=\sum_i O_ih_i,
\label{eq:Aexpand0}
\end{equation}
where $O_i=\mathrm{Tr}[h_i O]$,
according to the orthonormality
condition \eqref{eq:orthogonal0}.

There is a wide variety of sets
that can be used as $\mathcal{H}_n$.
To construct a
complete
set of matrices of dimension $n_t$
lets first build a
complete set of matrices $\mathcal{G}_{m}$ of
dimension $n_l$ that span the space
of the emitter energy levels. Altogether this set
should have $m=n_l^2$ elements.
The first $n_l$ matrices
$g_1$, $g_2$, $\dots$, $g_{n_l}$
are comprised of diagonal matrices
with only a $1$ in the diagonal.
These are explicitly given by
\begin{equation}
\left(g_k\right)_{m,n} = \delta_{m,k}\delta_{n,k},
\end{equation}
where $i,j,k = 1,2,\dots, n_l $.
The following matrices
are generalizations of the $\sigma_x$
Pauli matrix. These are given by
\begin{eqnarray}
(g_k)_{m,n} &=& \delta_{m,j}\delta_{n,l}+\delta_{m,l}\delta_{n,j},
\end{eqnarray}
where $j=1,2,\dots, n_{l}-1$, $l=m+1,m+2,\dots, n_l$
and $k=n_l+1, n_l+2,\dots, n_l(n_l+1)/2$.
The remaining matrices are generalizations of the
$\sigma_y$ Pauli matrix. Similarly to the
previous, these are given by
\begin{eqnarray}
(g_k)_{m,n} &=& i\left(\delta_{m,j}\delta_{n,l}
-\delta_{m,l}\delta_{n,j}\right),
\end{eqnarray}
where $j=1,2,\dots, n_{l}-1$, $l=m+1,m+2,\dots, n_l$
and $k=n_l(n_l+1)/2+1, n_l(n_l+1)/2+2, \dots, n_l^2$.
We can now build the complete basis as
a Kronecker product of the $q_k$ matrices
\begin{equation}
h_k=g_{k_{n_a}}\otimes g_{k_{n_a-1}}\otimes \dots \otimes g_{k_1}
\end{equation}
where $k_1, k_2, \dots, k_{n_a}=1,2,\dots, n_l^2$
and $k=\sum_{i=0}^{n_a-1}k_in_l^{i}$.
This basis has the advantage of being
composed of Hermitian matrices; therefore,
any Hermitian matrix expanded in this basis
will have exclusively real coefficients
considerably simplifying the system of differential 
equations.

It is easier to first outline the derivation
of the explicit form of the quantum master
equation \eqref{eq:rhodot2}, which is done in a straightforward
manner.
To do so, we start by expanding $\rho(t)$ in terms
of the elements of $\mathcal{H}_n$ as
\begin{equation}
\rho(t)=\sum_i\rho_i(t) h_i.
\end{equation}
Substituting this expansion into Eq. \eqref{eq:rhodot2}
and using the orthonormality condition \eqref{eq:orthogonal0}
we obtain the following differential equations
for the density matrix coefficients:
\begin{equation}
\dot{\rho_i}(t)=\sum_{j}L_{i,j}\rho_{j}(t)
\label{eq:liouvilleeq1}
\end{equation}
where
\begin{equation}
L_{i,j}=\Tr[h_i L h_j],
\label{eq:liouvillesuperoperator}
\end{equation}
are the
matrix elements of the Liouville superoperator.
It is important to explicitly express
the Liouville superoperator in terms of
its Hamiltonian and Lindbladian contributions
\begin{equation}
L_{i,j}=L^H_{i,j}+L^L_{i,j}.
\end{equation}
The Hamiltonian contribution is given by
\begin{equation}
L^H_{i,j}=-\frac{i}{\hbar}\Tr\left[h_i\left[H_C+H_{dd},h_j\right]\right],
\end{equation}
while the Lindbladian contribution reads
\begin{equation}
L^L_{i,j}=-\frac{1}{2}\sum_{i,j,\kappa_i=\kappa_j}\sum_{\alpha,\beta}
\gamma_{i,j,\alpha,\beta} \mathrm{e}^{i\delta_{i,j}t}
\left(
\Tr\left[h_i
\left\{\sigma_{i,\alpha}^\dagger\sigma_{j,\beta},h_j\right\}\right]
-2\Tr\left[h_i\sigma_{j,\beta} h_j\sigma_{i,\alpha}^\dagger\right]
\right).
\end{equation}

Equation \eqref{eq:liouvilleeq1} is a linear system
of $n$ ordinary differential equations
that can be easily solved by numerical methods.
The solution can then be leveraged to calculate
expected values of any operator $O$ by means of
\begin{equation}
\left\langle O\right\rangle =\Tr\left[O\rho(t)\right]
=\sum_iO_i\rho_i(t),
\end{equation}
where $O_i=\Tr[Oh_i]$.

A more general form of Eq. \eqref{eq:liouvilleeq1}
follows from expanding the initial density matrix
as $\rho(t)=\sum_j \rho_j(t)h_j$ in
\eqref{eq:rhodot2}. After some algebra, it is found that
\begin{equation}
\dot{V}_{i,j}(t) = \sum_k L_{i,k}V_{k,j}(t),
\label{eq:liouvilleV1}
\end{equation}
where $V_{i,j}(t)=\Tr\left[h_iV(t)h_j\right]$. We remember here that $V(t)$ was introduced in equation \eqref{eq:kraussum1}. The density matrix can thus be computed as
\begin{equation}
\rho_i(t)=\sum_jV_{i,j}(t)\rho_j.
\end{equation}
These equations represent the non-unitary counterpart, for an open quantum 
system, of the unitary dynamical equation governing the evolution operator 
in closed quantum systems, with the key difference that in this case
$V(t)$ is non-unitary.
This is a system of $n^2$ differential equations with
$n^2$ initial conditions of the form $V_{i,j}(0)=\delta_{i,j}$,
which is significantly more difficult to solve than 
Eq. \eqref{eq:liouvilleeq1} and is often impractical due to its large 
dimensionality.

The determination of the explicit form of the adjoint quantum
master equation is more intricate and demands careful attention.
By employing Eq. \eqref{eq:kraussum2}, we can express $n$
differential equations for each element of $\mathcal{H}_n$ as
\begin{equation}
\dot{h}_i(t)
=V^\dagger(t)Lh_i,
\label{eq:MLEexpand0}
\end{equation}
The explicit form of this differential equation can be derived by expanding
both sides in terms of the elements of $\mathcal{H}_n$.
More specifically, the left-hand side, which can be written as
\begin{equation}
\dot{h}_i(t) = \dot{V}^\dagger(t) h_i,
\end{equation}
can be expanded as
\begin{eqnarray}
\dot{h}_i(t) = \frac{d}{dt} \sum_j V_{j,i}^\dagger(t) h_j
= \sum_j \dot{V}_{j,i}^\dagger(t) h_j,
\label{eq:adjmaeqlhs}
\end{eqnarray}
where $V_{j,i}^\dagger(t)=\Tr[h_jV^\dagger(t)h_i]$.
The right hand-side can be expanded as
\begin{equation}
V^\dagger(t)L^\dagger h_i 
= \sum_{l,k} V_{k,l}^\dagger(t) L_{l,i}^\dagger h_k
\label{eq:adjmaeqrhs}
\end{equation}
where
\begin{equation}
L_{l, i}^\dagger = \Tr [h_lL^\dagger h_i]
\label{eq:adjointliouvillesuperoperator}
\end{equation}
are the matrix elements of the adjoint Liouville
superoperator.
It can readily be calculated since,
from Eq. \eqref{eq:Qdot8}, it can be inferred that
the action of the Liouville opertor on $h_i$ in the
Schr\"odinger picture is
\begin{equation}
L^\dagger h_i=\frac{i}{\hbar}
\left[H_E+H_{dd},h_i\right]+\mathcal{L}^\dagger h_i,
\end{equation}
where
\begin{equation}
\mathcal{L}^\dagger h_i=
-\frac{1}{2}\sum_{i,j,\kappa_i=\kappa_j}\sum_{\alpha,\beta}
\left[
\left\{\sigma_{i,\alpha}^\dagger\sigma_{j,\beta},h_i\right\}
-2\sigma_{i,\alpha}^\dagger h_i\sigma_{j,\beta}
\right].
\end{equation}
Note that, in contrast to Eq. \eqref{eq:Qdot8}, the two previous
expressions lack the time dependence that characterizes the Heisenberg
picture. This is because they are written in the Schr\"odinger picture.

By equating Eqs. \eqref{eq:adjmaeqlhs} and \eqref{eq:adjmaeqrhs},
multiplying by an arbitrary element of $\mathcal{H}_n$,
and taking the trace, we obtain
\begin{equation}
\dot{V}_{i,j}^\dagger(t) = \sum_k V_{i,k}^\dagger(t) L_{k,j}^\dagger.
\label{eq:eqdifforV}
\end{equation}
As with Eq. \eqref{eq:liouvilleV1}
this results in a system of $n^2$ ordinary differential equations
for $V_{j,i}^\dagger (t)$ with $n^2$ initial conditions
$V_{j,i}^\dagger (0)=\delta_{i,j}$.
The dimensionality of this system increases much more
rapidly than that of Eq. \eqref{eq:liouvilleeq1} as the number
of atoms and levels increases.
We can reduce the dimensionality of
the problem to
$n$ by restricting ourselves to calculating
only the expected values of the elements of
$\mathcal{H}_n$ for a given initial density matrix $\rho(0)$,
rather than the full evolution operator $V^\dagger (t)$ as
in Eq. \eqref{eq:eqdifforV}.
Then, multiplying \eqref{eq:eqdifforV} by $\rho_j(0)$,
summing over the index $j$
and considering that
the expected values of the elements of $\mathcal{H}_n$
are in general given by
\begin{multline}
\eta_i(t)=\Tr[h_i\rho(t)]=\Tr[ h_iV(t)\rho(0)]
=\Tr[h_i(t)\rho(0)]=\Tr[V^\dagger(t) h_i\rho(0)]\\
=\sum_j\rho_j(0)\Tr\left[h_jV^\dagger(t)h_i\right]
=\sum_j\rho_j(0)V_{j,i}^{\dagger}(t).
\label{eq:adjointmasterequationforq}
\end{multline}
we obtain the following equation
\begin{equation}
\dot{\eta}_i(t) = \sum_k \eta_k(t)L_{k,i}^\dagger.
\label{eq:eqdiffforq}
\end{equation}
Alternatively, one may also define these variables as
\begin{equation}
\eta_i(t)=\left\langle \psi_0 \left\vert
V^\dagger(t)h_i\right\vert \psi_0\right\rangle,
\end{equation}
where $\left\vert \psi_0\right\rangle$ is the initial pure
state of the system, thereby recovering Eq. \eqref{eq:eqdiffforq}.
This considerably reduces the number of differential
equations to $n$; however, the tradeoff is that
the initial state of the system must be fixed to $\rho(0)$.
Consequently, we must solve the system of differential
equations \eqref{eq:eqdiffforq} with the initial conditions
$\eta_i(0)=\Tr[h_i\rho(0)]=\rho_i(0)$ or
$\eta_i(0)=\left\langle \psi_0 \left\vert
h_i\right\vert \psi_0\right\rangle$. If a different initial condition is required, new equations need to be obtained.

\section{Rotating frame}\label{sec:rotframe}
The differential equations arising from
\eqref{eq:liouvilleeq1} or \eqref{eq:eqdiffforq}
might be difficult to solve given that $L_{i,j}$ and
$L_{i,j}^\dagger$ are, in general, time-dependent.
These time dependences come from the coherent light
sources of the Hamiltonian $H_C$ in
Eq. \eqref{eq:cohesource01} and the small mismatches
between transition energies in the dipole-dipole Hamiltonian $H_{dd}$
in Eq. \eqref{eq:hdd2}. The latter usually vanish or are negligible
($\delta_{i,j}\approx 0$).

For example, in the vicinity of the center of a magneto-optical trap,
the quadrupolar magnetic moments produce
almost vanishing magnetic fields, typically on the order
of tenths of a gauss. These weak fields only introduce
energy mismatches of a few megahertz
($\delta_{i,j}\sim$\,MHz), which can be neglected.
Under these conditions, the time dependences introduced
by $H_C$ can be removed from the master equation \eqref{eq:rhodot2}
by transforming from the lab frame into a rotating frame.
To enter the rotating frame, we apply the transformation
\begin{equation}
R(t) = \exp\left(-t\frac{i}{2}\sum_{\alpha}\sum_{j}\varphi_{j}
\sigma_{j,\alpha}^z\right),
\label{eq:rotframe0}
\end{equation}
where $j = \{j_1, j_2\}$ indexes the transition excited
by the coherent source. The corresponding $\sigma_z$–like Pauli operators are defined as
\begin{equation}
\sigma_{j, \alpha}^z 
= \ket{j_1}_\alpha \prescript{}{\alpha}{\bra{j_1}}
-\ket{j_2}_\alpha \prescript{}{\alpha}{\bra{j_2}},
\label{eq:rotframe1}
\end{equation}
and $\varphi_j$ denote the rotation parameters.
Choosing these parameters such that $H_C$ becomes time-independent
leads to a system of $N_C$ linear equations for the parameters $\varphi_j$,
where $N_C$ is the number of coherent sources driving the atomic system.
The condition $\delta_{i,j} = \Delta_i - \Delta_j \approx 0$ implies that
the difference between the transition energies $\Delta_i$ and $\Delta_j$
of coupled transitions is negligible. Therefore, the transformation
\eqref{eq:rotframe1} leaves both the Lindbladian in Eq. \eqref{eq:lindblad1}
and the dipole–dipole Hamiltonian $H_{dd}$ in Eq. \eqref{eq:hdd1}
invariant, regardless of the
particular solution obtained for the parameters $\varphi_j$.
The master equation \eqref{eq:atomicrhodot1} is thus transformed into
\begin{equation}
\dot{\rho}_R(t) = L_R\rho_R
= -\frac{i}{\hbar}\left[H_{R}+H_{dd},\rho_R(t)\right]
+\mathcal{L}\rho_R(t)
\end{equation}
where $L_R$ is the constant Liouville superoperator in the rotating frame.
The light–matter interaction Hamiltonian
\begin{equation}
H_R=R(t)H_CR^\dagger(t)+i\hbar \dot{R}(t)R^\dagger (t)
\end{equation}
is also constant and
\begin{equation}
\rho_R(t) = R(t)\rho(t) R^\dagger(t).
\end{equation}
The dipole-dipole Hamiltonian $H_{dd}$ and the Lindbladian $\mathcal{L}$
remain as given by Eqs. \eqref{eq:hdd1} and \eqref{eq:lindblad1},
since both are invariant under $R(t)$.

Just as in the lab frame, the Liouville equation in the rotating frame can
be expanded in terms of the operator basis $\mathcal{H}_n$.
This yields the following system of differential equations for
the density-matrix coefficients $\rho_{R,i}(t)$ in the rotating frame:
\begin{equation}
\dot{\rho}_{R,i}(t) = \sum_j L_{R,i,j}\rho_{R,j}(t),
\label{eq:liouvilleeq2}
\end{equation}
where $L_{R,i,j} = \Tr\left[h_i L_R h_j\right]$ are the constant matrix elements
of the Liouville operator in the rotating frame.
The formal solution to the previous equation is
\begin{equation}
\rho_R(t) = V_R(t)\rho_R,
\end{equation}
with
\begin{equation}
V_R(t) = \exp\left(L_R t\right).
\end{equation}
The relation between the solutions in the two frames depends on how the initial
state is represented.
If the initial coefficients $\rho$ are taken in the lab-frame basis, the evolution reads
\begin{equation}
\rho_i(t) = \sum_{j,k,l} U_{R,i,j}^\dagger(t)V_{R,j,k}(t)U_{R,k,l}(t)\rho_l\,\, ,
\end{equation}
where
\begin{equation}
U_{R,i,j}(t) = \Tr\left[h_i R(t) h_j R^\dagger(t)\right], \qquad
V_{R,j,k}(t) = \Tr\left[h_j V_R(t) h_k\right].
\end{equation}

Naturally, the same arguments apply to the adjoint master equation
\eqref{eq:Qdot9}, which can likewise be rendered time-independent
through the transformation \eqref{eq:rotframe0}. However, it is
worth noting that in this case all operators, including the
transformation $R(t)$, must be expressed in the Heisenberg picture.

\section{Far field}\label{sec:farfield}
At the detector position, the positive-frequency part of the
electric field operator in the Heisenberg picture, as given by 
Eq. \eqref{eq:posfreqfield0}, takes the form
\begin{equation}
\boldsymbol{E}^+(\boldsymbol{R},t_{R})
=\sum_{q,\epsilon}\mathcal{E}_{q,\epsilon}
\boldsymbol{e}_{q,\epsilon}
\mathrm{e}^{i\boldsymbol{k}_q\cdot \boldsymbol{R}}
a_{q,\epsilon}(t_R),
\label{eq:posfreqfield}
\end{equation}
where $\boldsymbol{R}$ is the position vector of the detector
relative to an arbitrary point within the system of emitters,
and $t_R=t+R/c$ is the time at which
the photon reaches the detector after travelling
from the ensemble of emitters.
Substituting the explicit expression of
the photon annihilation operator \eqref{eq:aq1}
into Eq. \eqref{eq:posfreqfield}
we get
\begin{equation}
\boldsymbol{E}^+(\boldsymbol{R},t_R)
=\boldsymbol{E}^{+}_0(\boldsymbol{R},t_R)
+i\sum_{q,\epsilon}\sum_\alpha\sum_{i}
\frac{\hbar \omega_q}{2\epsilon_0 V}
\boldsymbol{e}_{q,\epsilon}
\mathrm{e}^{i\boldsymbol{k}_q\cdot
(\boldsymbol{R}-\boldsymbol{r}_\alpha)}
\int_{0}^{t_R}dt^\prime
\mathrm{e}^{i\omega_q(t^\prime-t_R)}
\boldsymbol{e}_{q,\epsilon}^*\cdot\boldsymbol{P}_{i,\alpha}(s),
\end{equation}
where the purely photonic term
of the positive-frequency part of the
electric field operator
is given by
\begin{equation}
\boldsymbol{E}^{+}_0(\boldsymbol{R},t_R)
=\sum_{q,\epsilon}\mathcal{E}_{q,\epsilon}
\boldsymbol{e}_{q,\epsilon}
\mathrm{e}^{-i\boldsymbol{k}_q\cdot\boldsymbol{R}}
\mathrm{e}^{-i\omega_q t_R}a_{q,\epsilon}(0).
\end{equation}

By taking the continuous limit of the photon spectrum—i.e.,
letting  $V\rightarrow \infty$ and
$\boldsymbol{k}_q\rightarrow \boldsymbol{k}$-
and applying the Markov approximation \eqref{eq:markov1},
we arrive, after some lengthy calculations,
at the  expression for the electric field operator in the far field region
\begin{equation}
\boldsymbol{E}^+(\boldsymbol{R},t_R)
=\boldsymbol{E}_{0}^+(\boldsymbol{R},t_R)
+\frac{\hbar}{\epsilon_0c^2}
\sum_\alpha\sum_{j}
\frac{\boldsymbol{p}_{j}
-\hat{\boldsymbol{R}}_\alpha\hat{\boldsymbol{R}}_\alpha
\cdot \boldsymbol{p}_{j}}{R_\alpha}
\Delta_{j}^2 \exp\left[i\kappa_{j}\left(R_\alpha-R\right)\right]
\sigma_{j,\alpha}(t),
\end{equation}
where $\boldsymbol{R}_\alpha = \boldsymbol{R}-\boldsymbol{r}_\alpha$
and $\hat{\boldsymbol{R}}_\alpha$ is the corresponding
unit vector.

\section{Design principles}\label{sec:desprin}
The package \texttt{MulAtoLEG} is designed
according to the principles of modularity and clarity: each component
corresponds to a well-defined physical or mathematical structure,
enabling users to construct complex models systematically and
without ambiguity.

To achieve efficiency, the package leverages \texttt{Mathematica}’s
vectorization and sparse linear algebra capabilities,
thereby reducing both computational time and memory usage.

\begin{figure}
\centering
\includegraphics[width=1.1\textwidth]{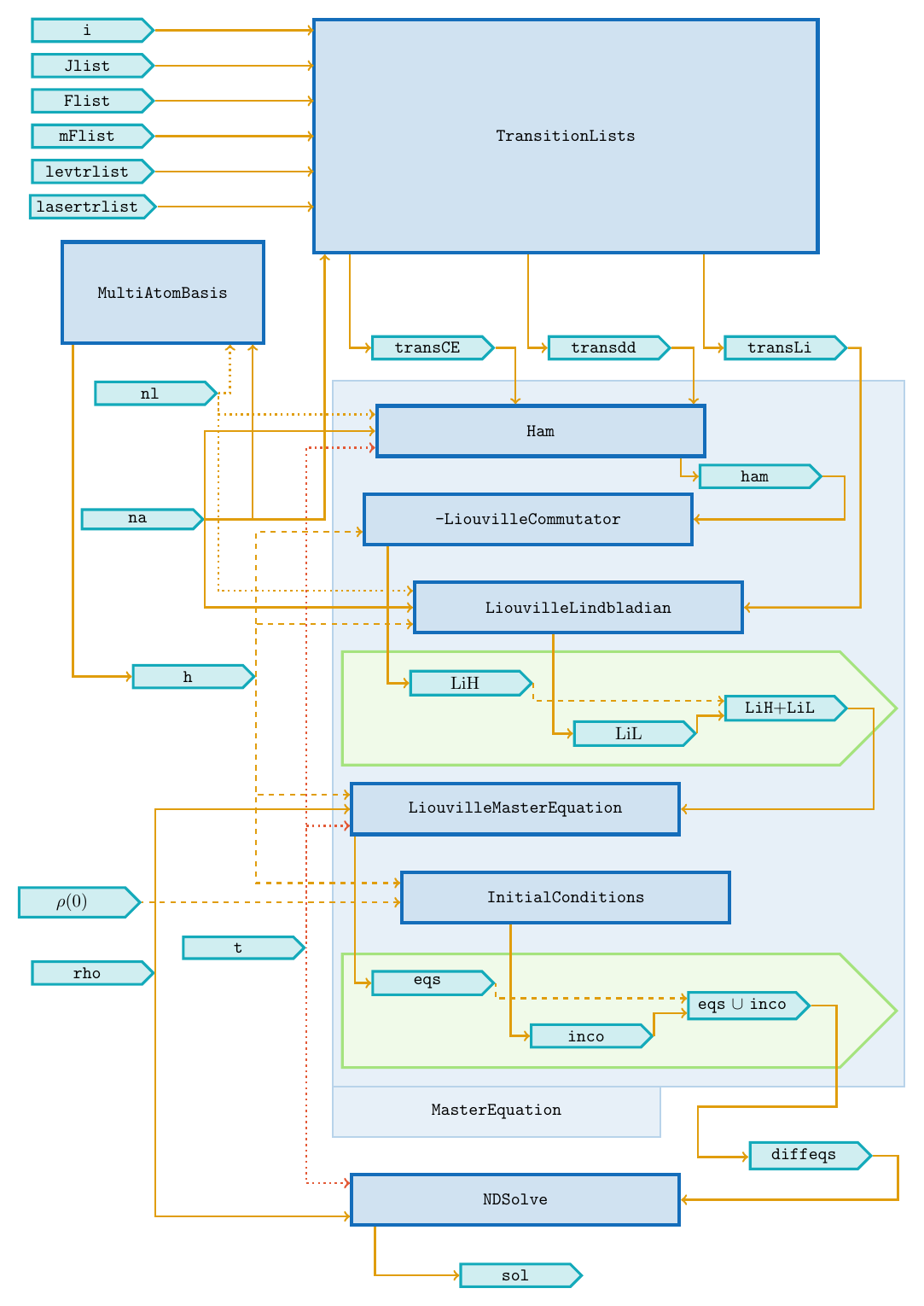}
\caption{Schematics of the \mathttt{MulAtoLEG}
package.}
\label{fig:1}
\end{figure}

From a general perspective, the algorithm consists of computing
the Liouville superoperator from Eq. \eqref{eq:liouvillesuperoperator}
and inserting it into Eq. \eqref{eq:liouvilleeq1} to construct the system
of differential equations for the density matrix coefficients.
This step of the algorithm is illustrated in Fig. \ref{fig:1},
within the light-blue rectangle labeled \texttt{MasterEquation}.
In a similar way, the adjoint Liouville superoperator can be obtained from
Eq. \eqref{eq:adjointliouvillesuperoperator} to derive the system of
 differential equations for $\eta_i(t)$, the expectation values of the
 elements of  $\mathcal{H}_n$
 through Eq. \eqref{eq:eqdiffforq}.

The two main contributions to the Liouville and adjoint Liouville
superoperators arise from the Hamiltonian and the Lindbladian.
The Hamiltonian part is computed from a given Hamiltonian
\mathttt{ham} using \texttt{LiouvilleCommutator}, which generates
the Liouville superoperator corresponding to the linear map defined
by the commutator of the Hamiltonian with the density matrix.
For efficiency, the Hamiltonian \mathttt{ham} should preferably
be provided in the form of a \texttt{SparseArray}, ensuring reduced
memory consumption and shorter computing times.
In particular, for atomic systems, the Hamiltonian \texttt{SparseArray}
can be generated using the function \mathttt{Ham},
which takes as arguments the number of quantum levels
\mathttt{nl}, the number of atoms \mathttt{na}, the matrix basis \mathttt{h},
and the configuration variables \mathttt{transCE} and \mathttt{transdd}.
The required structure of these two variables is explained in:
\\
\filettt{ReferenceGuideMultiAtomLiouvilleEquationGenerator.nb}
\\
and throughout the examples.
This is illustrated in Fig. \ref{fig:1}, at the top of the light-blue rectangle
labeled \texttt{MasterEquation}.
The Liouville superoperator \texttt{SparseArray} corresponding to
the Lindbladian’s linear map is obtained through
\mathttt{LiouvilleLindbladian}. This function takes as
arguments the number of quantum levels \mathttt{nl},
the number of atoms \mathttt{na}, the matrix basis \mathttt{h},
and the configuration variable \mathttt{transLi},
as shown in Fig. \ref{fig:1},
below \mathttt{LiouvilleCommutator},
within the light-blue \texttt{MasterEquation} rectangle.
Although this package is primarily specialized for atomic
systems, it can also compute the Liouville superoperators 
corresponding to arbitrary Hamiltonians and Lindbladians.
In this case, the Liouville superoperator for the Hamiltonian can be
obtained using the same function as for atomic systems,
\mathttt{LiouvilleCommutator}, by supplying the arbitrary Hamiltonian
through \mathttt{ham}.
A Liouville superoperator for a general Lindbladian of the
form \eqref{eq:genlindbladian},
may be generated with the function\\\\
\mathttt{LiouvilleLindbladianFromLindbladOperators},
which takes as input the \texttt{SparseArray}s containing
the Lindblad operators  $A_i$
and the decay rates $\gamma_{i,j}$
that characterize the arbitrary decay channels.

The function \mathttt{LiouvilleMasterEquation} takes as input the
overall Liouville superoperator—given by the sum of its Hamiltonian
and Lindbladian parts—and outputs the system of differential
equations \mathttt{eqs}. However, this system does not yet include
the initial conditions, which are provided by \mathttt{InitialConditions}.
This function takes the initial density matrix and converts it into a set
of initial conditions for the density matrix coefficients.

In the examples presented in the following sections, we adopt the
policy of computing the differential equations with symbolic free parameters.
Although this approach is slower than obtaining the equations directly with
numerical parameters, it has the advantage that the equations need to be
computed only once, after which numerical values can be assigned to
the parameters as needed. This is especially useful when solving the equations
for multiple sets of parameters.

The differential equations \mathttt{eqs} together with the initial
conditions \mathttt{inco} are combined and passed to the
numerical differential equation solver \texttt{NDSolve}.

The function \mathttt{MasterEquation}, shown as a light-blue rectangle
in Fig. \ref{fig:1}, serves as a wrapper for the functions \mathttt{Ham},
\mathttt{LiouvilleCommutator}, \mathttt{LiouvilleLindbladian},
\\
\mathttt{LiouvilleMasterEquation}, and \mathttt{InitialConditions}.

The resulting solution \mathttt{sol} can then be used,
in conjunction with \mathttt{ExpectationValue}, to compute
the expectation values of physically relevant observables.

For systems with a large number of quantum levels, it can be
challenging to manually write the configuration variables
 \mathttt{transCE}, \mathttt{transdd}, and \mathttt{transLi}.
This process is greatly simplified for alkali metals by using
the function \mathttt{TransitionLists}, which automatically generates these
three configuration variables. As shown at the top of Fig. \ref{fig:1},
\mathttt{TransitionLists} produces \mathttt{transCE}, \mathttt{transdd},
and \mathttt{transLi}, which are then passed as arguments
\\
to \mathttt{MasterEquation}.

Each of these variables plays a specific role in the construction
of the Liouville superoperators. The first two are required by \mathttt{Ham}.
The information about interactions with coherent sources - such as detuning,
Rabi parameters, and phases - is provided to \mathttt{H} through \mathttt{transCE}.
The information about dipole - dipole couplings between quantum
levels is supplied through \mathttt{transdd}.
Finally, the variable \mathttt{transLi} is required by \mathttt{LiouvilleLindbladian}
to produce the Liouville superoperator corresponding to the Lindbladian’s linear map.

\begin{enumerate}
\item Write the explicit form of the superoperators corresponding
to the commutator and Lindbladian linear maps.
\item In order for the algorithm to work properly
it is convenient to put most matrices in the form of
\mathttt{SparseArray}'s to ensure low memory consumption
and low computing times.
\end{enumerate}

\section{Installation}\label{sec:installation}
The \mathttt{MulAtoLEG}
package can be downloaded from
\href{https://github.com/alexkunold}{GitHub} .
Then, load the package in a Mathematica notebook using:
\begin{mathematicain}
\(\texttt{SetDirectory}[\texttt{PathToMultiAtomLiouvilleEquationGenerator}];\\
\texttt{Needs}[\texttt{{``}MultiAtomLiouvilleEquationGenerator$\grave{ }${''}}]\)
\end{mathematicain}
where the variable
\mathttt{PathToMultiAtomLiouvilleEquationGenerator}
must be set to the directory containing the file
\filettt{MultiAtomLiouvilleEquationGenerator.wl}.
A comprehensive list of the
\mathttt{PathToMultiAtomLiouvilleEquationGenerator}
Mathematica functions,
together with the seven examples presented in the next section,
can be found in the file
\filettt{ReferenceGuideMultiAtomLiouvilleEquationGenerator.nb}
that can also be downloaded from
\href{https://github.com/alexkunold/MultiAtomLiouvilleEquationGenerator}{GitHub}.

\section{Usage}\label{sec:usage}
Usage of the package is simple. While the underlying theory has been presented in detail, the user simply requires knowledge of the energy levels, dipole transitions and external fields involved in their problem. This can be a generic list of levels involving only one quantum number as a label or specific atomic levels involving angular momentum labels, as required. The physical information of the problem to be solved is encoded via lists of transitions and their parameters. First, the user needs the number of atoms and the number of levels, as this determines the dimensions of the system. Then the user needs a list of the allowed dipole transitions. For example, for atoms with three levels with transitions allowed between levels 1 and 2 and 2 and 3, the list would look like

\begin{mathematicain}
diptrans=\{\{1,2\},\{2,3\}\};
\end{mathematicain} This list sets the level labels. Note the transitions must be entered with the lower of the two energy levels coming first. Then, lists of the transitions involving external laser fields. These are given as the two levels in the transition and a vector indicating the polarization of the laser field. For a laser field connecting levels 1 and 2 with circular polarization 

\begin{mathematicain}
\text{lasertrlist}=\{ \{1,2\}, $\left\{\frac{1}{\sqrt{2}},-\frac{i}{\sqrt{2}},0\right\}$\};
\end{mathematicain} and so on for each laser field. Transitions that involve spontaneous decays are next. Here the transition structure is slightly different. In the case of multiple atoms, transitions in one atom couple to transitions in other atoms,so these lists involve atomic indexes. The list is structured as the two dipole transitions that are being coupled (usually the same transition), followed by the atomic indices, the decay rate and finally the energy gap between transitions (0 if the two transitions are the same). The regular spontaneous decay involving one transition is entered as the transition coupling to itself, so the atomic index is repeated, such as

\begin{mathematicain}
$\left\{\{1,2\},\{1,2\},1,1,\gamma_1,0\right\}$
\end{mathematicain} for the entry of the spontaneous decay from level 2 to 1 in atom 1 with decay rate $\gamma_1$. This entry would be repeated but with the atomic indices changed for every atom in the problem
\begin{mathematicain}
$\left\{\{1,2\},\{1,2\},2,2,\gamma_1,0\right\}$
\end{mathematicain} and again for every other transition. If dipole-dipole interactions are being considered for the problem, then cross transitions must also be entered. In the case of the atoms above, the following two entries would also be required

\begin{mathematicain}
$\left\{\{1,2\},\{1,2\},1,2,\gamma_1 F,0\right\},\left\{\{1,2\},\{1,2\},2,1,\gamma_1 F,0\right\};$
\end{mathematicain} here we account for atoms 1 and 2 being interchangeable and the factor $F$ would modify the cross coupling for the decay channels depending on atomic separation, dipole orientation etc as described in the theoretical discussion in the previous sections. Note that, in general, transitions only couple to transitions with the same energy gap or at least a very similar one. The items detailed above describe the transition from level 1 to level 2 in atom 1 coupling to itself and to the equivalent transition in atom 2. In cases where multiple transitions have very similar or equal energy gaps, all those couplings must be accounted for. Transitions that are coupled via coherent dipole-dipole interactions are also entered. These couplings only apply to multi-atom problems as self coupling of transitions is the Lamb shift and is considered accounted for. These transitions otherwise have the same structure as the spontaneous decay cross-transitions. All the transitions are put into a single list.

\begin{mathematicain}
$\left\{\left\{\{1,2\},\{1,2\},1,2,\gamma_1 G,0\right\},\left\{\{1,2\},\{1,2\},2,1,\gamma_1 G,0\right\}\right\}$;
\end{mathematicain} the $G$ factor here serves a similar purpose to the $F$ factor above and modulates the strength of the interaction based on dipole orientations and atomic separation. Every dipole-dipole coupling is put into a single list. Finally, the couplings of each transition to an external laser field are specified. These transitions are entered in the order of, dipole transition, atomic index, Rabi frequency, detuning from resonance and phase. The phase is only relevant for the case of multiple atoms. An example would be

\begin{mathematicain}
$\left\{\{1,2\},1,R_1,\delta_1,\phi _{1,1}\right\}$;
\end{mathematicain} All the laser field couplings are entered into a single list. In the case where the user is modelling a generic problem such as "two identical three level atoms interacting with two external laser fields" the user can input these lists manually. In the case where the user wishes to model a problem involving specific atomic orbitals, for example, "two Rubidium atoms where the $\ket{5S_{1/2}, F=1}$ level is coupled to the $\ket{5P_{3/2}, F=3}$ level by an external laser field with vertical polarization, including all hyper-fine levels" the transition lists can be quite lengthy. However, the user can employ the function \mathttt{TransitionLists} for this task. The function requires the ranges of the $F$, $m$, $J$ involved as well as the lists of allowed transitions and transitions coupled by lasers. This function will generate the relevant transition lists. Note that this function can be used to create transition lists for a generic problem if fed specific values of $F$, $m$ and $J$ but the resulting lists will involve the relevant Clebsch-Gordon coefficients. The user can take these lists, remove the coefficients and employ them for a generic problem, as desired. 

Once all the relevant transition lists have been established the user has a couple options. The next step is to generate the relevant differential equations for the problem. First, the user sets the basis with the function \mathttt{MultiAtomBasis} giving the number of atoms and the number of levels. Then the user generates a list of all of the density matrix elements that will be used as variables and an initial density matrix, this latter one in matrix form. The initial conditions are generated via the \mathttt{InitialConditions} function. The user then generates the Hamiltonian with the \mathttt{Ham} function, used the \mathttt{LiouvilleCommutator} function, generates the Lindbladian with the \mathttt{LiouvilleLindbladian} function and generates the Liouville master equation with the \mathttt{LiouvilleMasterEquation} function and finally joins it with the initial condition. These equations can be saved to a file with DumpSave and loaded for later use as this process can be time consuming for systems with a larger dimension. The code for this would be as follows:

\begin{mathematicain}
n = $\texttt{nl}^\texttt{na}$;\\
h = MultiAtomBasis[nl, na];\\
rho = Table[Subscript[$\rho$, ii][t], {ii, 1, $n^2$}];\\
rho0 = SparseArray[{{1, 1} -> 1}, {n, n}];\\
incon = InitialConditions[h, rho, rho0, t];\\
ham = Ham[transCE, transdd, nl, na, t];\\
LiH = LiouvilleCommutator[h, ham];\\
LiL = LiouvilleLindbladian[h, transLi, nl, na];\\
Li = LiH + LiL;\\
eqs = LiouvilleMasterEquation[rho, Li, t] // ExpToTrig;\\
diffeqs = Join[eqs, incon];\\
\end{mathematicain}

Alternatively, the user could use the code

\begin{mathematicain}
n = $\texttt{nl}^\texttt{na}$;\\
h = MultiAtomBasis[nl, na];\\
rho = Table[Subscript[$\rho$, ii][t], {ii, 1, $n^2$}];\\
rho0 = SparseArray[{{1, 1} -> 1}, {n, n}];\\
diffeqs = LiouvilleMasterEquation[h,rho,rho0,transCE,transdd,transLi,nl,na, t] // ExpToTrig;\\
\end{mathematicain} to generate the same equations, especially if having access to the Lindbladian or to the Hamiltonian is not necessary. The example section contains detailed examples where different physical systems are solved following these general guidelines.

\section{Comparison with existing packages}\label{sec:comparison}

Several software frameworks have been developed for the study of open quantum systems,
including \mathttt{QuTiP} (Python) \cite{JOHANSSON20121760}, 
\mathttt{QuantumOptics} (Julia) \cite{KRAMER2018109},
and Mathematica's 
\\ \mathttt{QuantumFramework}\cite{QuantumFramework}.
These packages are primarily designed for the
numerical integration of the master equation once the
Liouvillian operator has been explicitly defined by the user.
In contrast, \mathttt{MulAtoLEG} focuses on the
symbolic generation of the Liouville and adjoint master equations
for multilevel and multi-atom configurations, including
arbitrary Hamiltonians and Lindbladians.
It also provides routines to construct differential equations
in the dressed-state basis, from which the non-unitary evolution operator
can often be obtained analytically.
By exploiting Mathematica’s built-in vectorization and sparse linear algebra
capabilities, \mathttt{MulAtoLEG} achieves efficient manipulation of large symbolic
expressions while maintaining exactness.
Since the equations are produced without approximations, the achievable system
size is limited only by the available computational resources.

\begin{table}[h!]
\centering
\caption{Comparison of \mathttt{MulAtoLEG} with related
open-source packages for open quantum systems.}
\vspace{0.5em}
\begin{tabular}{lcc}
\hline
\textbf{Feature} & \textbf{MulAtoLEG} & \textbf{QuTiP}  \\
\hline
\shortstack{Symbolic Liouville\\ equation generation} & Yes & No  \\
\shortstack{Multilevel / multi-atom\\ systems} & Yes & Limited  \\
\shortstack{Arbitrary Hamiltonians\\ and Lindbladians} & Yes & Yes  \\
\shortstack{Dressed-state basis\\ equations} & Yes & No  \\
Sparse / vectorized algebra & Yes (Mathematica) & Yes (NumPy/SciPy)  \\
Main focus area & \shortstack{Atomic, optical and\\ general systems} & General open systems  \\
\hline
\end{tabular}

\vspace{20pt}

\begin{tabular}{lcc}
\hline
\textbf{Feature} & \textbf{QuantumFramework} & \textbf{Spinach} \\
\hline
\shortstack{Symbolic Liouville\\ equation generation} &  Partial & No \\
\shortstack{Multilevel / multi-atom\\ systems} &  Partial & Spin systems only \\
\shortstack{Arbitrary Hamiltonians\\ and Lindbladians} &  Yes & Yes \\
\shortstack{Dressed-state basis\\ equations}  &  No & No \\
Sparse / vectorized algebra &  Yes & Yes \\
Main focus area &  General quantum ops & NMR/ESR \\
\hline
\end{tabular}
\label{tab:comparison}
\end{table}

\section{Conclusion}\label{sec:conclusions}
We have presented an open-source Mathematica package
called \mathttt{MulAtoLEG}, designed to obtain the master and
adjoint master equations for atomic systems consisting of an arbitrary
number of atoms, each having an arbitrary number of quantum levels.
The underlying theory is based on an extension to multilevel atoms of
the adjoint master equation first developed by
Lehmberg \cite{PhysRevA.2.883} and later reformulated as a master
equation by Genes \cite{PRXQuantum.3.010201}.

The scheme includes tools that facilitate the generation of configuration
variables for complex arrangements of quantum transitions in alkali atoms.
Although this package is primarily intended for atomic systems, it can
also compute the master and adjoint master equations corresponding
to general Lindblad equations. Moreover, the framework provides
functionalities that simplify the calculation of master and adjoint
master equations in the dressed-state basis, where the
Liouville superoperator becomes time-independent and, consequently,
the corresponding evolution operator can be obtained exactly.

The package leverages Mathematica’s vectorization, sparse linear
algebra, and symbolic computation capabilities to optimize
computational efficiency. Since the package generates exact
equations without approximations, the tractability of the resulting system
is ultimately limited by the available computational resources.

\section*{Acknowledgements}
We thank LSCSC-LANMAC, where part of the computational
simulations reported in this work were performed using
their HPC server.
We are indebted to Francisco Ya\~nez for testing the examples.
We also thank Ma. Nieves Arias and Irvin Angeles for their help in checking the \mathttt{TransitionLists} module.


\paragraph{Funding information}
P.Y.-T. was financially supported by CONAHCYT through the Estancias Posdoctorales por M\'exico program (2022, grant No. 3). A.K. was financially supported by the Departamento de Ciencias B\'a'sicas, UAM-A, under Grants No. 2232218 and CB003-25 (No. 22322036). This work was partially supported by DGAPA-UNAM-PAPIIT Grants No. IG101826 and No. IN118823, SECIHTI Grant No. LNC-2023-51, and CONAHCYT-CB Grant No. A1-S-30934. D. S-S. received financial support from
projects PAPIIT-DGAPA-UNAM no. IN112624 and Ciencias de Frontera SECIHTI no. CF-2023-G-15. 

Correctly-provided data will be linked to funders listed in the
\href{https://www.crossref.org/services/funder-registry/}{\sf Fundref registry}.

\begin{appendix}
\numberwithin{equation}{section}

\section{Examples}\label{sec:examples}
In this appendix we present seven examples that illustrate the
use of  the \mathttt{MulAtoLEG} functions.

\subsection{A single two-level atom}\label{sec:example1}
A single two-level atom is characterized by a
ground level $\left\vert 0 \right\rangle$ and
an excited state $\left\vert 1 \right\rangle$ as
shown in Fig. \ref{fig:1}.
These two states are coupled to a coherent source characterized by
the Rabi parameter $R_1$ and a decay channel characterized
by the decay rate $\gamma_1$.

We start by setting the matrix basis, the density matrix elements,
the number of atoms and the number of quantum levels
\begin{mathematicain}
\(\texttt{na}=1;\\
\texttt{nl}=2;\\
n = \texttt{nl}^{2\texttt{na}};\\
h=\texttt{MultiAtomBasis}[\texttt{nl},\texttt{na}]\\
\texttt{rho}=\texttt{SparseArray}
\left[\texttt{Table}\left[\texttt{ii}\texttt{$\to$}\rho _{\texttt{ii}}[t],\{\texttt{ii},1,n\}\right]\right]\)
\end{mathematicain}
The states
$\left\vert 0 \right\rangle$ and
$\left\vert 1 \right\rangle$
are coupled by an external coherent field
and the incoherent interaction with the electromagnetic background.
The interaction
with the external coherent field is given by the Hamiltonian $H_C(t)$
of Eq. \eqref{eq:cohesource01}, whose parameters are encoded in the
variable \mathttt{transCE} given by
\begin{mathematicain}
\(\texttt{transCE}=\left\{\left\{\{1,2\},1, R_1, \Delta _1,0\right\}\right\}\) 
\end{mathematicain}
\begin{mathematicaout}
\(\left\{\left\{\{1,2\},1,R_1,\Delta _1,0\right\}\right\}\)   
\end{mathematicaout}

This variable indicates that there is only one laser coupling
the states
$\left\vert 0\right\rangle$ and $\left\vert 1\right\rangle$
\mathttt{\{1,2\}} of the single atom
(\mathttt{1}), with Rabi parameter $R_1$
and detuning $\Delta_1$.
The phase due to the atom's position 
$\boldsymbol{k}_i\cdot \boldsymbol{r}_\alpha$
is taken to be zero, assuming
the atom is located at the origin.
Since there is only one atom, the variable \mathttt{trandd},
which lists the coherent interactions among atoms, must be empty
\begin{mathematicain}
\texttt{transdd = $\{\}$.}
\end{mathematicain}
With these variables at hand, we compute the Hamiltonian
\begin{mathematicain}
\texttt{ham = Ham[transCE, transdd, nl, na, t];}
\texttt{MatrixForm[ham]}
\end{mathematicain}
\begin{mathematicaout}
\hspace*{20pt}
\(
\left(
\begin{array}{cc}
 0 & -e^{-i t \Delta _1} R_1 \\
 -e^{i t \Delta _1} R_1 & 0 \\
\end{array}
\right)
\)
\end{mathematicaout}
The Liouville operator corresponding to the Hamiltonian is given by
\begin{mathematicain}
\texttt{LiH = -LiouvilleCommutator[h, ham]}
\end{mathematicain}

The incoherent interaction with the electromagnetic
background is described by the Lindbladian term in Eq. \eqref{eq:lindblad2},
whose parameters are encoded in the variable
\begin{mathematicain}
\(\texttt{transLi}=\left\{\left\{\{1,2\},\{1,2\},1,1,\gamma _1,0\right\}\right\}\)
\end{mathematicain}

This entry implies that there is only decay channel between
the states $|$1$\rangle $ and $|$2$\rangle $ (\mathttt{$\{$1,2$\}$}).
This is a transition between the atom \mathttt{1} and itself,
therefore the two transition lists are followed by a \mathttt{1, 1}.
The parameter \mathttt{\(\gamma_1\)} is the dipole element of the transition
$|$1$\rangle $ $\rightarrow $ { }$|$2$\rangle $.
The final \mathttt{0} in \mathttt{transLi} represents the
energy difference between the two transitions. In this
case, both transitions are in fact the same, so the difference between the energy gaps is strictly zero.
Note that the list \mathttt{$\{$1,2$\}$} appears twice.
This is because transitions with the same energy gap are coupled, meaning
that any existing transition must be coupled with itself.
In more complex systems, however, there may be different transitions
with the same or nearly identical energy gaps that
are therefore coupled. To put it simply, in this case the two lists are the same
(\mathttt{$\{$1,2$\}$,$\{$1,2$\}$}) but in larger systems they could differ
- for example, (\mathttt{$\{$1,2$\}$,$\{$2,3$\}$}).

Having defined \mathttt{transLi}, we are ready to compute the Liouville operator corresponding to 
the Lindbladian
\begin{mathematicain}
\(\texttt{LiL} = \texttt{LiouvilleLindbladian}[h,\texttt{transLi},\texttt{nl},\texttt{na}]\)    
\end{mathematicain}
The overall Liouville operator is the sum of the Hamiltonian and Lindblad Liouville operators
\begin{mathematicain}
\(\texttt{Li} =\texttt{LiH}+\texttt{LiL}\)
\end{mathematicain}
At this point we are ready to generate the system of differential equations for the density
matrix coefficients of the variable \mathttt{rho}.
\begin{mathematicain}
\(\texttt{eqs}=\texttt{LiouvilleMasterEquation}[\texttt{rho},
\texttt{Li},t]\texttt{//}\texttt{ExpToTrig}\)
\end{mathematicain}
\begin{mathematicaout}
\(\bigg\{
\rho _1'[t]==\gamma _1 \rho _2[t]+\sqrt{2} \texttt{Sin}\left[t \Delta _1\right]
R_1 \rho _3[t]-\sqrt{2} \texttt{Cos}\left[t \Delta_1\right] R_1 \rho _4[t],
\\ \hspace*{20pt}
\rho _2'[t]==-\gamma _1 \rho _2[t]-\sqrt{2} \texttt{Sin}\left[t \Delta _1\right] R_1 \rho _3[t]
+\sqrt{2} \texttt{Cos}\left[t\Delta _1\right] R_1 \rho _4[t],
\\ \hspace*{20pt}
\rho _3'[t]==
-\sqrt{2} \texttt{Sin}\left[t \Delta _1\right] R_1 \rho _1[t]
+\sqrt{2} \texttt{Sin}\left[t \Delta _1\right]
R_1 \rho _2[t]-\frac{1}{2} \gamma _1 \rho _3[t],
\\ \hspace*{20pt}
\rho _4'[t]==\sqrt{2} \texttt{Cos}\left[t \Delta _1\right] R_1 \rho _1[t]
-\sqrt{2} \texttt{Cos}\left[t
\Delta _1\right] R_1 \rho _2[t]-\frac{1}{2} \gamma _1 \rho _4[t]\bigg\}\)
\end{mathematicaout}

For a numerical example, we set the physical parameters to
\begin{mathematicain}
\(\texttt{parmvals}=\left\{R_1\texttt{$\to$}12.0,
\Delta _1\texttt{$\to$}1.0,\gamma _1\texttt{$\to$}1.5\right\}\\
\texttt{eqsn}=\texttt{eqs}\texttt{/.}\texttt{parmvals}\)
\end{mathematicain}
\begin{mathematicaout}
\(\left\{R_1\to 12.,\Delta _1\to 1.,\gamma _1\to 1.5\right\}
\\ \\
\bigg\{\rho _1'[t]==1.5 \rho _2[t]+16.9706 \texttt{Sin}[1. t] \rho _3[t]
-16.9706 \texttt{Cos}[1. t] \rho _4[t],
\\ \hspace*{20pt}
\rho _2'[t]==-1.5 \rho _2[t]-16.9706\texttt{Sin}[1. t] \rho _3[t]
+16.9706 \texttt{Cos}[1. t] \rho _4[t],
\\ \hspace*{20pt}
\rho _3'[t]==-16.9706 \texttt{Sin}[1. t] \rho _1[t]
+16.9706 \texttt{Sin}[1. t] \rho _2[t]-0.75\rho _3[t],
\\ \hspace*{20pt}
\rho _4'[t]==16.9706 \texttt{Cos}[1. t] \rho _1[t]
-16.9706 \texttt{Cos}[1. t] \rho _2[t]-0.75 \rho _4[t]\bigg\}
\)
\end{mathematicaout}
We can assume that the system is initially in the lowest energy
level $|$1$\rangle $ . Therefore, we set the initial conditions to
\begin{mathematicain}
\(\texttt{rho0}=\{\{1,0\},\{0,0\}\};\\
\texttt{incon}=\texttt{InitialConditions}[h,\texttt{rho},\texttt{rho0},t]\)
\end{mathematicain}
\begin{mathematicaout}
\(\left\{\rho _1[0]==1,\hspace{10pt}\rho _2[0]==0,
\hspace{10pt}\rho _3[0]==0,\hspace{10pt}\rho _4[0]==0\right\}\)
\end{mathematicaout}
The final ODE is obtained by joining the differential equations and the initial conditions
\begin{mathematicain}
\(\texttt{diffeq}=\texttt{Join}[\texttt{eqsn},\texttt{incon}]\)
\end{mathematicain}
\begin{mathematicaout}
\(\bigg\{
\rho _1'[t]==1.5 \rho _2[t]+16.9706 \texttt{Sin}[1. t] \rho _3[t]-16.9706 \texttt{Cos}[1. t] \rho _4[t],
\\ \hspace*{20pt}
\rho _2'[t]==-1.5 \rho _2[t]-16.9706\texttt{Sin}[1. t] \rho _3[t]+16.9706 \texttt{Cos}[1. t] \rho _4[t],
\\ \hspace*{20pt}
\rho _3'[t]==-16.9706 \texttt{Sin}[1. t] \rho _1[t]+16.9706 \texttt{Sin}[1. t] \rho _2[t]
-0.75\rho _3[t],
\\ \hspace*{20pt}
\rho _4'[t]==16.9706 \texttt{Cos}[1. t] \rho _1[t]-16.9706 \texttt{Cos}[1. t] \rho _2[t]-0.75 \rho _4[t],
\\ \hspace*{20pt}
\rho _1[0]==1,\rho _2[0]==0,\rho_3[0]==0,\rho _4[0]==0\bigg\}\)
\end{mathematicaout}
We can finally obtain the solution of the ODE as
\begin{mathematicain}
\(\texttt{sol}=\texttt{NDSolve}[\texttt{diffeq},\texttt{Normal}[\texttt{rho}],\{t,0,5\}][[1]]\)
\end{mathematicain}
\begin{mathematicaout}
\(\bigg\{\rho _1[t]\to \texttt{InterpolatingFunction}\left[\Box\right][t],
\\ \hspace*{20pt}
\rho _2[t]\to \texttt{InterpolatingFunction}\left[\Box\right][t],
\\ \hspace*{20pt}
\rho_3[t]\to \texttt{InterpolatingFunction}\left[\Box\right][t],
\\ \hspace*{20pt}
\rho _4[t]\to \texttt{InterpolatingFunction}\left[\Box\right][t]\bigg\}\)
\end{mathematicaout}
Now we can observe the dynamics of different expectation values or
the density matrix coefficients. For example, to compute the
populations of $|$1$\rangle$ and $|$2$\rangle $ we calculate the expectation values
\begin{mathematicain}
\(\texttt{pop1}=\texttt{ExpectationValue}[h,\texttt{rho},\{\{1,0\},\{0,0\}\}]\\
\texttt{pop2}=\texttt{ExpectationValue}[h,\texttt{rho},\{\{0,0\},\{0,1\}\}]\\
\texttt{pop1n}=\texttt{pop1}\texttt{/.}\texttt{sol};\\
\texttt{pop2n}=\texttt{pop2}\texttt{/.}\texttt{sol};\)
\end{mathematicain}
\begin{mathematicaout}
\(\rho _1[t]\)\\
\(\rho _2[t]\)
\end{mathematicaout}
\begin{mathematicain}
\(\texttt{Plot}[\{\texttt{pop1n},\texttt{pop2n}\},\{t,0,5\},
\texttt{Axes}\texttt{$\to$}\texttt{False}, \texttt{Frame}\texttt{$\to$}\texttt{True},\)
\\ \hspace*{20pt}
\(\texttt{FrameLabel}\texttt{$\to$}\{\texttt{Style}[\texttt{t},20],
\texttt{Style}[\texttt{{``}Populations{''}},20]\}]\)
\end{mathematicain}
\includegraphics{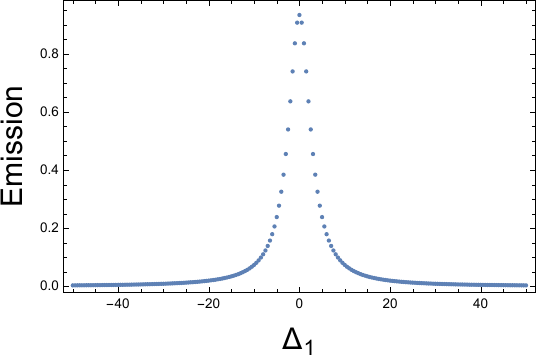}

\noindent With these ideas, we can also perform a spectroscopic analysis of the emission
of the two-level system by sweeping the detuning and evaluating the emission as
$\langle $\(\sigma _{1,2}^{\dagger }\)\(\sigma _{1,2}\)$\rangle $ where
\(\sigma _{1,2}\) = $|$1$\rangle \langle $2$|$
\begin{mathematicain}
\(\texttt{nd}=200;\\
\texttt{rho0}=\{\{1,0\},\{0,1\}\};\\
\texttt{incon}=\texttt{InitialConditions}[h,\texttt{rho},\texttt{rho0},t];\\
\texttt{$\sigma $12}=\texttt{SparseArray}[\{\{1,2\}\texttt{$\to$}1\},2];\\
\texttt{expval}=\texttt{ExpectationValue}[h,\texttt{rho},
\texttt{Transpose}[\texttt{$\sigma $12}].\texttt{$\sigma $12}];\\
\texttt{listplot}=\{\};\\
\texttt{Do}[
\\ \hspace*{20pt}
\texttt{det} =-50+100\texttt{jj}/\texttt{nd};
\\ \hspace*{20pt}
\texttt{parmvals}=\left\{R_1\texttt{$\to$}2.0,\Delta _1\texttt{$\to$}\texttt{det},
\gamma _1\texttt{$\to$}1.5\right\};
\\ \hspace*{20pt}
\texttt{eqsn}=\texttt{eqs}\texttt{/.}\texttt{parmvals};
\\ \hspace*{20pt}
\texttt{diffeq}=\texttt{Join}[\texttt{eqsn},\texttt{incon}];
\\ \hspace*{20pt}
\texttt{sol}=\texttt{NDSolve}[\texttt{diffeq},\texttt{Normal}[\texttt{rho}],\{t,0,10\}][[1]];
\\ \hspace*{20pt}
\texttt{expvaln}=\texttt{expval}\texttt{/.}\texttt{sol}\texttt{/.}\{t\texttt{$\to$}10\};
\\ \hspace*{20pt}
\texttt{AppendTo}[\texttt{listplot},\{\det ,\texttt{expvaln}\}];\\
,\{\texttt{jj},0,\texttt{nd}\}]\)\\
\\
\(\texttt{ListPlot}\big[\texttt{listplot},\texttt{Axes}\texttt{$\to$}\texttt{False},
\texttt{Frame}\texttt{$\to$}\texttt{True},
\\ \hspace*{20pt}
\texttt{FrameLabel}\texttt{$\to$}\left\{\texttt{Style}\left[\texttt{"}
\Delta_1\texttt{"},20\right],\texttt{Style}[\texttt{{``}Emission{''}},20]\right\},
\\ \hspace*{20pt}
\texttt{PlotRange}\texttt{$\to$}\texttt{All}\big]\)
\end{mathematicain}
\begin{mathematicaout}
\(\left\{\rho _1[0]==1,\rho _2[0]==1,\rho _3[0]==0,\rho _4[0]==0\right\}\)
\end{mathematicaout}

\includegraphics{example1_gr2-eps-converted-to.pdf}

\noindent It is possible to abbreviate the process of obtaining the differential
equations by using \mathttt{MasterEquation}, which is a wrapper for
\mathttt{LiouvilleMasterEquation},
\mathttt{LiouvilleLindbladian}, \mathttt{LiouvilleCommutator}, and
\mathttt{Ham}
\begin{mathematicain}
\(\texttt{rho0}=\texttt{SparseArray}[\{\{1,1\}\texttt{$\to$}1\},2];\\
\texttt{eqs}=\texttt{MasterEquation}[h,\texttt{rho},
\texttt{rho0},\texttt{transCE},\texttt{transdd},
\texttt{transLi},\texttt{nl},\texttt{na},t]\)
\end{mathematicain}
\begin{mathematicaout}
\(\bigg\{\rho _1'[t]==\gamma _1 \rho _2[t]+\sqrt{2}
\texttt{Sin}\left[t \Delta _1\right] R_1 \rho _3[t]
-\sqrt{2} \texttt{Cos}\left[t \Delta_1\right] R_1 \rho _4[t],\)
\\ \hspace*{10pt}
\(\rho _2'[t]==-\gamma _1 \rho _2[t]-\sqrt{2} \texttt{Sin}\left[t \Delta _1\right] R_1 \rho _3[t]
+\sqrt{2} \texttt{Cos}\left[t\Delta _1\right] R_1 \rho _4[t],\)
\\ \hspace*{10pt}
\(\rho _3'[t]==-\sqrt{2} \texttt{Sin}\left[t \Delta _1\right] R_1 \rho _1[t]
+\sqrt{2} \texttt{Sin}\left[t \Delta _1\right]R_1 \rho _2[t]
-\frac{1}{2} \gamma _1 \rho _3[t],\)
\\ \hspace*{10pt}
\(\rho _4'[t]==\sqrt{2} \texttt{Cos}\left[t \Delta _1\right] R_1 \rho _1[t]
-\sqrt{2} \texttt{Cos}\left[t\Delta _1\right] R_1 \rho _2[t]
-\frac{1}{2} \gamma _1 \rho _4[t],\)
\\ \hspace*{10pt}
\(\rho _1[t]==1,\rho _2[t]==0,\rho _3[t]==0,\rho _4[t]==0\bigg\}\)
\end{mathematicaout}

Note that in this case the initial condition was set using \mathttt{rho0}
in matrix form rather than as a list of coefficients.

In simpler cases one can attempt to obtain the analytical solution.
For example, if the drive is turned off and the system is initially set in the
excited state $|$2$\rangle $ , we readily obtain the typical equations for spontaneous emission
\begin{mathematicain}
\(\texttt{ode}=\texttt{Join}\left[\texttt{eqs}\texttt{/.}\left\{R_1\texttt{$\to$}0\right\},
\texttt{Table}\left[\rho _{\texttt{ii}}[0]\texttt{==}\texttt{KroneckerDelta}[\texttt{ii},2],
\{\texttt{ii},n\}\right]\right]\\
\texttt{rhon}=\texttt{Normal}[\texttt{rho}];\\
\texttt{DSolve}[\texttt{ode},\texttt{rhon},t]\)
\end{mathematicain}
\begin{mathematicaout}
\(\bigg\{\rho _1'[t]==\gamma _1 \rho _2[t],
\hspace{10pt}
\rho_2'[t]==-\gamma _1 \rho _2[t],\)\\
\hspace*{10pt}
\(\rho _3'[t]==-\frac{1}{2} \gamma _1 \rho _3[t],
\hspace{10pt}
\rho _4'[t]==-\frac{1}{2} \gamma _1 \rho _4[t],\)\\
\hspace*{10pt}
\(\rho _1[0]==0,\rho _2[0]==1,\rho _3[0]==0,\rho _4[0]==0\bigg\}\)\\
\(\bigg\{\left\{\rho _1[t]\to e^{-t \gamma _1} \left(-1+e^{t \gamma _1}\right),
\hspace{10pt}
\rho _2[t]\to e^{-t \gamma _1},
\hspace{10pt}
\rho _3[t]\to 0,
\hspace{10pt}
\rho _4[t]\to 0\right\}\bigg\}\)
\end{mathematicaout}

\begin{figure}
\includegraphics[width=0.25\textwidth]{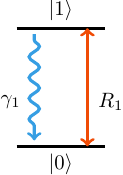}
\caption{Scheme of a single two-level atom. This system is characterized
by the ground state $\ket{0}$ and the excited state $\ket{1}$.
These two states are coupled to a coherent source characterized by
the Rabi parameter $R_1$ and a decay channel characterized
by the decay rate $\gamma_1$.
}\label{fig:2}
\end{figure}

\subsection{Two two-level atoms}\label{sec:example2}
The system of two two-level atoms is characterized by ground states
$\lvert 0 \rangle_1$ and $\lvert 0 \rangle_2$, and excited states
$\lvert 1 \rangle_1$ and $\lvert 1 \rangle_2$, corresponding to atoms 1 and 2,
as shown in Fig. \ref{fig:2}. The quantum levels of each atom are coupled
to a coherent source through the Rabi parameter $R_1$, which is common to
both atoms. These two levels are also connected to decay channels characterized
by the decay rate $\gamma_1$. In addition, the quantum levels of one atom are
coupled to those of the other via dipole–dipole interactions, mainly determined
by $\Omega_1$. Furthermore, the quantum levels are coupled through a collective
decay channel with decay rate $\gamma_{1,2}$.

\begin{figure}
\includegraphics[width=0.5\textwidth]{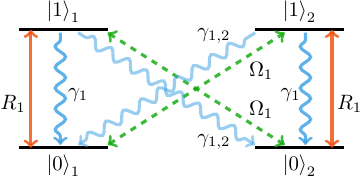}
\caption{Scheme of two two-level atoms. 
It is characterized by ground states
$\lvert 0 \rangle_1$ and $\lvert 0 \rangle_2$, and excited states
$\lvert 1 \rangle_1$ and $\lvert 1 \rangle_2$, corresponding to atoms $1$ and $2$.
The quantum levels of each atom are coupled
to a coherent source through the Rabi parameter $R_1$, which is common to
both atoms. These two levels are also connected to decay channels characterized
by the decay rate $\gamma_1$. In addition, the quantum levels of one atom are
coupled to those of the other via dipole–dipole interactions, mainly determined
by $\Omega_1$. Furthermore, the quantum levels are coupled through a collective
decay channel with decay rate $\gamma_{1,2}$.
}\label{fig:3}
\end{figure}

We start by setting the matrix basis, the density matrix elements,
the number of atoms and the number of quantum levels
\begin{mathematicain}
\(\texttt{na}=2;\\
\texttt{nl}=2;\\
n = \texttt{nl}^{2\texttt{na}};\\
h=\texttt{MultiAtomBasis}[\texttt{nl},\texttt{na}]\\
\texttt{rho}=\texttt{SparseArray}
\left[\texttt{Table}\left[\texttt{ii}\texttt{$\to$}\rho_{\texttt{ii}}[t],
\{\texttt{ii},1,n\}\right]\right]\)
\end{mathematicain}
\begin{mathematicaout}
\(\texttt{SparseArray}\left[\Box\right]\)\\
\(\texttt{SparseArray}\left[\Box\right]\)
\end{mathematicaout}
Next, we set the Hamiltonian parameters.
The \mathttt{transCE} configuration variable, which describes this
light-matter interaction, is
\begin{mathematicain}
\(\texttt{transCE}
=\left\{\left\{\{1,2\},1, R_1, \Delta _1,0\right\},
\left\{\{1,2\},2, R_1, \Delta _1,\phi \right\}\right\}\)
\end{mathematicain}
\begin{mathematicaout}
\(\left\{\left\{\{1,2\},1,R_1,\Delta _1,0\right\},\left\{\{1,2\},2,R_1,\Delta _1,\phi \right\}\right\}\)
\end{mathematicaout}
This variable indicates that there is only one laser coupling the states
$|$1$\rangle $ and $|$2$\rangle $ ($\{$1,2$\}$) of both atoms ($1$ and $2$),
with Rabi parameter \(R_1\) and detuning \(\Delta _1\).
The phase due to the atoms' positions is taken to be zero for the first atom and
$\phi$ for the second. Assuming that both atoms lie along the propagation direction
of the incident field, there must be a phase difference of $\phi $
between them. The variable \mathttt{trandd}, which lists the coherent interactions
among atoms, is given by
\begin{mathematicain}
\(\texttt{transdd}=\left\{\left\{\{1,2\},\{1,2\},1,2,\Omega _1,0\right\},
\left\{\{1,2\},\{1,2\},2,1,\Omega _1,0\right\}\right\}\)
\end{mathematicain}
\begin{mathematicaout}
\(\left\{\left\{\{1,2\},\{1,2\},1,2,\Omega _1,0\right\},
\left\{\{1,2\},\{1,2\},2,1,\Omega _1,0\right\}\right\}\)
\end{mathematicaout}
This variable can be read as follows. The first two elements are the coupled transitions.
The list \mathttt{$\{$1,2$\}$} appears twice. This is because transitions
with the same energy gap are coupled, meaning that any existing transition must
be coupled with itself. In more complex systems, however, there may
be different transitions with the same or nearly identical energy gaps
that are therefore coupled. To put it simply, in this case the two lists are
the same (\mathttt{$\{$1,2$\}$,$\{$1,2$\}$}) but in larger systems they could differ
- for example, (\mathttt{$\{$1,2$\}$,$\{$2,3$\}$}). The following two numbers
(1,2 in the first element and 2,1 in the second) label the atoms.
\(\Omega _1\) is the coupling constant between these two transtions.
The final $0$ in \mathttt{transdd} represents the energy difference between
the two transitions. In this case, both transitions are in fact the same,
so the difference between the energy gaps is strictly zero.
In the case of two different transitions with similar energies,
this number must be set equal to their energy difference.

With these variables at hand, we compute the Hamiltonian
\begin{mathematicain}
\(\texttt{ham}=\texttt{Ham}[\texttt{transCE},\texttt{transdd},\texttt{nl},\texttt{na},t];\\
\texttt{MatrixForm}[\texttt{ham}]\)
\end{mathematicain}
\begin{mathematicaout}
\(
\\ \hspace*{20pt}
\left(
\begin{array}{cccc}
 0 & -e^{i \left(\phi -t \Delta _1\right)} R_1 & -e^{-i t \Delta _1} R_1 & 0 \\
 -e^{i \left(-\phi +t \Delta _1\right)} R_1 & 0 & \Omega _1 & -e^{-i t \Delta _1} R_1 \\
 -e^{i t \Delta _1} R_1 & \Omega _1 & 0 & -e^{i \left(\phi -t \Delta _1\right)} R_1 \\
 0 & -e^{i t \Delta _1} R_1 & -e^{i \left(-\phi +t \Delta _1\right)} R_1 & 0 \\
\end{array}
\right)\)
\end{mathematicaout}

The Liouville operator corresponding to the Hamiltonian is given by
\begin{mathematicain}
\noindent\(\texttt{LiH}=-\texttt{LiouvilleCommutator}[h,\texttt{ham}]\)
\end{mathematicain}
\begin{mathematicaout}
\(\texttt{SparseArray}\left[\Box\right]\)
\end{mathematicaout}

In the following we set the Lindbladian parameters.
These are entered through the configuration variable \mathttt{transLi}
\begin{mathematicain}
\(\texttt{transLi}=\left\{\bigg\{\{1,2\},\{1,2\},1,1,\gamma _1,0\right\},
\left\{\{1,2\},\{1,2\},2,2,\gamma _1,0\right\},
\\ \hspace*{20pt}
\left\{\{1,2\},\{1,2\},1,2,\gamma_{1,2},0\right\},
\left\{\{1,2\},\{1,2\},2,1,\gamma _{1,2},0\right\}\bigg\}\)
\end{mathematicain}
\begin{mathematicaout}
\(\bigg\{\left\{\{1,2\},\{1,2\},1,1,\gamma _1,0\right\},
\left\{\{1,2\},\{1,2\},2,2,\gamma _1,0\right\},\)
\\ \hspace*{20pt}
\(\left\{\{1,2\},\{1,2\},1,2,\gamma_{1,2},0\right\},
\left\{\{1,2\},\{1,2\},2,1,\gamma _{1,2},0\right\}\bigg\}\)
\end{mathematicaout}
The first two entries of \mathttt{transLi} list the decay channels between
the states $|$1$\rangle $ and $|$2$\rangle $ ($\{$1,2$\}$) for each atom. These
transitions occur internally within each atom, so the next two transition entries
are $1$, $1$ and $2$, $2$. The parameter \(\gamma _1\) corresponds to the
dipole element of this transition. The last two terms give rise to the collective
channels formed between the transitions $|$1$\rangle $ $\rightarrow
$ { }$|$2$\rangle $ between different atoms.
The parameter \(\gamma _{1,2}\) is the dipole element of this transition.
The final $0$ in \mathttt{transLi} represents the energy difference between
the two transitions. In this case, both transitions are in fact the same,
so the difference between the energy gaps is strictly zero.

Note that the list \mathttt{$\{$1,2$\}$} appears twice.
This is because transitions with the same energy gap are coupled,
meaning that any existing transition must be coupled with itself.
In more complex systems, however, there may be different transitions
with the same or nearly identical energy gaps that are therefore coupled.
To put it simply, in this case the two lists are the same
(\mathttt{$\{$1,2$\}$,$\{$1,2$\}$}) but in larger systems they could
differ - for example, (\mathttt{$\{$1,2$\}$,$\{$2,3$\}$}).
If the energy difference between these two transitions is nonzero,
the last entry of the corresponding element of \mathttt{transLi}
should be exactly this energy difference.
Having defined \mathttt{transLi}, we are ready to compute
the Liouville operator corresponding to the Lindbladian
\begin{mathematicain}
\(\texttt{LiL} = \texttt{LiouvilleLindbladian}[h,\texttt{transLi},\texttt{nl},\texttt{na}]\)
\end{mathematicain}
\begin{mathematicaout}
\(\texttt{SparseArray}\left[\Box\right]\)
\end{mathematicaout}

The overall Liouville operator is the sum of the Hamiltonian and Lindblad Liouville operators
\begin{mathematicain}
\(\texttt{Li} =\texttt{LiH}+\texttt{LiL}\)
\end{mathematicain}
\begin{mathematicaout}
\noindent\(\texttt{SparseArray}\left[\Box\right]\)
\end{mathematicaout}

At this point we are ready to generate the system of differential equations
for the density matrix coefficients of the variable rho
\begin{mathematicain}
\(\texttt{eqs}=\texttt{LiouvilleMasterEquation}[\texttt{rho},\texttt{Li},t]
\texttt{//}\texttt{ExpToTrig};\\
\texttt{eqs}[[1\texttt{;;}2]]\)
\end{mathematicain}
\begin{mathematicaout}
\(\bigg\{\rho _1'[t]==\gamma _1 \rho _2[t]
-\sqrt{2} \texttt{Sin}\left[\phi -t \Delta _1\right] R_1 \rho _3[t]
-\sqrt{2} \texttt{Cos}\left[\phi-t \Delta _1\right] R_1 \rho _4[t]\)
\\ \hspace*{20pt}
\(+\gamma _1 \rho _5[t]
+\sqrt{2} \texttt{Sin}\left[t \Delta _1\right] R_1 \rho _9[t]
+\gamma _{1,2} \rho _{11}[t]
\\ \hspace*{20pt}
-\sqrt{2}\texttt{Cos}\left[t \Delta _1\right] R_1 \rho _{13}[t]
+\gamma _{1,2} \rho _{16}[t],\)
\\ \hspace*{10pt}
\(\rho _2'[t]==-\gamma _1 \rho _2[t]
+\sqrt{2} \texttt{Sin}\left[\phi-t \Delta _1\right] R_1 \rho _3[t]
+\sqrt{2} \texttt{Cos}\left[\phi -t \Delta _1\right] R_1 \rho _4[t]\)
\\ \hspace*{20pt}
\(+\gamma _1 \rho _6[t]+\sqrt{2} \texttt{Sin}\left[t\Delta _1\right] R_1 \rho _{10}[t]
-\frac{1}{2} \gamma _{1,2} \rho _{11}[t]-\Omega _1 \rho _{12}[t]
\)
\\ \hspace*{20pt}
\(-\sqrt{2} \texttt{Cos}\left[t \Delta _1\right] R_1\rho _{14}[t]
+\Omega _1 \rho _{15}[t]-\frac{1}{2} \gamma _{1,2} \rho _{16}[t]\bigg\}\)
\end{mathematicaout}

For a numerical example, we set the physical parameters to
\begin{mathematicain}
\(\texttt{parmvals}=\left\{R_1\texttt{$\to$}12.0,
\Delta _1\texttt{$\to$}1.0,\phi \texttt{$\to$}1.1,\gamma _1\texttt{$\to$}1.5,
\gamma _{1,2}\texttt{$\to$}0.5,\Omega_1\texttt{$\to$}0.7\right\}\\
\texttt{eqsn}=\texttt{eqs}\texttt{/.}\texttt{parmvals};\\
\texttt{eqsn}[[1\texttt{;;}2]]\)
\end{mathematicain}
\begin{mathematicaout}
\(\left\{R_1\to 12.,\Delta _1\to 1.,\phi \to 1.1,\gamma _1\to 1.5,
\gamma _{1,2}\to 0.5,\Omega _1\to 0.7\right\}\)\\
\\
\(\bigg\{\rho _1'[t]==1.5 \rho _2[t]-16.9706 \texttt{Sin}[1.1\, -1. t] \rho _3[t]
-16.9706 \texttt{Cos}[1.1\, -1. t] \rho _4[t]\)
\\ \hspace*{20pt}
\(+1.5 \rho _5[t]
+16.9706\texttt{Sin}[1. t] \rho _9[t]+0.5 \rho _{11}[t]
\\ \hspace*{20pt}
-16.9706 \texttt{Cos}[1. t] \rho _{13}[t]
+0.5 \rho _{16}[t],\)
\\ \hspace*{10pt}
\(\rho _2'[t]==-1.5 \rho _2[t]
+16.9706 \texttt{Sin}[1.1\,-1. t] \rho _3[t]
+16.9706 \texttt{Cos}[1.1\, -1. t] \rho _4[t]\)
\\ \hspace*{20pt}
\(+1.5 \rho _6[t]+16.9706 \texttt{Sin}[1. t] \rho _{10}[t]
-0.25 \rho _{11}[t]-0.7 \rho _{12}[t]
\\ \hspace*{20pt}
-16.9706\texttt{Cos}[1. t] \rho _{14}[t]\)
\(+0.7 \rho _{15}[t]-0.25 \rho _{16}[t]\bigg\}\)
\end{mathematicaout}

We can assume that the atoms are both initially in the lowest energy level
$|$1$\rangle $ . Therefore, we set the initial conditions to \(\rho
_0\)=\(\eta _0\)\(\otimes\)\(\eta _0\) where \(\eta _0\)=$|$0$\rangle \langle $0$|$
\begin{mathematicain}
\(\texttt{$\eta $0}=\texttt{SparseArray}[\{\{1,1\}\texttt{$\to$}1\},2]\\
\texttt{rho0}=\texttt{KroneckerProduct}[\texttt{$\eta $0},\texttt{$\eta $0}]\\
\texttt{incon}=\texttt{InitialConditions}[h,\texttt{rho},\texttt{rho0},t]\)
\end{mathematicain}
\begin{mathematicaout}
\(\texttt{SparseArray}\left[\Box\right]\)\\
\(\texttt{SparseArray}\left[\Box\right]\)\\
\(\bigg\{\rho _1[0]==1,\rho _2[0]==0,\rho _3[0]==0,\rho _4[0]==0,\rho _5[0]==0,\rho _6[0]==0,\)
\\ \hspace*{20pt}
\(\rho _7[0]==0,\rho _8[0]==0,\rho _9[0]==0,\rho_{10}[0]==0,\rho _{11}[0]==0,\rho _{12}[0]==0,
\\ \hspace*{20pt}
\rho _{13}[0]==0,\)
\(\rho _{14}[0]==0, \rho _{15}[0]==0,\rho _{16}[0]==0\bigg\}\)
\end{mathematicaout}
The final ODE is obtained by joining the differential equations and the initial conditions
\begin{mathematicain}
\(\texttt{diffeq}=\texttt{Join}[\texttt{eqsn},\texttt{incon}];\\
\texttt{diffeq}[[\{1,2,17,18\}]]\)
\end{mathematicain}
\begin{mathematicaout}
\(\bigg\{
\rho _1'[t]==1.5 \rho _2[t]-16.9706 \texttt{Sin}[1.1\, -1. t] \rho _3[t]
-16.9706 \texttt{Cos}[1.1\, -1. t] \rho _4[t]\)
\\ \hspace*{20pt}
\(+1.5 \rho _5[t]+16.9706\texttt{Sin}[1. t] \rho _9[t]
+0.5 \rho _{11}[t]-16.9706 \texttt{Cos}[1. t] \rho _{13}[t]
\\ \hspace*{20pt}
+0.5 \rho _{16}[t],\)\\
\hspace*{10pt}
\(\rho _2'[t]==-1.5 \rho _2[t]+16.9706 \texttt{Sin}[1.1\,-1. t] \rho _3[t]
+16.9706 \texttt{Cos}[1.1\, -1. t] \rho _4[t]\)
\\ \hspace*{20pt}
\(+1.5 \rho _6[t]
+16.9706 \texttt{Sin}[1. t] \rho _{10}[t]-0.25 \rho _{11}[t]-0.7 \rho _{12}[t]
\\ \hspace*{20pt}
-16.9706\texttt{Cos}[1. t] \rho _{14}[t]\)
\(+0.7 \rho _{15}[t] -0.25 \rho _{16}[t],
\\ \hspace*{20pt}
\rho _1[0]==1,\rho _2[0]==0\bigg\}\)
\end{mathematicaout}

We can finally obtain the solution of the ODE as

\begin{mathematicain}
\(\mathcomm{\texttt{sol}=\texttt{NDSolve}[\texttt{diffeq},
\texttt{Normal}[\texttt{rho}],\{t,0,5\}][[1]];}\\
\mathcomm{\texttt{sol}[[1\texttt{;;}2]]}\)
\end{mathematicain}
\begin{mathematicaout}
\(\bigg\{\rho _1[t]\to \texttt{InterpolatingFunction}\left[\Box\right][t],\)\\
\hspace*{30pt}
\(\rho _2[t]\to \texttt{InterpolatingFunction}\left[\Box\right][t]\bigg\}\)
\end{mathematicaout}

Now we can observe the dynamics of different expectation values or the density
matrix coefficients. For example, to compute the populations of $|$1$\rangle$
and $|$2$\rangle $ for both atoms we calculate the expectation values
\begin{mathematicain}
\(\mathcomm{\texttt{p1}=\texttt{SparseArray}[\{\{1,1\}\texttt{$\to$}1\},2];}\\
\mathcomm{\texttt{p2}=\texttt{SparseArray}[\{\{2,2\}\texttt{$\to$}1\},2];}\\
\mathcomm{\texttt{one}=\texttt{SparseArray}[\{\{1,1\}\texttt{$\to$}1,\{2,2\}\texttt{$\to$}1\},2];}\\
\mathcomm{\texttt{pop11}=\texttt{ExpectationValue}[h,\texttt{rho},
\texttt{KroneckerProduct}[\texttt{p1},\texttt{one}]]}\\
\mathcomm{\texttt{pop21}=\texttt{ExpectationValue}[h,\texttt{rho},
\texttt{KroneckerProduct}[\texttt{p2},\texttt{one}]]}\\
\mathcomm{\texttt{pop12}=\texttt{ExpectationValue}[h,\texttt{rho},
\texttt{KroneckerProduct}[\texttt{one},\texttt{p1}]]}\\
\mathcomm{\texttt{pop22}=\texttt{ExpectationValue}[h,\texttt{rho},
\texttt{KroneckerProduct}[\texttt{one},\texttt{p2}]]}\\
\mathcomm{}\\
\mathcomm{\texttt{pop11n}=\texttt{pop11}\texttt{/.}\texttt{sol};}\\
\mathcomm{\texttt{pop21n}=\texttt{pop21}\texttt{/.}\texttt{sol};}\\
\mathcomm{\texttt{pop12n}=\texttt{pop12}\texttt{/.}\texttt{sol};}\\
\mathcomm{\texttt{pop22n}=\texttt{pop22}\texttt{/.}\texttt{sol};}\)
\end{mathematicain}
\begin{mathematicaout}
\(\rho _1[t]+\rho _2[t]\)\\
\(\rho _5[t]+\rho _6[t]\)\\
\(\rho _1[t]+\rho _5[t]\)\\
\(\rho _2[t]+\rho _6[t]\)
\end{mathematicaout}
\begin{mathematicain}
\(\texttt{Plot}[\{\texttt{pop11n},\texttt{pop12n}\},\{t,0,5\},
\texttt{Axes}\texttt{$\to$}\texttt{False},\texttt{Frame}\texttt{$\to$}\texttt{True},\)
\\ \hspace*{30pt}
\(\texttt{FrameLabel}\texttt{$\to$}\{\texttt{Style}[\texttt{t},20],
\texttt{Style}[\texttt{{``}Populations{''}},20]\}]\)
\end{mathematicain}

\includegraphics{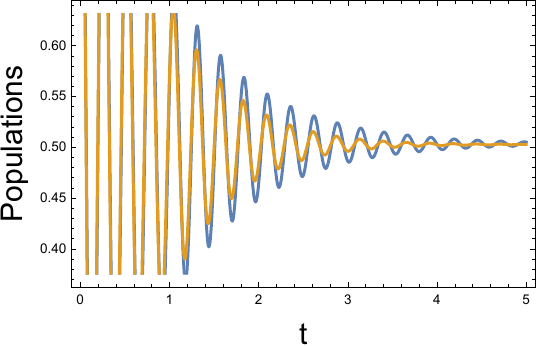}

\begin{mathematicain}
\noindent\(\texttt{Plot}[\{\texttt{pop21n},\texttt{pop22n}\},\{t,0,5\},
\texttt{Axes}\texttt{$\to$}\texttt{False},\texttt{Frame}\texttt{$\to$}\texttt{True},\)\\
\hspace*{30pt}
\(\texttt{FrameLabel}\texttt{$\to$}\{\texttt{Style}[\texttt{t},20],
\texttt{Style}[\texttt{{``}Populations{''}},20]\}]\)
\end{mathematicain}

\includegraphics{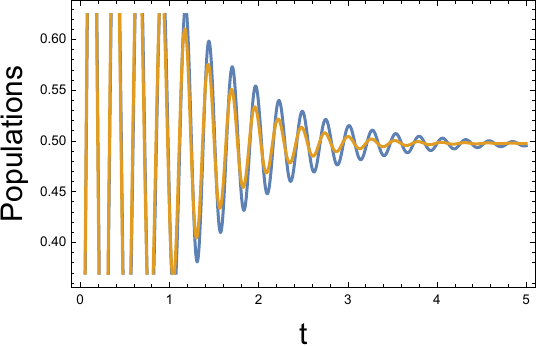}

With these ideas, we can also perform a spectroscopic analysis of the emission
of the two-level system by sweeping the detuning and evaluating the
correlation
\mathttt{corr} = $\langle $\(\sigma _{2,1}^1\)\(\sigma _{1,2}^2\)
+\(\sigma _{2,1}^2\)\(\sigma _{1,2}^1\)$\rangle $
where \(\sigma _{1,2}^1\) =1$\otimes$ $|$1$\rangle \langle $2$|$
and \(\sigma _{1,2}^2\) = $|$1$\rangle \langle $2$|\otimes $1
\begin{mathematicain}
\(\texttt{nd}=200;\\
\texttt{$\eta $0}=\texttt{SparseArray}[\{\{1,1\}\texttt{$\to$}1\},2];\\
\texttt{rho0}=\texttt{KroneckerProduct}[\texttt{$\eta $0},\texttt{$\eta $0}];\\
\texttt{incon}=\texttt{InitialConditions}[h,\texttt{rho},\texttt{rho0},t];\\
\texttt{corr}=\texttt{KroneckerProduct}[\texttt{SparseArray}[\{\{1,2\}\texttt{$\to$}1\},2],
\\ \hspace*{40pt}
\texttt{SparseArray}[\{\{2,1\}\texttt{$\to$}1\},2]]
\\ \hspace*{20pt}
+\texttt{KroneckerProduct}[\texttt{SparseArray}[\{\{2,1\}\texttt{$\to$}1\},2],
\\ \hspace*{40pt}
\texttt{SparseArray}[\{\{1,2\}\texttt{$\to$}1\},2]];\\
\texttt{expval}=\texttt{ExpectationValue}[h,\texttt{rho},\texttt{corr}];\\
\texttt{listplot}=\{\};\\
\texttt{Do}[\\
\hspace*{20pt}\texttt{det} =-150+300\texttt{jj}/\texttt{nd};\\
\hspace*{20pt}\texttt{parmvals}=\left\{R_1\texttt{$\to$}12.0,\Delta _1\texttt{$\to$}\texttt{det} ,
\phi \texttt{$\to$}1.1,\gamma _1\texttt{$\to$}1.5,\gamma _{1,2}\texttt{$\to$}0.5,
\Omega_1\texttt{$\to$}0.7\right\};\\
\hspace*{20pt}\texttt{eqsn}=\texttt{eqs}\texttt{/.}\texttt{parmvals};\\
\hspace*{20pt}\texttt{diffeq}=\texttt{Join}[\texttt{eqsn},\texttt{incon}];\\
\hspace*{20pt}\texttt{sol}=\texttt{NDSolve}[\texttt{diffeq},\texttt{Normal}[\texttt{rho}],\{t,0,10\}][[1]];\\
\hspace*{20pt}\texttt{expvaln}=\texttt{expval}\texttt{/.}\texttt{sol}\texttt{/.}\{t\texttt{$\to$}10\};\\
\hspace*{20pt}\texttt{AppendTo}[\texttt{listplot},\{\det ,\texttt{expvaln}\}];\\
,\{\texttt{jj},0,\texttt{nd}\}]\\
\)\\
\(\texttt{ListPlot}\big[\texttt{listplot},\texttt{Axes}\texttt{$\to$}\texttt{False},
\texttt{Frame}\texttt{$\to$}\texttt{True},\)\\
\hspace*{30pt}
\(\texttt{FrameLabel}\texttt{$\to$}
\left\{\texttt{Style}\left[\texttt{"}\Delta_1\texttt{"},20\right],
\texttt{Style}\left[\texttt{"}\langle +\rangle \texttt{"},20\right]\right\},\)\\
\hspace*{30pt}
\(\texttt{PlotRange}\texttt{$\to$}\texttt{All}]\)
\end{mathematicain}

\includegraphics{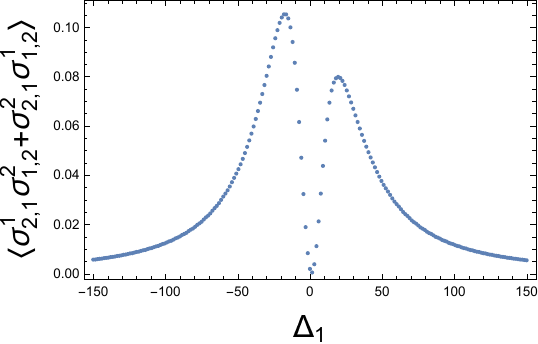}

It is possible to abbreviate the process of obtaining the differential equations
by using \mathttt{MasterEquation}, which is a wrapper for \mathttt{LiouvilleMasterEquation},
\mathttt{LiouvilleLindbladian}, \mathttt{LiouvilleCommutator}, and \mathttt{Ham}
\begin{mathematicain}
\(\mathcomm{\texttt{$\eta $0}=\texttt{SparseArray}[\{\{1,1\}\texttt{$\to$}1\},2]}\\
\mathcomm{\texttt{rho0}=\texttt{KroneckerProduct}[\texttt{$\eta $0},\texttt{$\eta $0}]}\\
\mathcomm{\texttt{eqs}=\texttt{MasterEquation}[h,\texttt{rho},\texttt{rho0},
\texttt{transCE},\texttt{transdd},\texttt{transLi},\texttt{nl},\texttt{na},t];}\\
\mathcomm{\texttt{eqs}[[\{1,2,17,18\}]]}\)
\end{mathematicain}
\begin{mathematicaout}
\noindent\(\texttt{SparseArray}\left[\Box\right]\)\\
\noindent\(\texttt{SparseArray}\left[\Box\right]\)\\
\noindent\(\bigg\{\rho _1'[t]==\gamma _1 \rho _2[t]
-\sqrt{2} \texttt{Sin}\left[\phi -t \Delta _1\right] R_1 \rho _3[t]
-\sqrt{2} \texttt{Cos}\left[\phi-t \Delta _1\right] R_1 \rho _4[t]
+\gamma _1 \rho _5[t]\)
\\ \hspace*{20pt}
\(+\sqrt{2} \texttt{Sin}\left[t \Delta _1\right] R_1 \rho _9[t]
+\gamma _{1,2} \rho _{11}[t]-\sqrt{2}\texttt{Cos}\left[t \Delta _1\right] R_1 \rho _{13}[t]
+\gamma _{1,2} \rho _{16}[t],\)
\\ \hspace*{10pt}
\(\rho _2'[t]==-\gamma _1 \rho _2[t]
+\sqrt{2} \texttt{Sin}\left[\phi-t \Delta _1\right] R_1 \rho _3[t]
+\sqrt{2} \texttt{Cos}\left[\phi -t \Delta _1\right] R_1 \rho _4[t]
\)
\\ \hspace*{20pt}
\(+\gamma _1 \rho _6[t]+\sqrt{2} \texttt{Sin}\left[t\Delta _1\right] R_1 \rho _{10}[t]
-\frac{1}{2} \gamma _{1,2} \rho _{11}[t]-\Omega _1 \rho _{12}[t]
-\sqrt{2} \texttt{Cos}\left[t \Delta _1\right] R_1\rho _{14}[t]
\)
\\ \hspace*{20pt}
\(+\Omega _1 \rho _{15}[t]-\frac{1}{2} \gamma _{1,2} \rho _{16}[t],\)
\\ \hspace*{10pt}
\(\rho _1[t]==1,\rho _2[t]==0\bigg\}\)
\end{mathematicaout}
Note that in this case the initial condition was set using \mathttt{rho0}
in matrix form rather than as a list of coefficients.

\subsection{Five two-level atoms}\label{sec:example3}
The system of five two-level atoms is characterized by ground states
$\lvert 0 \rangle_\alpha$ for $\alpha = 1,2,3,4,5$, and excited states
$\lvert 1 \rangle_\alpha$ for $\alpha = 1,2,3,4,5$, corresponding to
atoms $1$ through $5$, as shown in Fig. \ref{fig:3}.
The quantum levels of each atom are coupled to a coherent source
through the Rabi parameter $R_1$, which is common to all five atoms.
These levels are also connected to decay channels characterized by
the decay rate $\gamma_1$. In addition, the quantum levels of one atom
are coupled to those of its neighbors via dipole–dipole interactions,
mainly determined by $\Omega_1$.
Moreover, neighboring atoms are coupled through a collective
decay channel with decay rate $\gamma_{1,2}$.

\begin{figure}
\includegraphics[width=0.99\textwidth]{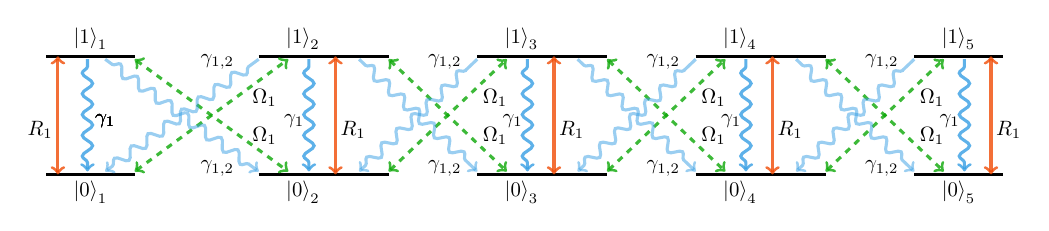}
\caption{Scheme of five two-level atoms. 
It is characterized by ground states
$\lvert 0 \rangle_\alpha$ for $\alpha = 1,2,3,4,5$, and excited states
$\lvert 1 \rangle_\alpha$ for $\alpha = 1,2,3,4,5$, corresponding to
atoms $1$ through $5$.
The quantum levels of each atom are coupled to a coherent source
through the Rabi parameter $R_1$, which is common to all five atoms.
These levels are also connected to decay channels characterized by
the decay rate $\gamma_1$. In addition, the quantum levels of one atom
are coupled to those of its neighbors via dipole–dipole interactions,
mainly determined by $\Omega_1$.
Moreover, neighboring atoms are coupled through a collective
decay channel with decay rate $\gamma_{1,2}$.
}\label{fig:4}
\end{figure}

We start by setting the matrix basis, the density matrix elements,
the number of atoms and the number of quantum levels

\begin{mathematicain}
\noindent\(\mathcomm{\texttt{na}=5;}\\
\mathcomm{\texttt{nl}=2;}\\
\mathcomm{n = \texttt{nl}^{2\texttt{na}};}\\
\mathcomm{h=\texttt{MultiAtomBasis}[\texttt{nl},\texttt{na}]}\\
\mathcomm{\texttt{rho}=\texttt{SparseArray}\left[\texttt{Table}\left[\texttt{ii}\texttt{$\to$}\rho _{\texttt{ii}}[t],\{\texttt{ii},1,n\}\right]\right]}\)
\end{mathematicain}
\begin{mathematicaout}
\(\texttt{SparseArray}[\Box]\)\\
\(\texttt{SparseArray}\left[\Box\right]\)
\end{mathematicaout}
Next, we set the Hamiltonian parameters.
The \mathttt{transCE} configuration variable, which describes this light-matter
interaction, is
\begin{mathematicain}
\noindent\(\mathcomm{\texttt{transCE}=\texttt{Table}\left[\left\{\{1,2\},\texttt{ii}, R_1, \Delta _1,(\texttt{ii}-1)\phi \right\},\{\texttt{ii},1,5\}\right]}\)
\end{mathematicain}
\begin{mathematicaout}
\(\bigg\{\left\{\{1,2\},1,R_1,\Delta _1,0\right\},
\left\{\{1,2\},2,R_1,\Delta_1,\phi \right\},\)
\\ \hspace*{20pt}
\(\left\{\{1,2\},3,R_1,\Delta _1,2 \phi \right\},
\left\{\{1,2\},4,R_1,\Delta_1,3 \phi \right\},
\left\{\{1,2\},5,R_1,\Delta _1,4 \phi \right\}\bigg\}\)
\end{mathematicaout}
This variable indicates that there is only one laser coupling the states
$\lvert 1 \rangle$ and $\lvert 2 \rangle$ ($\{1,2\}$) of the five atoms,
with Rabi parameter $R_{1}$ and detuning $\Delta_{1}$.
The phase arising from the atoms’ positions is taken to be
zero for the first atom and to increase by $\phi$ for each subsequent atom.
Assuming that the five atoms lie along the propagation direction of the
incident field, there is a phase difference of $\phi$ between each
consecutive pair of atoms. If the interaction occurs only between
nearest neighbors, the variable \mathttt{trandd}, which lists the
coherent interactions among atoms, is given by
\begin{mathematicain}
\(\mathcomm{\texttt{transdd}=\texttt{Join}
\left[\texttt{Table}\left[\left\{\{1,2\},\{1,2\},\texttt{ii},
\texttt{ii}+1,\Omega _1,0\right\},\{\texttt{ii},1,\texttt{na}-1\}\right],\right.}\\
\mathcomm{\left.\texttt{Table}\left[\left\{\{1,2\},\{1,2\},\texttt{ii}+1,\texttt{ii},\Omega _1,0\right\},\{\texttt{ii},1,\texttt{na}-1\}\right]\right]}\)
\end{mathematicain}
\begin{mathematicaout}
\(\bigg\{\left\{\{1,2\},\{1,2\},1,2,\Omega _1,0\right\},
\left\{\{1,2\},\{1,2\},2,3,\Omega _1,0\right\},
\left\{\{1,2\},\{1,2\},3,4,\Omega_1,0\right\},\)
\\ \hspace*{20pt}
\(\left\{\{1,2\},\{1,2\},4,5,\Omega _1,0\right\},
\left\{\{1,2\},\{1,2\},2,1,\Omega _1,0\right\},
\left\{\{1,2\},\{1,2\},3,2,\Omega _1,0\right\},\)
\\ \hspace*{10pt}
\(\left\{\{1,2\},\{1,2\},4,3,\Omega_1,0\right\},
\left\{\{1,2\},\{1,2\},5,4,\Omega _1,0\right\}\bigg\}\)
\end{mathematicaout}
This variable can be read as follows. The first two elements are the coupled transitions.
The list $\{$1,2$\}$ appears twice. This is because transitions
with the same energy gap are coupled, meaning that any existing
transition must be coupled with itself. In more complex systems,
however, there may be different transitions with the same or nearly
identical energy gaps that are therefore coupled. To put it simply,
in this case the two lists are the same (\mathttt{$\{$1,2$\}$,$\{$1,2$\}$})
but in larger systems they could differ - for example,
(\mathttt{$\{$1,2$\}$,$\{$2,3$\}$}).
The following two numbers ($1$, $2$) in the first element and $2$, $1$ in the second)
label the atoms. \(\Omega _1\) is the coupling constant between these two transtions.
The final $0$ in \mathttt{transdd} represents the energy difference between
the two transitions. In this case, both transitions are in fact the same,
so the difference between the energy gaps is strictly zero.
In the case of two different transitions with similar energies,
this number must be set equal to their energy difference.

With these variables at hand, we compute the Hamiltonian
\begin{mathematicain}
\(\mathcomm{\texttt{ham}=\texttt{Ham}[\texttt{transCE},\texttt{transdd},\texttt{nl},\texttt{na},t]}\)
\end{mathematicain}
\begin{mathematicaout}
\(\texttt{SparseArray}\left[\Box\right]\)
\end{mathematicaout}

The Liouville operator corresponding to the Hamiltonian is given by
(this might take longer than the previous examples)
\begin{mathematicain}
\(\mathcomm{\texttt{LiH}=-\texttt{LiouvilleCommutator}[h,\texttt{ham}]}\)
\end{mathematicain}
\begin{mathematicaout}
\(\texttt{SparseArray}[]\)
\end{mathematicaout}

In the following we set the Lindbladian parameters.
These are entered through the configuration variable \mathttt{transLi}
(this might take longer than the previous examples)
\begin{mathematicain}
\(\texttt{transLi}=\texttt{Join}
\left[\texttt{Table}
\big[\left\{\{1,2\},\{1,2\},\texttt{ii},\texttt{ii},\gamma _1,0\right\},
\{\texttt{ii},\texttt{na}\}\right],\)\\
\(\texttt{Table}\left[\left\{\{1,2\},\{1,2\},\texttt{ii},\texttt{ii}+1,\gamma_{1,2},0\right\},
\{\texttt{ii},\texttt{na}-1\}\right],\\
\texttt{Table}\left[\left\{\{1,2\},\{1,2\},\texttt{ii}+1,\texttt{ii},\gamma _{1,2},0\right\},\{\texttt{ii},\texttt{na}-1\}\right]\big]\)
\end{mathematicain}
\begin{mathematicaout}
\(\bigg\{
\left\{\{1,2\},\{1,2\},1,1,\gamma _1,0\right\},
\left\{\{1,2\},\{1,2\},2,2,\gamma_1,0\right\},
\left\{\{1,2\},\{1,2\},3,3,\gamma_1,0\right\},\)
\\ \hspace*{20pt}
\(\left\{\{1,2\},\{1,2\},4,4,\gamma_1,0\right\},
\left\{\{1,2\},\{1,2\},5,5,\gamma_1,0\right\},
\left\{\{1,2\},\{1,2\},1,2,\gamma_{1,2},0\right\},\)
\\ \hspace*{20pt}
\(\left\{\{1,2\},\{1,2\},2,3,\gamma_{1,2},0\right\},
\left\{\{1,2\},\{1,2\},3,4,\gamma_{1,2},0\right\},
\left\{\{1,2\},\{1,2\},4,5,\gamma_{1,2},0\right\},\)
\\ \hspace*{20pt}
\(\left\{\{1,2\},\{1,2\},2,1,\gamma_{1,2},0\right\},
\left\{\{1,2\},\{1,2\},3,2,\gamma_{1,2},0\right\},
\left\{\{1,2\},\{1,2\},4,3,\gamma_{1,2},0\right\},\)
\\ \hspace*{10pt}
\(\left\{\{1,2\},\{1,2\},5,4,\gamma_{1,2},0\right\}
\bigg\}\)
\end{mathematicaout}
The first five entries of \mathttt{transLi} list the decay channels between the states
$|$1$\rangle $ and $|$2$\rangle $ ($\{$1,2$\}$) for each atom. These
transitions occur internally within each atom, so the next two transition entries
are $(1, 1)$, $(2, 2)$, $(3, 3)$, $(4, 4)$ and $(5, 5)$.
The parameter \(\gamma_1\) corresponds to the dipole element of this transition.
The last eight terms give rise to the collective channels formed between the transitions
$|$1$\rangle $ $\rightarrow $ { }$|$2$\rangle $ between neighbouring atoms.
The parameter \(\gamma _{1,2}\) is the dipole element of this kind
of transition. The final $0$ in \mathttt{transLi} represents the energy
difference between the two transitions. In this case, all these transitions are in
fact the same, so the difference between the energy gaps is strictly zero.

Having defined \mathttt{transLi}, we are ready to compute the Liouville operator
corresponding to the Lindbladian
(this might take longer than the previous examples)
\begin{mathematicain}
\(\mathcomm{\texttt{LiL} = \texttt{LiouvilleLindbladian}[h,\texttt{transLi},\texttt{nl},\texttt{na}]}\)
\end{mathematicain}
\begin{mathematicaout}
\(\texttt{SparseArray}[]\)
\end{mathematicaout}

The overall Liouville operator is the sum of the Hamiltonian and Lindblad Liouville operators
\begin{mathematicain}
\(\mathcomm{\texttt{Li} =\texttt{LiH}+\texttt{LiL}}\)
\end{mathematicain}
\begin{mathematicaout}
\(\texttt{SparseArray}[]\)
\end{mathematicaout}

At this point we are ready to generate the system of differential
equations for the density matrix coefficients of the variable \mathttt{rho}
\begin{mathematicain}
\(\mathcomm{\texttt{eqs}
=\texttt{LiouvilleMasterEquation}[\texttt{rho},\texttt{Li},t]\texttt{//}\texttt{ExpToTrig};}\\
\mathcomm{\texttt{eqs}[[1\texttt{;;}2]]}\)
\end{mathematicain}
\begin{mathematicaout}
\(\bigg\{\rho _1'[t]==\gamma _1 \rho _2[t]
-\sqrt{2} \texttt{Sin}\left[4 \phi -t \Delta _1\right] R_1 \rho _3[t]
-\sqrt{2} \texttt{Cos}\left[4\phi -t \Delta _1\right] R_1 \rho _4[t]\)
\\ \hspace*{20pt}
\(+\gamma _1 \rho _5[t]
-\sqrt{2} \texttt{Sin}\left[3 \phi -t \Delta _1\right] R_1 \rho _9[t]
+\gamma _{1,2} \rho_{11}[t]-\sqrt{2} \texttt{Cos}\left[3 \phi -t \Delta _1\right] R_1 \rho _{13}[t]\)
\\ \hspace*{20pt}
\(+\gamma _{1,2} \rho _{16}[t]+\gamma _1 \rho _{17}[t]
-\sqrt{2} \texttt{Sin}\left[2\phi -t \Delta _1\right] R_1 \rho _{33}[t]
+\gamma _{1,2} \rho _{41}[t]\)
\\ \hspace*{20pt}
\(-\sqrt{2} \texttt{Cos}\left[2 \phi -t \Delta _1\right] R_1 \rho_{49}[t]
+\gamma_{1,2} \rho _{61}[t]+\gamma _1 \rho _{65}[t]
-\sqrt{2} \texttt{Sin}\left[\phi -t \Delta _1\right] R_1 \rho_{129}[t]\)
\\ \hspace*{20pt}
\(+\gamma _{1,2} \rho _{161}[t]
-\sqrt{2}\texttt{Cos}\left[\phi -t \Delta _1\right] R_1 \rho _{193}[t]
+\gamma _{1,2} \rho _{241}[t]+\gamma _1 \rho _{257}[t]\)
\\ \hspace*{20pt}
\(+\sqrt{2} \texttt{Sin}\left[t \Delta_1\right] R_1 \rho _{513}[t]
+\gamma _{1,2} \rho _{641}[t]-\sqrt{2} \texttt{Cos}\left[t \Delta _1\right] R_1 \rho _{769}[t]
+\gamma _{1,2} \rho _{961}[t],\)
\\ \hspace*{10pt}
\(\rho_2'[t]==-\gamma _1 \rho _2[t]+\sqrt{2} \texttt{Sin}\left[4 \phi -t \Delta _1\right] R_1 \rho _3[t]
+\sqrt{2} \texttt{Cos}\left[4 \phi -t \Delta _1\right]R_1 \rho _4[t]
\)
\\ \hspace*{20pt}
\(+\gamma _1 \rho _6[t]-\sqrt{2} \texttt{Sin}\left[3 \phi -t \Delta _1\right] R_1 \rho _{10}[t]
-\frac{1}{2} \gamma_{1,2} \rho _{11}[t]-\Omega_1 \rho _{12}[t]
\\ \hspace*{20pt}
-\sqrt{2} \texttt{Cos}\left[3 \phi -t \Delta _1\right] R_1 \rho _{14}[t]\)
\(+\Omega _1 \rho _{15}[t]-\frac{1}{2} \gamma _{1,2} \rho _{16}[t]+\gamma_1 \rho _{18}[t]
\\ \hspace*{20pt}
-\sqrt{2} \texttt{Sin}\left[2 \phi -t \Delta _1\right] R_1 \rho _{34}[t]
+\gamma _{1,2} \rho _{42}[t]\)
\(-\sqrt{2} \texttt{Cos}\left[2 \phi-t \Delta _1\right] R_1 \rho _{50}[t]
+\gamma _{1,2} \rho _{62}[t]
\\ \hspace*{20pt}
+\gamma _1 \rho _{66}[t]
-\sqrt{2} \texttt{Sin}\left[\phi -t \Delta _1\right] R_1 \rho_{130}[t]\)
\(+\gamma_{1,2} \rho_{162}[t]
-\sqrt{2} \texttt{Cos}\left[\phi -t \Delta _1\right] R_1 \rho_{194}[t]
\\ \hspace*{20pt}
+\gamma_{1,2} \rho_{242}[t]+\gamma _1\rho _{258}[t]\)
\(+\sqrt{2} \texttt{Sin}\left[t \Delta _1\right] R_1 \rho _{514}[t]
+\gamma _{1,2} \rho _{642}[t]
\\ \hspace*{20pt}
-\sqrt{2} \texttt{Cos}\left[t \Delta _1\right]R_1 \rho _{770}[t]
+\gamma _{1,2} \rho _{962}[t]\bigg\}\)
\end{mathematicaout}

For a numerical example, we set the physical parameters to
\begin{mathematicain}
\(\texttt{parmvals}=\left\{R_1\texttt{$\to$}12.0,
\Delta _1\texttt{$\to$}10.0,\phi \texttt{$\to$}1.1,
\gamma _1\texttt{$\to$}1.5,\gamma _{1,2}\texttt{$\to$}0.5,
\Omega_1\texttt{$\to$}0.7\right\}\\
\texttt{eqsn}=\texttt{eqs}\texttt{/.}\texttt{parmvals};\\
\texttt{eqsn}[[1\texttt{;;}2]]\)
\end{mathematicain}
\begin{mathematicaout}
\(\left\{R_1\to 12.,\Delta _1\to 10.,\phi \to 1.1,\gamma_1\to 1.5,\gamma_{1,2}\to 0.5,\Omega_1\to 0.7\right\}\)
\end{mathematicaout}
\begin{mathematicaout}
\(\bigg\{
\rho _1'[t]==1.5 \rho _2[t]-16.9706 \texttt{Sin}[4.4\, -10. t] \rho _3[t]
-16.9706 \texttt{Cos}[4.4\, -10. t] \rho _4[t]\)
\\ \hspace*{20pt}
\(+1.5 \rho _5[t]-16.9706\texttt{Sin}[3.3\, -10. t] \rho _9[t]+0.5 \rho _{11}[t]
\\ \hspace*{20pt}
-16.9706 \texttt{Cos}[3.3\, -10. t] \rho_{13}[t]\)
\(+0.5 \rho _{16}[t]+1.5 \rho _{17}[t]
\\ \hspace*{20pt}
-16.9706\texttt{Sin}[2.2\, -10. t] \rho _{33}[t]
+0.5 \rho _{41}[t]\)
\(-16.9706 \texttt{Cos}[2.2\, -10. t] \rho_{49}[t]
\\ \hspace*{20pt}
+0.5 \rho _{61}[t]+1.5 \rho _{65}[t]
-16.9706\texttt{Sin}[1.1\, -10. t] \rho_{129}[t]
+0.5 \rho_{161}[t]\)
\\ \hspace*{20pt}
\(-16.9706 \texttt{Cos}[1.1\, -10. t] \rho _{193}[t]+0.5 \rho _{241}[t]
+1.5 \rho _{257}[t]+0.5 \rho _{961}[t]\)
\\ \hspace*{20pt}
\(+16.9706\texttt{Sin}[10. t] \rho _{513}[t]
+0.5 \rho_{641}[t]-16.9706 \texttt{Cos}[10. t] \rho _{769}[t],\)
\\ \hspace*{10pt}
\(\rho_2'[t]==-1.5 \rho _2[t]+16.9706\texttt{Sin}[4.4\, -10. t] \rho _3[t]
+16.9706 \texttt{Cos}[4.4\, -10. t] \rho _4[t]\)
\\ \hspace*{20pt}
\(+1.5 \rho _6[t]-16.9706 \texttt{Sin}[3.3\, -10. t] \rho _{10}[t]-0.25\rho _{11}[t]-0.7 \rho _{12}[t]\)
\\ \hspace*{20pt}
\(-16.9706 \texttt{Cos}[3.3\, -10. t] \rho _{14}[t]+0.7 \rho _{15}[t]-0.25 \rho _{16}[t]
+1.5 \rho _{18}[t]\)
\\ \hspace*{20pt}
\(-16.9706 \texttt{Sin}[2.2\,-10. t] \rho _{34}[t]+0.5 \rho _{42}[t]
-16.9706 \texttt{Cos}[2.2\, -10. t] \rho _{50}[t]\)
\\ \hspace*{20pt}
\(+0.5 \rho _{62}[t]+1.5 \rho _{66}[t]
-16.9706 \texttt{Sin}[1.1\,-10. t] \rho _{130}[t]+0.5 \rho_{162}[t]
\\ \hspace*{20pt}
-16.9706 \texttt{Cos}[1.1\, -10. t] \rho _{194}[t]\)
\(+0.5 \rho _{242}[t]+1.5 \rho _{258}[t]
\\ \hspace*{20pt}
+16.9706 \texttt{Sin}[10.t] \rho _{514}[t]+0.5 \rho _{642}[t]\)
\(-16.9706 \texttt{Cos}[10. t] \rho_{770}[t]
+0.5 \rho _{962}[t]\bigg\}\)
\end{mathematicaout}
We can assume that the atoms are initially in the lowest energy level 
$|$1$\rangle $ . Therefore, we set the initial conditions to
\(\rho _0\)=\(\eta_0\)\(\otimes\)\(\eta _0\)\(\otimes\)\(\eta _0\)\(\otimes\)\(\eta _0\)\(\otimes\)\(\eta _0\)
where \(\eta _0\)=$|$0$\rangle \langle $0$|$
\begin{mathematicain}
\(\texttt{$\eta $0}=\texttt{SparseArray}[\{\{1,1\}\texttt{$\to$}1\},2]\\
\texttt{rho0}=\texttt{KroneckerProduct}[\texttt{$\eta $0},\texttt{$\eta $0},
\texttt{$\eta $0},\texttt{$\eta $0},\texttt{$\eta $0}]\\
\texttt{incon}=\texttt{InitialConditions}[h,\texttt{rho},\texttt{rho0},t];\\
\texttt{incon}[[1\texttt{;;}5]]\)
\end{mathematicain}
\begin{mathematicaout}
\(\texttt{SparseArray}\left[\Box\right]\)\\
\(\texttt{SparseArray}\left[\Box\right]\)\\
\(\left\{\rho _1[0]==1,\rho _2[0]==0,\rho _3[0]==0,\rho _4[0]==0,\rho _5[0]==0\right\}\)
\end{mathematicaout}
The final ODE is obtained by joining the differential equations and the initial conditions
\begin{mathematicain}
\(\texttt{diffeq}=\texttt{Join}[\texttt{eqsn},\texttt{incon}];\\
\texttt{diffeq}[[\{1,2,1025,1026\}]]\)
\end{mathematicain}
\begin{mathematicaout}
\(\bigg\{
\rho _1'[t]==1.5 \rho _2[t]-16.9706 \texttt{Sin}[4.4\, -10. t] \rho _3[t]
-16.9706 \texttt{Cos}[4.4\, -10. t] \rho _4[t]\\
\hspace*{20pt}+1.5 \rho _5[t]
-16.9706\texttt{Sin}[3.3\, -10. t] \rho _9[t]
+0.5 \rho _{11}[t]
\\ \hspace*{20pt}
-16.9706 \texttt{Cos}[3.3\, -10. t] \rho _{13}[t]+0.5 \rho _{16}[t]
+1.5 \rho _{17}[t]
\\ \hspace*{20pt}
-16.9706\texttt{Sin}[2.2\, -10. t] \rho _{33}[t]+0.5 \rho _{41}[t]-
16.9706 \texttt{Cos}[2.2\, -10. t] \rho _{49}[t]
\\ \hspace*{20pt}
+0.5 \rho _{61}[t]+1.5 \rho _{65}[t]
-16.9706\texttt{Sin}[1.1\, -10. t] \rho _{129}[t]+0.5 \rho _{161}[t]
\\ \hspace*{20pt}
-16.9706 \texttt{Cos}[1.1\, -10. t] \rho _{193}[t]+0.5 \rho _{241}[t]
+1.5 \rho _{257}[t]+0.5 \rho _{961}[t]
\\ \hspace*{20pt}
+16.9706\texttt{Sin}[10. t] \rho _{513}[t]
+0.5 \rho _{641}[t]
\\ \hspace*{20pt}
-16.9706 \texttt{Cos}[10. t] \rho _{769}[t]
,\\
\hspace*{10pt}\rho _2'[t]==-1.5 \rho _2[t]+16.9706\texttt{Sin}[4.4\, -10. t] \rho _3[t]
\\ \hspace*{20pt}
+16.9706 \texttt{Cos}[4.4\, -10. t] \rho _4[t]+1.5 \rho _6[t]
-16.9706 \texttt{Sin}[3.3\, -10. t] \rho _{10}[t]
\\ \hspace*{20pt}
-0.25\rho _{11}[t]-0.7 \rho _{12}[t]-16.9706 \texttt{Cos}[3.3\, -10. t] \rho _{14}[t]
+0.7 \rho _{15}[t]
\\ \hspace*{20pt}
-0.25 \rho _{16}[t]+1.5 \rho _{18}[t]
-16.9706 \texttt{Sin}[2.2\,-10. t] \rho _{34}[t]+0.5 \rho _{42}[t]
\\ \hspace*{20pt}
-16.9706 \texttt{Cos}[2.2\, -10. t] \rho _{50}[t]+0.5 \rho _{62}[t]+1.5 \rho _{66}[t]
\\ \hspace*{20pt}
-16.9706 \texttt{Sin}[1.1\,-10. t] \rho _{130}[t]
+0.5 \rho _{162}[t]
-16.9706 \texttt{Cos}[1.1\, -10. t] \rho _{194}[t]
\\ \hspace*{20pt}
+0.5 \rho _{242}[t]+1.5 \rho _{258}[t]
+16.9706 \texttt{Sin}[10.t] \rho _{514}[t]+0.5 \rho _{642}[t]
\\ \hspace*{20pt}
-16.9706 \texttt{Cos}[10. t] \rho _{770}[t]
+0.5 \rho _{962}[t],
\\ \hspace*{10pt}
\rho _1[0]==1,\rho _2[0]==0\bigg\}\)
\end{mathematicaout}
We can finally obtain the solution of the ODE as
\begin{mathematicain}
\(\texttt{sol}=\texttt{NDSolve}[\texttt{diffeq},\texttt{Normal}[\texttt{rho}],\{t,0,20\}][[1]];\\
\texttt{sol}[[1\texttt{;;}2]]\)
\end{mathematicain}
\begin{mathematicaout}
\(\bigg\{\rho _1[t]\to \texttt{InterpolatingFunction}\left[\Box\right][t],\\
\hspace*{20pt}
\rho _2[t]\to \texttt{InterpolatingFunction}\left[\Box\right][t]\bigg\}\)
\end{mathematicaout}

Now we can observe the dynamics of different expectation values or the density matrix coefficients.
For example, to compute the correlations $\langle
$\(\sigma _{2,1}^1\)\(\sigma _{1,2}^5\)+\(\sigma _{2,1}^5\)\(\sigma _{1,2}^1\)$\rangle $
and $\langle $\(\sigma _{2,1}^1\)\(\sigma _{1,2}^2\)+\(\sigma_{2,1}^2\)\(\sigma _{1,2}^1\)$\rangle$
where \(\sigma _{1,2}^1\) =1$\otimes $1$\otimes $1$\otimes $1$\otimes $ $|$1$\rangle \langle $2$|$,
\(\sigma_{1,2}^2\) = 1$\otimes $1$\otimes $1$\otimes |$1$\rangle \langle $2$|\otimes $1
and \(\sigma _{1,2}^5\) = $|$1$\rangle \langle $2$|\otimes $1$\otimes $1$\otimes $1$\otimes $1
\begin{mathematicain}
\(\texttt{$\sigma $12}=\texttt{SparseArray}[\{\{1,2\}\texttt{$\to$}1\},2];\\
\texttt{$\sigma $21}=\texttt{SparseArray}[\{\{2,1\}\texttt{$\to$}1\},2];\\
\texttt{one}=\texttt{SparseArray}[\{\{1,1\}\texttt{$\to$}1,\{2,2\}\texttt{$\to$}1\},2];\\
\texttt{corr1}=\texttt{ExpectationValue}[h,\texttt{rho},\\
\hspace*{20pt}\texttt{KroneckerProduct}[\texttt{$\sigma $12},
\texttt{one},\texttt{one},\texttt{one},\texttt{$\sigma$21}]\\
\hspace*{30pt}
+\texttt{KroneckerProduct}[\texttt{$\sigma $21},\texttt{one},\texttt{one},\texttt{one},\texttt{$\sigma $12}]]\\
\texttt{corr2}=\texttt{ExpectationValue}[h,\texttt{rho},\\
\hspace*{20pt}
\texttt{KroneckerProduct}[\texttt{one},\texttt{one},\texttt{one},\texttt{$\sigma $12},\texttt{$\sigma$21}]\\
\hspace*{30pt}
+\texttt{KroneckerProduct}[\texttt{one},\texttt{one},\texttt{one},\texttt{$\sigma $21},\texttt{$\sigma $12}]]\\
\texttt{corr1n}=\texttt{corr1}\texttt{/.}\texttt{sol};\\
\texttt{corr2n}=\texttt{corr2}\texttt{/.}\texttt{sol};\)
\end{mathematicain}
\begin{mathematicaout}
\(\rho _{515}[t]+\rho _{519}[t]+\rho _{531}[t]+\rho _{535}[t]+\rho _{579}[t]
+\rho _{583}[t]+\rho _{595}[t]+\rho _{599}[t]+\rho _{772}[t]
\\
\hspace*{20pt}
+\rho_{776}[t]+\rho _{788}[t]+\rho _{792}[t]+\rho _{836}[t]+\rho _{840}[t]+\rho _{852}[t]+\rho _{856}[t]\)\\

\(\rho_{11}[t]+\rho _{16}[t]+\rho _{27}[t]+\rho _{32}[t]+\rho _{75}[t]+\rho _{80}[t]
+\rho_{91}[t]+\rho _{96}[t]+\rho_{267}[t]+\rho_{272}[t]\\
\hspace*{20pt}+\rho_{283}[t]+\rho _{288}[t]
+\rho_{331}[t]+\rho_{336}[t]+\rho_{347}[t]+\rho _{352}[t]\)
\end{mathematicaout}
\begin{mathematicain}
\(\texttt{Plot}[\{\texttt{Re}[\texttt{corr1n}],\texttt{Re}[\texttt{corr2n}]\},\{t,0,10\},
\texttt{Axes}\texttt{$\to$}\texttt{False},\texttt{Frame}\texttt{$\to$}\texttt{True},\\
\texttt{FrameLabel}\texttt{$\to$}\{\texttt{Style}[\texttt{t}, 20],
\texttt{Style}[\texttt{{``}Correlations{''}},20]\},\texttt{PlotRange}\texttt{$\to$}\texttt{All}]\)
\end{mathematicain}

\includegraphics{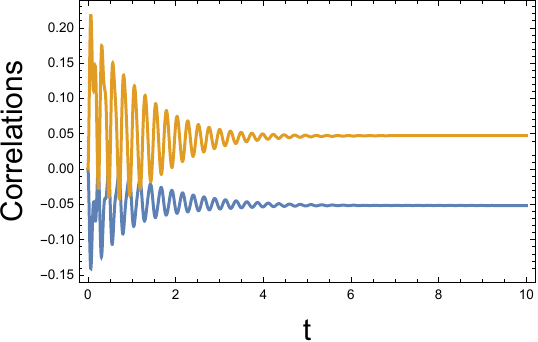}

With these ideas, we can also perform a spectroscopic analysis of the emission
of the two-level system by sweeping the detuning and evaluating the
correlations $\langle $\(\sigma _{2,1}^1\)\(\sigma _{1,2}^2\)
+\(\sigma _{2,1}^2\)\(\sigma _{1,2}^1\)$\rangle $
and $\langle $\(\sigma _{2,1}^1\)\(\sigma_{1,2}^5\)
+\(\sigma _{2,1}^5\)\(\sigma_{1,2}^1\)$\rangle$

\begin{mathematicain}
\(\texttt{nd}=150;\\
\texttt{$\eta $0}=\texttt{SparseArray}[\{\{1,1\}\texttt{$\to$}1\},2];\\
\texttt{rho0}=\texttt{KroneckerProduct}[\texttt{$\eta $0},
\texttt{$\eta $0},\texttt{$\eta $0},\texttt{$\eta $0},\texttt{$\eta $0}];\\
\texttt{incon}=\texttt{InitialConditions}[h,\texttt{rho},\texttt{rho0},t];\\
\texttt{$\sigma $12}=\texttt{SparseArray}[\{\{1,2\}\texttt{$\to$}1\},2];\\
\texttt{$\sigma $21}=\texttt{SparseArray}[\{\{2,1\}\texttt{$\to$}1\},2];\\
\texttt{corr1}=\texttt{ExpectationValue}[h,\texttt{rho},
\texttt{KroneckerProduct}[\texttt{$\sigma $12},\texttt{one},\texttt{one},\texttt{one},\texttt{$\sigma$21}]\\
\hspace*{60pt}
+\texttt{KroneckerProduct}[\texttt{$\sigma $21},\texttt{one},\texttt{one},\texttt{one},\texttt{$\sigma $12}]];\\
\texttt{corr2}=\texttt{ExpectationValue}[h,\texttt{rho},
\texttt{KroneckerProduct}[\texttt{one},\texttt{one},\texttt{one},\texttt{$\sigma $12},\texttt{$\sigma$21}]\\
\hspace*{60pt}
+\texttt{KroneckerProduct}[\texttt{one},\texttt{one},\texttt{one},\texttt{$\sigma $21},\texttt{$\sigma $12}]];\\
\texttt{listplot1}=\{\};\\
\texttt{listplot2}=\{\};\\
\texttt{Print}[\texttt{Dynamic}[\texttt{jj}],\texttt{{``}/{''}},\texttt{nd}]\\
\texttt{Do}[\\
\hspace*{20pt}\texttt{det} =-150+300\texttt{jj}/\texttt{nd};\\
\hspace*{20pt}\texttt{parmvals}=\left\{R_1\texttt{$\to$}12.0,\Delta _1\texttt{$\to$}\det ,
\phi \texttt{$\to$}1.1,\gamma _1\texttt{$\to$}1.5,\gamma _{1,2}\texttt{$\to$}0.5,
\Omega_1\texttt{$\to$}0.7\right\};\\
\hspace*{20pt}\texttt{eqsn}=\texttt{eqs}\texttt{/.}\texttt{parmvals};\\
\hspace*{20pt}\texttt{diffeq}=\texttt{Join}[\texttt{eqsn},\texttt{incon}];\\
\hspace*{20pt}\texttt{sol}=\texttt{NDSolve}[\texttt{diffeq},\texttt{Normal}[\texttt{rho}],\{t,0,10\}][[1]];\\
\hspace*{20pt}\texttt{corr1n}=\texttt{corr1}\texttt{/.}\texttt{sol}\texttt{/.}\{t\texttt{$\to$}10\};\\
\hspace*{20pt}\texttt{corr2n}=\texttt{corr2}\texttt{/.}\texttt{sol}\texttt{/.}\{t\texttt{$\to$}10\};\\
\hspace*{20pt}\texttt{AppendTo}[\texttt{listplot1},\{\det ,\texttt{corr1n}\}];\\
\hspace*{20pt}\texttt{AppendTo}[\texttt{listplot2},\{\det ,\texttt{corr2n}\}];\\
,\{\texttt{jj},0,\texttt{nd}\}]\\
\\
\texttt{ListPlot}[\{\texttt{listplot1},\texttt{listplot2}\},\texttt{Axes}\texttt{$\to$}\texttt{False},
\texttt{Frame}\texttt{$\to$}\texttt{True},\\
\left.\texttt{FrameLabel}\texttt{$\to$}\left\{\texttt{Style}\left[\texttt{"}\Delta _1\texttt{"},20\right],
\texttt{Style}\left[\texttt{"}\langle +\rangle
\texttt{"},20\right]\right\},\texttt{PlotRange}\texttt{$\to$}\texttt{All}\right]\)
\end{mathematicain}

\includegraphics{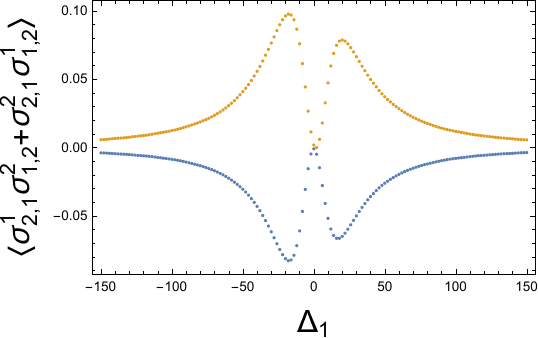}

It is possible to abbreviate the process of obtaining the differential equations by
using \mathttt{MasterEquation}, which is a wrapper for \mathttt{LiouvilleMasterEquation},
\mathttt{LiouvilleLindbladian}, \mathttt{LiouvilleCommutator}, and \mathttt{Ham}
\begin{mathematicain}
\noindent\(\mathcomm{\texttt{$\eta $0}=\texttt{SparseArray}[\{\{1,1\}\texttt{$\to$}1\},2]}\\
\mathcomm{\texttt{rho0}=\texttt{KroneckerProduct}[\texttt{$\eta $0},\texttt{$\eta $0},\texttt{$\eta $0},\texttt{$\eta $0},\texttt{$\eta $0}]}\\
\mathcomm{\texttt{eqs}=\texttt{MasterEquation}[h,\texttt{rho},\texttt{rho0},\texttt{transCE},\texttt{transdd},\texttt{transLi},\texttt{nl},\texttt{na},t];}\\
\mathcomm{\texttt{eqs}[[\{1,2,1025,1026\}]]}\)
\end{mathematicain}
\begin{mathematicaout}
\(\texttt{SparseArray}\left[\Box\right]\)\\
\(\texttt{SparseArray}\left[\Box\right]\)\\
\\
\(\bigg\{\rho _1'[t]==\gamma _1 \rho _2[t]-\sqrt{2} \texttt{Sin}\left[4 \phi -t \Delta _1\right] R_1 \rho _3[t]
-\sqrt{2} \texttt{Cos}\left[4\phi -t \Delta _1\right] R_1 \rho _4[t]+\gamma _1 \rho _5[t]
\\ \hspace*{20pt}
-\sqrt{2} \texttt{Sin}\left[3 \phi -t \Delta _1\right] R_1 \rho _9[t]
+\gamma _{1,2} \rho_{11}[t]-\sqrt{2} \texttt{Cos}\left[3 \phi -t \Delta _1\right] R_1 \rho _{13}[t]
+\gamma _{1,2} \rho _{16}[t]
\\ \hspace*{20pt}
+\gamma _1 \rho_{17}[t]
-\sqrt{2} \texttt{Sin}\left[2\phi -t \Delta _1\right] R_1 \rho _{33}[t]
+\gamma _{1,2} \rho _{41}[t]-\sqrt{2} \texttt{Cos}\left[2 \phi -t \Delta _1\right] R_1 \rho _{49}[t]
\\ \hspace*{20pt}
+\gamma_{1,2} \rho _{61}[t]+\gamma _1 \rho _{65}[t]
-\sqrt{2} \texttt{Sin}\left[\phi -t \Delta _1\right] R_1 \rho _{129}[t]+\gamma _{1,2} \rho _{161}[t]
\\ \hspace*{20pt}
-\sqrt{2}\texttt{Cos}\left[\phi -t \Delta _1\right] R_1 \rho _{193}[t]+\gamma _{1,2} \rho _{241}[t]
+\gamma _1 \rho _{257}[t]+\sqrt{2} \texttt{Sin}\left[t \Delta_1\right] R_1 \rho _{513}[t]
\\ \hspace*{20pt}
+\gamma _{1,2} \rho _{641}[t]
-\sqrt{2} \texttt{Cos}\left[t \Delta _1\right] R_1 \rho _{769}[t]
+\gamma _{1,2} \rho _{961}[t],
\\ \hspace*{10pt}
\rho_2'[t]==-\gamma _1 \rho _2[t]+\sqrt{2} \texttt{Sin}\left[4 \phi -t \Delta _1\right] R_1 \rho _3[t]
+\sqrt{2} \texttt{Cos}\left[4 \phi -t \Delta _1\right]R_1 \rho _4[t]+\gamma _1 \rho _6[t]
\\ \hspace*{20pt}
-\sqrt{2} \texttt{Sin}\left[3 \phi -t \Delta _1\right] R_1 \rho _{10}[t]
-\frac{1}{2} \gamma _{1,2} \rho _{11}[t]-\Omega_1 \rho _{12}[t]
-\sqrt{2} \texttt{Cos}\left[3 \phi -t \Delta _1\right] R_1 \rho _{14}[t]
\\ \hspace*{20pt}
+\Omega _1 \rho_{15}[t]
-\frac{1}{2} \gamma _{1,2} \rho _{16}[t]+\gamma_1 \rho _{18}[t]
-\sqrt{2} \texttt{Sin}\left[2 \phi -t \Delta _1\right] R_1 \rho _{34}[t]+\gamma _{1,2} \rho _{42}[t]
\\ \hspace*{20pt}
-\sqrt{2} \texttt{Cos}\left[2 \phi-t \Delta _1\right] R_1 \rho _{50}[t]+\gamma _{1,2} \rho _{62}[t]
+\gamma _1 \rho _{66}[t]-\sqrt{2} \texttt{Sin}\left[\phi -t \Delta _1\right] R_1 \rho_{130}[t]
\\ \hspace*{20pt}
+\gamma _{1,2} \rho _{162}[t]-\sqrt{2} \texttt{Cos}\left[\phi -t \Delta _1\right] R_1 \rho _{194}[t]
+\gamma _{1,2} \rho _{242}[t]+\gamma _1\rho _{258}[t]
\\ \hspace*{20pt}
+\sqrt{2} \texttt{Sin}\left[t \Delta _1\right] R_1 \rho _{514}[t]+\gamma _{1,2} \rho _{642}[t]
-\sqrt{2} \texttt{Cos}\left[t \Delta _1\right]R_1 \rho _{770}[t]
+\gamma _{1,2} \rho _{962}[t],
\\ \hspace*{10pt}
\rho _1[t]==1,\rho _2[t]==0\bigg\}\)
\end{mathematicaout}
Note that in this case the initial condition was set using \mathttt{rho0}
in matrix form rather than as a list of coefficients.

\subsection{Optical pumping}\label{sec:example4}
Optical pumping is the process by which resonant light redistributes
the population of atomic energy levels, creating a non-equilibrium
population distribution or spin polarization
\cite{Kastler:57,RevModPhys.44.169,happer2010optically}.
In this example, we set up the master equation for a $^{87}$Rb
atom in a four-wave-mixing configuration and examine
its behavior under optical pumping conditions.

We start by generating
the master equation for a single \(\, ^{87}\texttt{Rb}\)
atom, explicitly taking into account the orbitals
5 \(S_{1/2}\), F=2; 5 \(P_{3/2}\), F=3 and
5 \(D_{3/2}\), F=3. . As an initial step in this construction,
we obtain the configuration files produced by \mathttt{TransitionLists},
which provide the necessary information about the allowed
transitions among these levels.
\begin{mathematicain}
\(\texttt{na}=1;\\
\texttt{Jlist} = \left\{\frac{1}{2},\frac{3}{2},\frac{3}{2}\right\};\\
\texttt{Flist}=\{2,3,3\};\\
\texttt{mFlist}=\{\texttt{Range}[-2,2],\texttt{Range}[-3,3], \texttt{Range}[-3,3]\};\\
\texttt{levtrlist}=\{\{1,2\},\{2,3\}\};\\
\texttt{lasertrlist}=\left\{\left\{\{1,2\},
\left\{\frac{1}{\sqrt{2}},-\frac{i}{\sqrt{2}},0\right\}\right\},
\left\{\{2,3\},\left\{\frac{1}{\sqrt{2}},-\frac{i}{\sqrt{2}},0\right\}\right\}\right\};\\
\{\texttt{levels}, \texttt{transitions}, \texttt{transdd}, \texttt{transLi}, \texttt{transCE}\}\\
\hspace*{20pt}=\texttt{TransitionLists}\big[\frac{3}{2},\texttt{Jlist},
\texttt{Flist},\texttt{mFlist},\texttt{levtrlist},\\
\hspace*{80pt}\texttt{lasertrlist},\texttt{na}, R, \delta ,\phi
, \gamma , F, G\big];\)
\end{mathematicain}
At this point, it is helpful to inspect the variable \mathttt{levels},
as this provides a clearer understanding of how the atomic orbitals
are organized and how they will enter into the master equation.
\begin{mathematicain}
\(\texttt{levels}\)
\end{mathematicain}
\begin{mathematicaout}
\(\bigg\{
\big\{\left\{1,-2,\frac{1}{2},2\right\},
\left\{2,-1,\frac{1}{2},2\right\},
\left\{3,0,\frac{1}{2},2\right\},
\left\{4,1,\frac{1}{2},2\right\},
\left\{5,2,\frac{1}{2},2\right\}\big\},
\\ \hspace*{20pt}
\big\{\left\{6,-3,\frac{3}{2},3\right\},
\left\{7,-2,\frac{3}{2},3\right\},
\left\{8,-1,\frac{3}{2},3\right\},
\left\{9,0,\frac{3}{2},3\right\},
\left\{10,1,\frac{3}{2},3\right\},
\\ \hspace*{40pt}
\left\{11,2,\frac{3}{2},3\right\},
\left\{12,3,\frac{3}{2},3\right\}\big\},
\\ \hspace*{20pt}
\big\{\left\{13,-3,\frac{3}{2},3\right\},
\left\{14,-2,\frac{3}{2},3\right\},
\left\{15,-1,\frac{3}{2},3\right\},
\left\{16,0,\frac{3}{2},3\right\},
\left\{17,1,\frac{3}{2},3\right\},
\\ \hspace*{40pt}
\left\{18,2,\frac{3}{2},3\right\},
\left\{19,3,\frac{3}{2},3\right\}\big\}
\bigg\}\)
\end{mathematicaout}
The variable \mathttt{levels} is a three-level
nested list. It is organized into three groups corresponding
to the orbitals
5 \(S_{1/2}\), F=2; 5 \(P_{3/2}\), F=3 and
5 \(D_{3/2}\), F=3.
We can extract the the number of levels from
the variable \mathttt{levels}
\begin{mathematicain}
\(\texttt{nl}=\texttt{levels}[[-1,-1,1]]\)
\end{mathematicain}
\begin{mathematicaout}
\(19\)
\end{mathematicaout}

We now get the matrix basis, define the density matrix and set
the initial conditions so that
the 5 orbitals of 5 \(S_{1/2}\)F=2 are equally occupied
\begin{mathematicain}
\noindent\(h=\texttt{MultiAtomBasis}[\texttt{nl},\texttt{na}]\\
\texttt{n}=\texttt{Length}[h]\\
\texttt{rho}=\texttt{SparseArray}[\texttt{Table}[\texttt{ii}\texttt{$\to$}
\texttt{Subscript}[\rho ,\texttt{ii}][t],\{\texttt{ii},n\}]]\\
\texttt{rho0}=\texttt{SparseArray}
[\texttt{Table}[\{\texttt{ii},\texttt{ii}\}\texttt{$\to$}1/5,\{\texttt{ii},5\}],\{\texttt{nl},\texttt{nl}\}]\)
\end{mathematicain}
\begin{mathematicaout}
\(\texttt{SparseArray}\left[\Box\right]\)\\
\(361\)\\
\(\texttt{SparseArray}\left[\Box\right]\)\\
\(\texttt{SparseArray}\left[\Box\right]\)
\end{mathematicaout}
We use \mathttt{MasterEquation} to obtain the master equation and set the
system parameters in the variable \mathttt{consts}
\begin{mathematicain}
\(\texttt{diffeqs}=\texttt{MasterEquation}[h,\texttt{rho},\texttt{rho0},
\texttt{transCE},\texttt{transdd},\texttt{transLi},\texttt{nl},\texttt{na},t];\\
\texttt{consts}=\big\{\gamma _1\texttt{$\to$}0.07573,\gamma _2\texttt{$\to$}0.001343,
R_1\texttt{$\to$}0.1,R_2\texttt{$\to$}0.1,\delta _1\texttt{$\to$}0.0,
\\ \hspace*{20pt}
\delta_2\texttt{$\to$}0.0,\phi _{1,1}\texttt{$\to$}0,
\phi _{2,1}\texttt{$\to$}0\big\};\\
\texttt{diffeqsn}=\texttt{diffeqs}\texttt{/.}\texttt{consts};\)
\end{mathematicain}
The constants in consts correspond to \(^{87}\)Rb and are given in
units of ns$^{-1}$. Finally, we solve the system of differential
equations
\begin{mathematicain}
\(\texttt{tmax}=100000;\\
\texttt{sol}=\texttt{NDSolve}[\texttt{diffeqsn},
\texttt{Normal}[\texttt{rho}],\{t,0,\texttt{tmax}\}][[1]];\)
\end{mathematicain}
We plot the populations of 5 \(S_{1/2}\), F=2; 5 \(P_{3/2}\), F=3 and 5 \(D_{3/2}\)F=3,
confirming that optical pumping conditions are reached
\begin{mathematicain}
\(\texttt{groups}=\left\{\texttt{"}5^2S^{1/2}\texttt{ F=2$\texttt{"}$},
\texttt{$\texttt{"}$5 }\texttt{F=3$\texttt{"}$},
\texttt{$\texttt{"}$5}\texttt{F=3$\texttt{"}$}\right\};\\
\texttt{pl}=\{\};\\
\texttt{con}=0;\\
\texttt{Do}[\\
\hspace*{20pt}\texttt{con}\texttt{+=}1;\\
\hspace*{20pt}\texttt{pops}=\texttt{Table}[\texttt{Subscript}[\rho ,
\texttt{ii}][t],\{\texttt{ii},\texttt{lev}[[\texttt{;;},1]]\}]\texttt{/.}\texttt{sol};\\
\hspace*{20pt}\texttt{AppendTo}[\texttt{pl},\texttt{Plot}[\texttt{pops},\{t,0,\texttt{tmax}\},
\texttt{PlotRange}\texttt{$\to$}\texttt{All},\texttt{Frame}\texttt{$\to$}\texttt{True},\\
\hspace*{40pt}
\texttt{Axes}\texttt{$\to$}\texttt{False},\\
\hspace*{40pt}
\texttt{FrameLabel}\texttt{$\to$}\{\texttt{Style}[\texttt{{``}t (ns){''}},20],
\texttt{Style}[\texttt{groups}[[\texttt{con}]],20]\},\\
\hspace*{20pt}\texttt{PlotLabels}\texttt{$\to$}
\texttt{Table}[\texttt{StringJoin}[\texttt{{``}m={''}},
\texttt{ToString}[m]],\{m,\texttt{lev}[[\texttt{;;},2]]\}]]];\\
,\{\texttt{lev},\texttt{levels}\}];\\
\texttt{GraphicsColumn}[\texttt{pl}]\)
\end{mathematicain}

\includegraphics{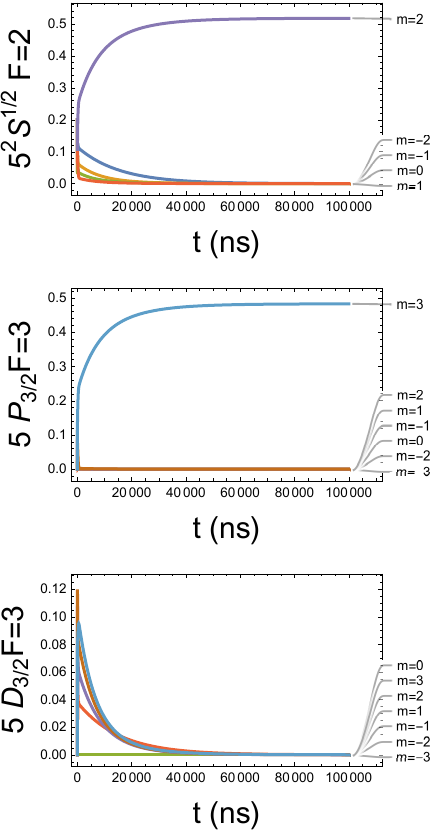}

To distinguish the dynamics of all the populations,
we repeat the previous example by slightly shifting the polarization vectors
We set the parameters and build the master equation
\begin{mathematicain}
\(\texttt{na}=1;\\
\texttt{Jlist} = \left\{\frac{1}{2},\frac{3}{2},\frac{3}{2}\right\};\\
\texttt{Flist}=\{2,3,3\};\\
\texttt{mFlist}=\{\texttt{Range}[-2,2],\texttt{Range}[-3,3], \texttt{Range}[-3,3]\};\\
\texttt{levtrlist}=\{\{1,2\},\{2,3\}\};\\
\\
\theta =0.5;\\
\texttt{lasertrlist}=\big\{
\left\{\{1,2\},\left\{\frac{1}{\sqrt{2}},-\frac{i}{\sqrt{2}},0\right\}\texttt{Cos}[\theta ]+\left\{\frac{1}{\sqrt{2}},
\frac{i}{\sqrt{2}},0\right\}\texttt{Sin}[\theta]\right\},\\
\hspace*{20pt}
\left\{\{2,3\},\left\{\frac{1}{\sqrt{2}},-\frac{i}{\sqrt{2}},0\right\}\texttt{Cos}[\theta ]
+\left\{\frac{1}{\sqrt{2}},\frac{i}{\sqrt{2}},0\right\}\texttt{Sin}[\theta]\right\}\big\};\\
\{\texttt{levels}, \texttt{transitions}, \texttt{transdd}, \texttt{transLi}, \texttt{transCE}\}\\
\hspace*{20pt}
=\texttt{TransitionLists}\big[\frac{3}{2},\texttt{Jlist},\texttt{Flist},\texttt{mFlist},\texttt{levtrlist},\\
\hspace*{40pt}\texttt{lasertrlist},\texttt{na}, R, \delta ,\phi, \gamma , F, G\big];\\
\\
h=\texttt{MultiAtomBasis}[\texttt{nl},\texttt{na}];\\
\texttt{n}=\texttt{Length}[h];\\
\texttt{rho}=\texttt{SparseArray}[\texttt{Table}[\texttt{ii}\texttt{$\to$}
\texttt{Subscript}[\rho ,\texttt{ii}][t],\{\texttt{ii},n\}]];\\
\texttt{rho0}=\texttt{SparseArray}
[\texttt{Table}[\{\texttt{ii},\texttt{ii}\}\texttt{$\to$}1/5,
\{\texttt{ii},5\}],\{\texttt{nl},\texttt{nl}\}];\\
\\
\texttt{diffeqs}=\texttt{MasterEquation}[h,\texttt{rho},\texttt{rho0},
\texttt{transCE},\texttt{transdd},\texttt{transLi},\texttt{nl},\texttt{na},t];\\
\texttt{consts}=\big\{\gamma _1\texttt{$\to$}0.07573,\gamma_2\texttt{$\to$}0.001343,
R_1\texttt{$\to$}0.1,R_2\texttt{$\to$}0.1,\delta _1\texttt{$\to$}0.0,\\
\hspace*{20pt}
\delta_2\texttt{$\to$}0.0,\phi _{1,1}\texttt{$\to$}0,
\phi _{2,1}\texttt{$\to$}0\big\};\\
\texttt{diffeqsn}=\texttt{diffeqs}\texttt{/.}\texttt{consts};\\
\\
\texttt{tmax}=100000;\\
\texttt{sol}=\texttt{NDSolve}[\texttt{diffeqsn},\texttt{Normal}[\texttt{rho}],\{t,0,\texttt{tmax}\}][[1]];\)
\end{mathematicain}
When plotting the populations of 5 \(S_{1/2}\), F=2; 5 \(P_{3/2}\), F=3 and 5 \(D_{3/2}\)F=3,
we confirm once again that the optical pumping conditions
are reached. However, orbitals that were unpopulated in the
previous example now acquire a small population
\begin{mathematicain}
\(\texttt{groups}=\left\{\texttt{"}5^2S^{1/2}\texttt{ F=2$\texttt{"}$},
\texttt{$\texttt{"}$5 }\texttt{F=3$\texttt{"}$},
\texttt{$\texttt{"}$5}\texttt{F=3$\texttt{"}$}\right\};\\
\texttt{pl}=\{\};\\
\texttt{con}=0;\\
\texttt{Do}[\\
\hspace*{20pt}\texttt{con}\texttt{+=}1;\\
\hspace*{20pt}\texttt{pops}=\texttt{Table}[\texttt{Subscript}[\rho ,\texttt{ii}][t],
\{\texttt{ii},\texttt{lev}[[\texttt{;;},1]]\}]\texttt{/.}\texttt{sol};\\
\hspace*{20pt}\texttt{AppendTo}[\texttt{pl},\texttt{Plot}[\texttt{pops},\{t,0,\texttt{tmax}\},
\texttt{PlotRange}\texttt{$\to$}\texttt{All},\texttt{Frame}\texttt{$\to$}\texttt{True},\\
\hspace*{40pt}
\texttt{Axes}\texttt{$\to$}\texttt{False},\\
\hspace*{40pt}
\texttt{FrameLabel}\texttt{$\to$}\{\texttt{Style}[\texttt{{``}t (ns){''}},20],
\texttt{Style}[\texttt{groups}[[\texttt{con}]],20]\},\\
\hspace*{40pt}
\texttt{PlotLabels}\texttt{$\to$}
\texttt{Table}[\\
\hspace*{60pt}
\texttt{StringJoin}[\texttt{{``}m={''}},
\texttt{ToString}[m]],\{m,\texttt{lev}[[\texttt{;;},2]]\}]]];\\
,\{\texttt{lev},\texttt{levels}\}];\\
\texttt{GraphicsColumn}[\texttt{pl}]\)
\end{mathematicain}

\includegraphics{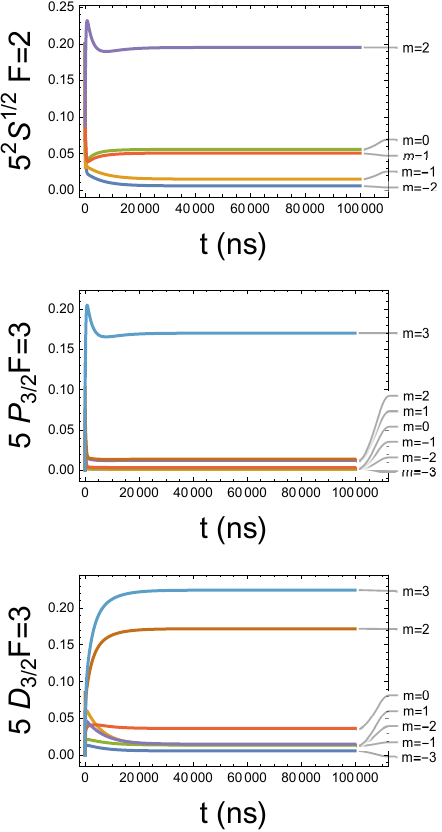}

\subsection{Superradiance}
We illustrate the phenomenon of superradiance by comparing the spontaneous
emission of a single atom with that of five coupled atoms.
Superradiance \cite{Dicke_1954, xcxr-sm9c} is a collective quantum effect where a group of emitters
radiate light much more intensely and faster than they would if each one emitted independently.

Load \mathttt{MulAtoLEG}
\begin{mathematicain}
\(\texttt{SetDirectory}[\texttt{PathToMultiAtomLiouvilleEquationGenerator}];\\
\texttt{Needs}[\texttt{{``}MultiAtomLiouvilleEquationGenerator$\grave{ }${''}}]\)
\end{mathematicain}
We do this example with five two-level atoms
\begin{mathematicain}
\(\texttt{na}=5;\\
\texttt{nl}=2;\\
n = \texttt{nl}^{\texttt{na}};\)
\end{mathematicain}
Set the matrix basis, density matrix and initial density matrix
\begin{mathematicain}
\(h=\texttt{MultiAtomBasis}[\texttt{nl},\texttt{na}];\\
\texttt{rho}=\texttt{Table}\left[\texttt{Subscript}[\rho ,\texttt{ii}][t],\left\{\texttt{ii},1,n^2\right\}\right];\\
\\
\texttt{$\eta $0}=\{\{0,0\},\{0,1\}\};\\
\texttt{rho0}=\texttt{KroneckerProduct}[\texttt{$\eta $0},\texttt{$\eta $0},\texttt{$\eta $0},\texttt{$\eta $0},\texttt{$\eta $0}];\)
\end{mathematicain}

The configuration files for the Hamiltonian and Lindbladian Liouville superoperators are defined such that all atoms are equally coupled. Two distinct
atoms are assigned coupling parameters M$\gamma $ and W$\gamma $ for the Hamiltonian and Lindbladian Liouville superoperators, respectively. No self-coupling
terms are included in the transdd part of the Hamiltonian. The configuration files are therefore specified as follows:
\begin{mathematicain}
\(\texttt{transCE}=\{\};\\
\texttt{transLi}=\{\};\\
\texttt{transdd}=\{\};\\
\texttt{Do}[\\
\texttt{If}[\alpha \texttt{==}\beta ,\\
\texttt{AppendTo}[\texttt{transLi},\{\{1,2\},\{1,2\},\alpha ,\beta ,\gamma ,0\}];,\\
\texttt{AppendTo}[\texttt{transLi},\{\{1,2\},\{1,2\},\alpha ,\beta ,W \gamma ,0\}];\\
\texttt{AppendTo}[\texttt{transdd},\{\{1,2\},\{1,2\},\alpha ,\beta ,M \gamma ,0\}];\\
];\\
,\{\alpha ,1,\texttt{na}\},\{\beta ,1,\texttt{na}\}];\)
\end{mathematicain}
At this point, we are ready to derive the master equation in terms of the differential equations for the density matrix coefficients (this might
take some time)
\begin{mathematicain}
\(\texttt{diffeqs}=\texttt{MasterEquation}[h,\texttt{rho},\texttt{rho0},\texttt{transCE},\texttt{transdd},\texttt{transLi},\texttt{nl},\texttt{na},t];\)
\end{mathematicain}
For comparison, we first calculate the emission from five independent atoms. 
To this end, we set the coupling parameters to zero, compute the numerical
form of the differential equations, and obtain the corresponding solution
\begin{mathematicain}
\(\texttt{consts}=\{\gamma \to 0.151458,W\to 0.0,M\texttt{$\to$}0.0\};\\
\texttt{diffeqsn}=\texttt{diffeqs}\texttt{/.}\texttt{consts};\\
\\
\texttt{tmin}=0;\\
\texttt{tmax}=15;\\
\texttt{sol}=\texttt{NDSolve}[\texttt{diffeqsn},\texttt{rho},\{t,\texttt{tmin},\texttt{tmax}\}][[1]];\)
\end{mathematicain}
The field of the emitted light contains the contributions from the five atoms
\begin{equation}
 E  \propto \sum _{\alpha =1}^5 \sigma _{1,\alpha }
 \end{equation}
where $\sigma_{1,\alpha }$ denotes the atomic lowering operator 
$\left\vert 1\right\rangle _{\alpha } {}_{\alpha }\left\langle 2 \right\vert$ for atom $\alpha$,
hence $E^{\dagger}E$, which is proportional to the emission power, is given by
\begin{mathematicain}
\(\texttt{EtE}=\texttt{Sum}[\texttt{SigmaT}[\{1,2\},\alpha ,\texttt{nl},\texttt{na}].\texttt{Sigma}[\{1,2\},\beta ,\texttt{nl},\texttt{na}],\{\alpha
,\texttt{na}\},\{\beta ,\texttt{na}\}];\)
\end{mathematicain}
Using the solution we compute $\left\langle E^{\dagger }E\right\rangle$,
 the expected value of $E^{\dagger }E$
\begin{mathematicain}
\(\texttt{evi}=\texttt{Re}[\texttt{ExpectationValue}[h,\texttt{rho},\texttt{EtE}]]\texttt{/.}\texttt{sol};\)
\end{mathematicain}
calculate a list of 50 points of time vs. emission power and fit it to \(I_0\exp (-t/\tau )\) to obtain the decay time $\tau $ 
\begin{mathematicain}
\(\texttt{nd}=50;\\
\texttt{datai}=\texttt{Table}[\{\texttt{tmin}+(\texttt{tmax}-\texttt{tmin})\texttt{ii}/\texttt{nd},\\
\hspace*{20pt}
\texttt{evi}\texttt{/.}\{t\texttt{$\to$}\texttt{tmin}+(\texttt{tmax}-\texttt{tmin})\texttt{ii}/\texttt{nd}\}\},
\{\texttt{ii},0,\texttt{nd}\}];\\
\texttt{fit}=\texttt{FindFit}[\texttt{datai},\texttt{I0} \texttt{Exp}[- t/\tau ],\{\{\texttt{I0},1\},\{\tau ,5\}\},t]\\
\texttt{funi}=\texttt{I0} \texttt{Exp}[- t/\tau ]\texttt{/.}\texttt{fit};\)
\end{mathematicain}
\begin{mathematicaout}
\(\{\texttt{I0}\to 5.,\tau \to 6.60249\}\)
\end{mathematicaout}
Note that $\tau  = 1/\gamma = 6.60249$ 
\begin{mathematicain}
\(1/\gamma \texttt{/.}\texttt{consts}\)
\end{mathematicain}
\begin{mathematicaout}
\(6.60249\)
\end{mathematicaout}
This is expected, since the atoms are independent.
We plot the fitted function, the data points and the
numerical solution for the expected value
\begin{mathematicain}
\(\texttt{Show}[\{\\
\hspace*{20pt}\texttt{Plot}[\texttt{funi},\{t,\texttt{tmin},\texttt{tmax}\},
\texttt{PlotStyle}\texttt{$\to$}\{\texttt{Thickness}[0.02],\texttt{Opacity}[0.3]\},\\
\hspace*{40pt}\texttt{Axes}\texttt{$\to$}\texttt{False},\texttt{Frame}\texttt{$\to$}\texttt{True},\\
\hspace*{40pt}\left.\texttt{FrameLabel}\texttt{$\to$}\left\{\texttt{Style}[\texttt{{``}t (ns){''}},20],
\texttt{Style}\left[\texttt{"}\langle E^{\dagger }\texttt{E$\rangle
\texttt{"}$},20\right]\right\}\right],\\
\hspace*{20pt}\texttt{Plot}[\texttt{evi},\{t,\texttt{tmin},\texttt{tmax}\}],\\
\hspace*{20pt}\texttt{ListPlot}[\texttt{datai},\texttt{PlotStyle}\texttt{$\to$}\{\texttt{Opacity}[0.7],\texttt{PointSize}[0.02]\}]\\
\}]\)
\end{mathematicain}

\includegraphics{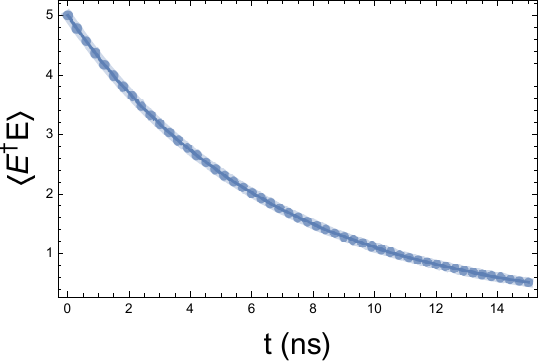}

\noindent Now we calculate the emission of five strongly coupled atoms.
To this end, we set the coupling parameters $W$ and $M$ to $0.752$ and $0.45$, compute the
numerical form of the differential equations, and obtain the corresponding solution

\begin{mathematicain}
\(\texttt{consts}=\{\gamma \to 0.151458,W\to 0.752,M\texttt{$\to$}0.45\};\\
\texttt{diffeqsn}=\texttt{diffeqs}\texttt{/.}\texttt{consts};\\
\\
\texttt{sol}=\texttt{NDSolve}[\texttt{diffeqsn},\texttt{rho},\{t,\texttt{tmin},\texttt{tmax}\}][[1]];\)
\end{mathematicain}
Using the solution, we compute $\left\langle E^{\dagger }E\right\rangle $,
the expectation value of $E^{\dagger }E$, which was obtained previously
\begin{mathematicain}
\(\texttt{evd}=\texttt{Re}[\texttt{ExpectationValue}[h,\texttt{rho},\texttt{EtE}]]\texttt{/.}\texttt{sol};\)
\end{mathematicain}
We then calculate a list of 60 points of emission power as a function of time and fit it to
$I_0\exp (-t/\tau )$ in order to extract the decay
time $\tau $ 
\begin{mathematicain}
\(\texttt{nd}=60;\\
\texttt{datad}=\texttt{Table}[\{\texttt{tmin}+(\texttt{tmax}-\texttt{tmin})\texttt{ii}/\texttt{nd},\\
\hspace*{20pt}\texttt{evd}\texttt{/.}\{t\texttt{$\to$}\texttt{tmin}+(\texttt{tmax}-\texttt{tmin})\texttt{ii}/\texttt{nd}\}\},
\{\texttt{ii},0,\texttt{nd}\}];\\
\text{fit}=\texttt{FindFit}[\texttt{datad}[[15\texttt{;;}]],\texttt{I0} \texttt{Exp}[- t/\tau ],\{\{\texttt{I0},1\},\{\tau ,5\}\},t]\\
\texttt{fund}=\texttt{I0} \texttt{Exp}[- t/\tau ]\texttt{/.}\texttt{fit};\)
\end{mathematicain}
\begin{mathematicaout}
\(\{\texttt{I0}\to 17.3229,\tau \to 2.90696\}\)
\end{mathematicaout}
In this case, we used only the last 45 points, since the first 15 do not exhibit an exponential
decay. Note that the decay time for coupled atoms is shorter.
We plot the fitted function together with the data points and the numerical solution for the expectation value
\begin{mathematicain}
\(\texttt{Show}[\{\\
\hspace*{20pt}\texttt{Plot}[\texttt{fund},\{t,\texttt{tmin},\texttt{tmax}\},
\texttt{PlotStyle}\texttt{$\to$}\{\texttt{Thickness}[0.02],\texttt{Opacity}[0.3]\},\\
\hspace*{40pt}\texttt{FrameLabel}\texttt{$\to$}\left\{\texttt{Style}[\texttt{{``}t (ns){''}},20],
\texttt{Style}\left[\texttt{"}\langle E^{\dagger }\texttt{E$\rangle\texttt{"}$},20\right]\right\},\\
\hspace*{40pt}\texttt{Axes}\texttt{$\to$}\texttt{False},\texttt{Frame}\texttt{$\to$}\texttt{True},
\texttt{PlotRange}\texttt{$\to$}\texttt{All}],\\
\hspace*{20pt}\texttt{Plot}[\texttt{evd},\{t,\texttt{tmin},\texttt{tmax}\},\texttt{PlotRange}\texttt{$\to$}\texttt{All}],\\
\hspace*{20pt}
\texttt{ListPlot}[\texttt{datad},\texttt{PlotStyle}\texttt{$\to$}\{\texttt{Opacity}[0.7],\texttt{PointSize}[0.02]\}]\\
\}]\)
\end{mathematicain}

\includegraphics{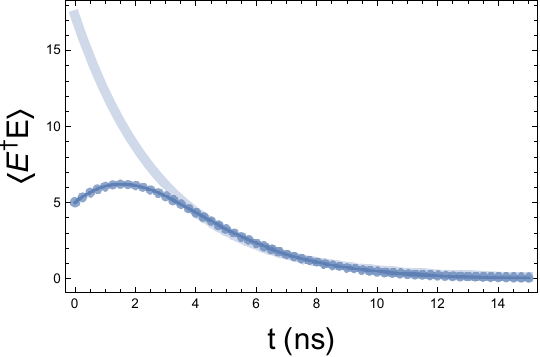}\\

\noindent The emission curve exhibits the typical characteristics of superradiance:
it starts at the same value as the emission from independent atoms, then
rises to a maximum above the initial value, and finally decays twice
as fast as in the independent case. Finally, both examples are plotted together
for comparison
\begin{mathematicain}
\(\texttt{Show}[\{ \\
\hspace*{20pt}\texttt{Plot}[\{\texttt{funi},\texttt{fund}\},\{t,\texttt{tmin},\texttt{tmax}\},\\
\hspace*{40pt}\texttt{PlotStyle}\texttt{$\to$}\{\{\texttt{Thickness}[0.02],\texttt{Opacity}[0.3]\},\\
\hspace*{60pt}\{\texttt{Thickness}[0.02],\texttt{Opacity}[0.3]\}\},\\
\hspace*{40pt}\texttt{Axes}\texttt{$\to$}\texttt{False},\texttt{Frame}\texttt{$\to$}\texttt{True},\\
\hspace*{40pt}\texttt{FrameLabel}\texttt{$\to$}\left\{\texttt{Style}[\texttt{{``}t (ns){''}},20],\texttt{Style}\left[\texttt{"}\langle
E^{\dagger }\texttt{E$\rangle \texttt{"}$},20\right]\right\}],\\
\hspace*{20pt}\texttt{Plot}[\{\texttt{evi},\texttt{evd}\},\{t,\texttt{tmin},\texttt{tmax}\},\texttt{PlotRange}\texttt{$\to$}\texttt{All}],\\
\hspace*{20pt}\texttt{ListPlot}[\{\texttt{datai},\texttt{datad}\},\\
\hspace*{40pt}\texttt{PlotStyle}\texttt{$\to$}\{\{\texttt{Opacity}[0.7],
\texttt{PointSize}[0.02]\},\\
\hspace*{60pt}\{\texttt{Opacity}[0.7],\texttt{PointSize}[0.02]\}\}]\}]\)
\end{mathematicain}

\includegraphics{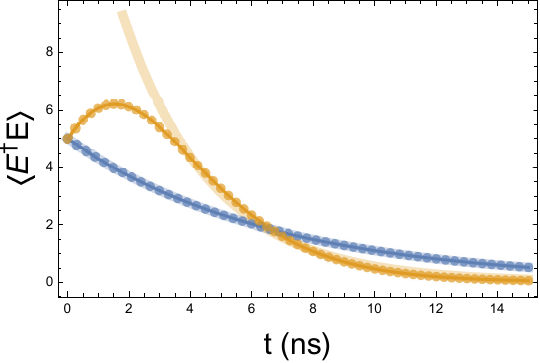}

\subsection{The dressed state evolution operator}\label{sec:example5}
In this example we illustrate how to transform the solution of the master equation from
the dressed-state basis to the standard-state basis for a system of two three-level atoms.
This method provides a way to obtain the explicit form of the evolution operator.
The resulting solution is shown to be consistent with that obtained directly
from the master equation in the standard basis.

We calculate the evolution operator for a system consisting of two
($na=2$) three level ($nl=3$) atoms subject to two lasers with Rabi
parameters \(R_1\) and \(R_2\) and detuning parameters \(\Delta _1\)
and \(\Delta _2\), respectively.
The first laser couples the states $|$1$\rangle $ and $|$2$\rangle$
with a Rabi parameter \(R_1\) and the second couples $|$2$\rangle $
and $|$3$\rangle $ with a Rabi parameter \(R_2\) for both atoms.
Both atoms are at different positions along the light-propagation
direction and therefore have a phase difference $\phi $.
First we set the numerical values of the system parameters
\begin{mathematicain}
\(\texttt{parmvals}=\big\{R_1\texttt{$\to$}12.0,R_2\texttt{$\to$}8.0,\phi \texttt{$\to$}0,
\Delta_1\texttt{$\to$}1.0,\Delta _2\texttt{$\to$}-1.0,\gamma_1\texttt{$\to$}1.5,
\gamma_2\texttt{$\to$}5.0,\\
\hspace*{20pt}
\gamma_{1,1}\texttt{$\to$}0.3,\gamma _{2,2}\texttt{$\to$}0.1,\Omega _1\texttt{$\to$}0.7,
\Omega_2\texttt{$\to$}0.43\big\}\)
\end{mathematicain}
\begin{mathematicaout}
\(\big\{R_1\to 12.,R_2\to 8.,\phi \to 0,\Delta _1\to 1.,\Delta _2\to -1.,\gamma _1\to 1.5,
\gamma _2\to 5.,\gamma _{1,1}\to 0.3,\\
\hspace*{20pt}
\gamma _{2,2}\to0.1,\Omega _1\to 0.7,\Omega _2\to 0.43\big\}\)
\end{mathematicaout}

This is an important step because, although the evolution operator can be obtained symbolically
for very sparse Liouville operators, in most cases it must be calculated numerically.
The \mathttt{transCE} configuration variable, which describes this light-matter interaction, is
\begin{mathematicain}
\(\texttt{transCE}=
\big\{\left\{ \{1,2\},1, R_1, \Delta _1,0\right\},\left\{\{1,2\},2, R_1, \Delta _1,0\right\},
\left\{\{2,3\},1, R_2, \Delta _2,\phi \right\},\\
\hspace*{20pt}\left\{\{2,3\},2, R_2, \Delta _2,\phi \right\}\big\}\texttt{/.}\texttt{parmvals}\)
\end{mathematicain}
\begin{mathematicaout}
\(\{\{\{1,2\},1,12.,1.,0\},\{\{1,2\},2,12.,1.,0\},\{\{2,3\},1,8.,-1.,0\},\{\{2,3\},2,8.,-1.,0\}\}\)
\end{mathematicaout}

Assuming that the transitions $|$1$\rangle $ $\rightarrow $ $|$2$\rangle $ and
$|$2$\rangle $ $\rightarrow $ $|$3$\rangle $ have distinct energy
gaps and therefore do not interact with each other, the transdd parameter is given by
\begin{mathematicain}
\(\texttt{transdd}=
\big\{\left\{\{1,2\},\{1,2\},1,2,\Omega _1,0\right\},
\left\{\{2,3\}, \{2,3\},1,2,\Omega _2,0\right\},\\
\hspace*{20pt}\left\{\{1,2\},\{1,2\},2,1,\Omega _1,0\right\},
\left\{\{2,3\}, \{2,3\},2,1,\Omega _2,0\right\}\big\}\texttt{/.}\texttt{parmvals}\)
\end{mathematicain}
\begin{mathematicaout}
\(\{\{\{1,2\},\{1,2\},1,2,0.7,0\},\{\{2,3\},\{2,3\},1,2,0.43,0\},\\
\hspace*{20pt}
\{\{1,2\},\{1,2\},2,1,0.7,0\},\{\{2,3\},\{2,3\},2,1,0.43,0\}\}\)
\end{mathematicaout}

The only transitions with non-vanishing dipole matrix elements are
$|$1$\rangle $ $\rightarrow |$2$\rangle $ and
$|$2$\rangle $ $\rightarrow|$3$\rangle $.
The transition $|$1$\rangle $ $\rightarrow |$2$\rangle $ is labeled as transition
1, and the transition { }$|$2$\rangle $ $\rightarrow |$3$\rangle $ is labeled as
transition 2. The strengths of the corresponding dipole couplings are denoted
accordingly: \(\gamma _1\) is the coupling
due to the dipole element of the transition 1 if a single atom is involved,
\(\gamma _{1,1}\) is the coupling due to the dipole element of the transition
1 when two different atoms are involved, { }\(\gamma _2\) is the coupling due
to the dipole element of the transition 2 if a single atom is involved
and \(\gamma _{2,2}\) is the coupling due to the dipole element of the
transition 2 when two different atoms are involved. The transLi configuration
variable is thus given by
\begin{mathematicain}
\(\texttt{transLi}=
\big\{\left\{\{1,2\},\{1,2\},1,1,\gamma_1,0\right\},
\left\{\{1,2\},\{1,2\},2,2,\gamma _1,0\right\},\\
\hspace*{20pt}\left\{\{1,2\},\{1,2\},1,2,\gamma_{1,1},0\right\},
\left\{\{1,2\},\{1,2\},2,1,\gamma _{1,1},0\right\},
\left\{\{2,3\},\{2,3\},1,1,\gamma _2,0\right\},\\
\hspace*{20pt}
\left\{\{2,3\},\{2,3\},2,2,\gamma_2,0\right\},
\left\{\{1,2\},\{1,2\},1,2,\gamma_{2,2},0\right\},\\
\hspace*{20pt}
\left\{\{1,2\},\{1,2\},2,1,\gamma_{2,2},0\right\}\big\}\texttt{/.}\texttt{parmvals}\)
\end{mathematicain}
\begin{mathematicaout}
\(\{\{\{1,2\},\{1,2\},1,1,1.5,0\},
\{\{1,2\},\{1,2\},2,2,1.5,0\},
\{\{1,2\},\{1,2\},1,2,0.3,0\},\\
\hspace*{20pt}\{\{1,2\},\{1,2\},2,1,0.3,0\},
\{\{2,3\},\{2,3\},1,1,5.,0\},
\{\{2,3\},\{2,3\},2,2,5.,0\},\\
\hspace*{20pt}
\{\{1,2\},\{1,2\},1,2,0.1,0\},
\{\{1,2\},\{1,2\},2,1,0.1,0\}\}\)
\end{mathematicaout}
\begin{mathematicain}
\(\texttt{nl}= 3;\\
\texttt{na} =2;\\
\texttt{rho}=\texttt{SparseArray}
\left[\texttt{Table}\left[\{\texttt{ii}\}\texttt{$\to$}\rho_{\texttt{ii}}[t],
\left\{\texttt{ii},\texttt{nl}^{2\texttt{na}}\right\}\right]\right]\\
h = \texttt{MultiAtomBasis}[\texttt{nl},\texttt{na}]\)
\end{mathematicain}
\begin{mathematicaout}
\(\texttt{SparseArray}\left[\Box\right]\)\\
\(\texttt{SparseArray}\left[\Box\right]\)
\end{mathematicaout}

We calculate the dressed state Hamiltonian and the corresponding Liouville operator
\begin{mathematicain}
\(\texttt{hamd}=\texttt{HamDressed}[\texttt{transCE}, \texttt{transdd}, \texttt{nl}, \texttt{na}];\\
\texttt{LiouvilleHR} = -\texttt{CleanUp}[\texttt{LiouvilleCommutator}[h,\texttt{hamd}]];\)
\end{mathematicain}
We calculate the Lindbladian superoperator
\begin{mathematicain}
\(\texttt{LiouvilleL}=\texttt{CleanUp}
[\texttt{LiouvilleLindbladian}[h, \texttt{transLi},\texttt{nl},\texttt{na}]];\)
\end{mathematicain}
It is important to remember that the Lindbladian superoperator remains invariant under
rotating-frame transformations and, therefore, is calculated
just as it would be in the standard basis.
The full Liouville operator is the sum of \mathttt{LiouvilleHR}
and \mathttt{LiouvilleL}
\begin{mathematicain}
\(\texttt{LiouvilleR}=\texttt{Re}
[\texttt{Developer$\grave{ }$ToPackedArray}@(\texttt{LiouvilleHR}+\texttt{LiouvilleL})];\)\\
\(\texttt{VRD}=\texttt{MatrixExp}[\texttt{LiouvilleR}\, t];\)
\end{mathematicain}
Now we obtain the rotating-frame transformations
\begin{mathematicain}
\(\{U, \texttt{UT},\texttt{UR}, \texttt{URT}\} 
= \texttt{RotatingFrameTrans}[h,\texttt{transCE},\texttt{nl}, \texttt{na}, t]\)
\end{mathematicain}
\begin{mathematicaout}
\(\left\{\texttt{SparseArray}\left[\Box\right],\texttt{SparseArray}\left[\Box\right],\texttt{SparseArray}[],\texttt{SparseArray}[]\right\}\)
\end{mathematicaout}
and use them to convert the results from the rotating frame to the standard basis
\begin{mathematicain}
\(\texttt{$\rho $0}=\texttt{SparseArray}[\{1,1\}\texttt{$\to$}1,9];\\
\texttt{rhot}=\texttt{URT}.\texttt{VRD}.\texttt{UR}.\texttt{Proj}[h,\texttt{$\rho $0}];\)
\end{mathematicain}
To illustrate the procedure for calculating expectation values, we construct the operator
\texttt{op} = $|$1$\rangle \langle $2$|\otimes $1+ $|$2$\rangle \langle $1$|\otimes $1
and then compute its expectation value
\begin{mathematicain}
\(\texttt{op}=\texttt{KroneckerProduct}[\texttt{SparseArray}[\{1,2\}\texttt{$\to$}1,3],\texttt{IdentityMatrix}[3]]\\
\hspace*{20pt}+\texttt{KroneckerProduct}[\texttt{SparseArray}[\{2,1\}\texttt{$\to$}1,3],\texttt{IdentityMatrix}[3]];\\
\texttt{rexpval} = \texttt{ExpectationValue}[h,\texttt{rhot},\texttt{op}];\)
\end{mathematicain}
To compare, we also find the system's evolution through the master equation 
\begin{mathematicain}
\(\texttt{eqs}
=\texttt{MasterEquation}[h,\texttt{rho},\texttt{rho0},\texttt{transCE},
\texttt{transdd}, \texttt{transLi},\texttt{nl},\texttt{na},t];\)
\end{mathematicain}

numerically solve the differential equations
\begin{mathematicain}
\(\texttt{sol}=\texttt{NDSolve}[\texttt{deqs},\texttt{Normal}[\texttt{rho}],\{t,0,10.0\}];\)
\end{mathematicain}
and find the expectation value of op using this solution
\begin{mathematicain}
\(\texttt{expval}=\texttt{ExpectationValue}[h,\texttt{rho},\texttt{op}]\texttt{/.}\texttt{sol};\)
\end{mathematicain}
To verify that both solutions match, we
superimpose the solutions obtained from the dressed state and the standard base
\begin{mathematicain}
\(\texttt{Plot}[\{\texttt{Re}[\texttt{rexpval}],\texttt{Re}[\texttt{expval}]\},\{t,0,5\},
\texttt{PlotRange}\texttt{$\to$}\texttt{All},\texttt{Axes}\texttt{$\to$}\texttt{False},\\
\hspace*{20pt}\texttt{Frame}\texttt{$\to$}\texttt{True},\\
\hspace*{20pt}\texttt{FrameLabel}\texttt{$\to$}\{\texttt{Style}[\texttt{{``}t (ns){''}},20],
\texttt{Style}[\texttt{{``}Expectation value{''}},20]\},\\
\hspace*{20pt}\texttt{PlotStyle}
\texttt{$\to$}\{\texttt{Directive}[\texttt{Thickness}[0.02],\texttt{LightBlue}],\\
\hspace*{40pt}
\texttt{Directive}[\texttt{Thickness}[0.005],\texttt{Blue}]\}]\)
\end{mathematicain}

\includegraphics{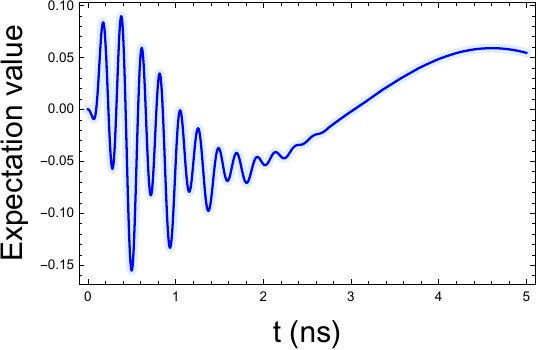}

\subsection{Time evolution of transmons}\label{sec:example6}

To illustrate the use of 
\hspace{20pt}\mathttt{LiouvilleLindbladianFromLindbladOperators}
\hspace{20pt}
and 
\mathttt{LiouvilleDaggerLindbladianFromLindbladOperators} in obtaining the dynamics
of an arbitrary system, we analyze the time evolution of a transmon
subject to both a coherent drive and dissipative mechanisms. 

The Hamiltonian of a transmon qubit is given by
\begin{equation}
H = \hbar \omega  \left(a^{\dagger }a + \frac{1}{2}\right)
+\frac{\alpha }{12}\left(a^{\dagger}+ a\right)^4+\lambda(t) (a^{\dagger }+ a),
\end{equation}
where $a^{\dagger}$ and $a$ are the usual ascent and descent operators,
$\omega $ is the transmon frequency, $\alpha $ is the anharmonicity parameter
and $\lambda(t)$ represents the drive. We assume that the transmon is coupled
to a thermal bath, which is modeled by the Lindbladian
\begin{equation}
L = -\frac{\gamma }{2}\left(\left\{a^{\dagger }a, \rho \right\}
-2 a \rho  a^{\dagger}\right).
\end{equation}
To build the Hamiltonian and Lindbladian we define the ascent and descent operators
\begin{mathematicain}
\(\texttt{aop}[\texttt{ns$\_$}]\texttt{:=}
\texttt{SparseArray}\left[\texttt{Table}\left[\{n+1,n\}\texttt{$\to$}\sqrt{n}
,\{n,1,\texttt{ns}-1\}\right],\texttt{ns}\right];\\
\texttt{dop}[\texttt{ns$\_$}]\texttt{:=}
\texttt{SparseArray}\left[\texttt{Table}\left[\{n+1,n+2\}\texttt{$\to$}\sqrt{n+1}
,\{n,0,\texttt{ns}-2\}\right],\texttt{ns}\right];\)
\end{mathematicain}
where \mathttt{ns} is the number of states. We analyze the behaviour of the first
8 states of a transmon. We thus start by generating a matrix basis with 8
states with
\begin{mathematicain}
\(h=\texttt{MultiAtomBasis}[8,1]\)
\end{mathematicain}
\begin{mathematicaout}
\(\texttt{SparseArray}\left[\Box\right]\)
\end{mathematicaout}
We also define the ascent and descent operators for such a number of states
and a \mathttt{SparseArray} called one for the unitary matrix
\begin{mathematicain}
\(\texttt{ns}=8;\\
\texttt{a}=\texttt{dop}[\texttt{ns}];\\
\texttt{at}=\texttt{aop}[\texttt{ns}];\\
\texttt{one}=\texttt{SparseArray}[\texttt{Table}
[\{\texttt{ii},\texttt{ii}\}\texttt{$\to$}1,\{\texttt{ii},1,\texttt{ns}\}]];\)
\end{mathematicain}
The Hamiltonian is thus given by
\begin{mathematicain}
\( H=\hbar  \omega  \left(\texttt{at}.\texttt{a}+\frac{1}{2} \texttt{one}\right)
-\frac{\alpha }{12}(\texttt{a}+\texttt{at}).(\texttt{a}+\texttt{at})
.(\texttt{a}+\texttt{at}).(\texttt{a}+\texttt{at})
+\lambda  (\texttt{a}+\texttt{at});\)
\end{mathematicain}
We set the constants of the system
\begin{mathematicain}
\(\texttt{consts}=\{\hbar \texttt{$\to$}1.0,\omega \texttt{$\to$}4.5,\alpha \texttt{$\to$}-0.30,
\lambda \texttt{$\to$}0\}\)
\end{mathematicain}
\begin{mathematicaout}
\(\{\hbar \to 1.,\omega \to 4.5,\alpha \to -0.3,\lambda \to 0\}\)
\end{mathematicaout}
and calculate the eigenvalues of $H$ in order to know the excitation energy
$\Delta $ between the ground and the first excited states
\begin{mathematicain}
\(\texttt{Hn}=\texttt{Normal}[H]\texttt{/.}\texttt{consts};\\
\texttt{eval}=\texttt{Eigenvalues}[\texttt{Hn}]\\
\Delta =\texttt{eval}[[7]]-\texttt{eval}[[8]]\)
\end{mathematicain}
\begin{mathematicaout}
\(\{37.2165,35.0278,28.4791,22.6551,17.3162,12.0971,7.08815,2.31999\}\)\\
\(4.76816\)
\end{mathematicaout}
We calculate the Liouville operator for the Hamiltonian
\begin{mathematicain}
\(\texttt{LiH}=\texttt{CleanUp}[-\texttt{LiouvilleCommutator}[h,H]]\)
\end{mathematicain}
\begin{mathematicaout}
\(\texttt{SparseArray}\left[\Box\right]\)
\end{mathematicaout}
The Lindbladian Liouville operator is calculated from just one Lindblad operator,
namely $a$, and just one decay coefficient $\gamma $
\begin{mathematicain}
\(\texttt{liops} = \texttt{SparseArray}[\{a\}]\\
\texttt{gammas} = \texttt{SparseArray}[\{\gamma \}]\\
\texttt{LiL} =\texttt{LiouvilleLindbladianFromLindbladOperators}[h,\texttt{liops}, \texttt{gammas}]\)
\end{mathematicain}
\begin{mathematicaout}
\(\texttt{SparseArray}\left[\Box\right]\)\\
\(\texttt{SparseArray}\left[\Box\right]\)\\
\(\texttt{SparseArray}\left[\Box\right]\)
\end{mathematicaout}
We compute the overall Liouville operator and build the differential equations: 
\begin{mathematicain}
\noindent\(\mathcomm{\texttt{Li}=\texttt{CleanUp}[\texttt{LiH}+\texttt{LiL}]}\)
\end{mathematicain}
\begin{mathematicaout}
\(\texttt{SparseArray}\left[\Box\right]\)
\end{mathematicaout}
Using this operator, we then build the system of differential equations for the density matrix coefficients
\begin{mathematicain}
\(\texttt{rho}=\texttt{Table}\left[\rho _{\texttt{ii}}[t],\{\texttt{ii},\texttt{Length}[h]\}\right];\\
\texttt{eqs}=\texttt{LiouvilleMasterEquation}[\texttt{rho}, \texttt{Li}, t];\)
\end{mathematicain}
We reset the \mathttt{consts} variable to include the decay coefficient $\gamma$ and a
drive with frequency $\Delta$ that excites the transition from the ground
state to the first excited state. We then evaluate the equations with these
updated constants and set the initial conditions such that the transmon
starts in the ground state. Finally, the initial conditions are joined to the equations.
\begin{mathematicain}
\(\texttt{consts}=\{\hbar \texttt{$\to$}1.0,\omega \texttt{$\to$}4.5,
\alpha \texttt{$\to$}-0.30,\lambda \texttt{$\to$}0.1 \texttt{Sin}[\Delta  t],
\gamma\texttt{$\to$}0.02\}\\
\texttt{eqsn}=\texttt{eqs}\texttt{/.}\texttt{consts};\\
\texttt{inconds}=\texttt{Table}\left[\rho _{\texttt{ii}}[0]\texttt{==}\texttt{KroneckerDelta}[\texttt{ii},1],\{\texttt{ii},\texttt{Length}[h]\}\right];\\
\texttt{difeqs}=\texttt{Join}[\texttt{eqsn},\texttt{inconds}];\)
\end{mathematicain}
\begin{mathematicaout}
\(\{\hbar \to 1.,\omega \to 4.5,\alpha \to -0.3,\lambda \to 0.1 \texttt{Sin}[4.76816 t],\gamma \to 0.02\}\)
\end{mathematicaout}
Having established both the constants and the initial state,
the next step is to solve the system of differential equations
that govern the transmon’s dynamics.
\begin{mathematicain}
\(\texttt{sol}=\texttt{NDSolve}[\texttt{difeqs},\texttt{rho},\{t,0,100\}][[1]];\)
\end{mathematicain}
The numerical solution allows us to extract the populations
of all eight transmon levels, which we represent graphically
to illustrate their dynamical behavior
\begin{mathematicain}
\(\texttt{plots}=\{\};\\
\texttt{Do}[\\
\hspace*{20pt}\texttt{op}
=\texttt{SparseArray}[\{\{\texttt{ii},\texttt{ii}\}\texttt{$\to$}1\},\texttt{ns}];\\
\hspace*{20pt}\texttt{expval}=\texttt{ExpectationValue}[h,\texttt{rho},\texttt{op}];\\
\hspace*{20pt}\texttt{expvaln}=\texttt{expval}\texttt{/.}\texttt{sol};\\
\hspace*{20pt}\texttt{AppendTo}[\texttt{plots},\texttt{expvaln}];\\
,\{\texttt{ii},1,8\}]\)
\end{mathematicain}

\begin{mathematicain}
\(\texttt{Plot}[\texttt{plots},\{t,0,32\},\texttt{PlotRange}\texttt{$\to$}\texttt{All},
\texttt{Axes}\texttt{$\to$}\texttt{False},\texttt{Frame}\texttt{$\to$}\texttt{True},\\
\hspace*{20pt}\texttt{FrameLabel}\texttt{$\to$}\{\texttt{Style}[\texttt{{``}t (ns){''}},20],
\texttt{Style}[\texttt{{``}Populations{''}},20]\},\\
\hspace*{40pt}
\texttt{PlotLabels}\texttt{$\to$}\texttt{Table}\left[\texttt{Style}\left[\rho_{\texttt{ii}},16\right],
\{\texttt{ii},1,8\}\right],\texttt{ImageSize}\texttt{$\to$}450]\)
\end{mathematicain}
\includegraphics{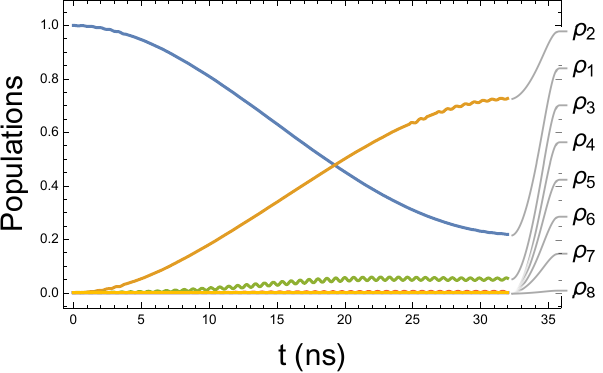}

\noindent We observe that as the ground state (blue) is gradually depopulated,
the first excited state (orange) becomes populated due to the driving field.
In addition, a small fraction of the population leaks into the higher excited
states (green and subsequent levels), reflecting the multi-level structure
of the transmon and the fact that the drive is not perfectly restricted to
a two-level subspace.

\end{appendix}






\bibliography{main}

\end{document}